\documentclass[
   aps,prl,twocolumn,
   reprint,
   superscriptaddress,
   floatfix,
   amssymb,
  longbibliography
]{revtex4-1}

\usepackage{amsmath}
\usepackage{amsfonts}
\usepackage{graphicx}
\usepackage[utf8]{inputenc}
\usepackage{capt-of} % Used to include Fig3
\usepackage{nicefrac} % Default
\usepackage[normalem]{ulem}
\usepackage[dvipsnames]{xcolor}
\definecolor{linkColor}{rgb}{0,0.3,0.7}
\usepackage{hyperref}
\hypersetup{colorlinks=true,
           allcolors=linkColor,
           pdfborder={0 0 0},
           pdfencoding=auto}

\definecolor{brickred}{rgb}{0.8, 0.25, 0.33}

\usepackage{color, colortbl}

\begin{document}

% Title
\title{Anomalous collective dynamics of auto-chemotactic populations}

% Authors
\author{Jasper van der Kolk} 
\thanks{JK, FR and RS (alphabetically) contributed equally.}
\author{Florian Raßhofer}
\thanks{JK, FR and RS (alphabetically) contributed equally.}
\author{Richard Swiderski}
\thanks{JK, FR and RS (alphabetically) contributed equally.}
\affiliation{Arnold Sommerfeld Center for Theoretical Physics and Center for NanoScience, Department of Physics, Ludwig-Maximilians-Universit\"at M\"unchen, Theresienstra\ss e 37, D-80333 Munich, Germany}
\author{Astik Haldar}
\author{Abhik Basu}
\affiliation{Theory Division, Saha Institute of Nuclear Physics, HBNI, 1/AF Bidhannagar, Calcutta 700 064, West Bengal, India}
\author{Erwin Frey}
\email[Corresponding author: ]{frey@lmu.de}
\affiliation{Arnold Sommerfeld Center for Theoretical Physics and Center for NanoScience, Department of Physics, Ludwig-Maximilians-Universit\"at M\"unchen, Theresienstra\ss e 37, D-80333 Munich, Germany}
\affiliation{Max Planck School Matter to Life, Hofgartenstraße 8, 80539 Munich, Germany}

% Abstract
\begin{abstract}
    While the role of local interactions in nonequilibrium phase transitions is well studied, a fundamental understanding of the effects of long-range interactions is lacking. 
    We study the critical dynamics of reproducing agents subject to auto-chemotactic interactions and limited resources.
    A renormalization group analysis reveals distinct scaling regimes for fast (attractive or repulsive) interactions; for slow signal transduction the dynamics is dominated by a diffusive fixed point.
    Further, we present a correction to the Keller-Segel nonlinearity emerging close to the extinction threshold and a novel nonlinear mechanism that stabilizes the continuous transition against the emergence of a characteristic length scale due to a chemotactic collapse.
\end{abstract}
\maketitle

% Introduction
Nonequilibrium phase transitions encompass a broad class of systems, including absorbing-state phase transitions~\cite{Hinrichsen2000, JanssenTauber2005}, roughening transitions~\cite{Halpin1995, Kardar1986}, and ordering transitions in active matter~\cite{Ramaswamy2010, Marchetti2013}. 
Most theoretical studies of these paradigmatic model systems focus on the role of local interactions.
However, in addition to short-ranged interactions, several biological and synthetic systems exhibit many-body long-range interactions between agents \cite{Ziepke2022}. 
For example, the social amoeba \textit{Dictyostelium discoideum} uses chemical signaling and chemotaxis to control aggregation under harsh conditions~\cite{Parent1999}, signaling molecules mediate intercellular communication in microbial populations~\cite{Bauer2017}, and microrobots and robotic fish use infrared, electrical, and acoustic signals to communicate~\cite{Katzschmann2018}.

Studying long-ranged interactions has a longstanding history in the context of equilibrium continuous phase transitions \cite{Fisher1972, Frey1994b, Truong1999}. 
Their nonequilibrium counterparts are, however, less well explored. 
Most attention has been paid to systems where the long-rangedness results from L\'evy-flight-like motion, nonlocal effects due to an underlying network architecture or spatially-dependent reaction rates \cite{Janssen1999b, Hinrichsen2007, Lyra2013, Fontanari2016}. 
There, the additional interactions may lead to a new universality class \cite{Janssen1999b, Lyra2013} or change the nature of the phase transition \cite{Fontanari2016}. 
Here, we are interested in the role of long-range chemical signaling on classical models of population dynamics.

For this purpose, we consider agents emitting a signal in the form of a chemical substance which spreads by diffusion and can be sensed by other agents that respond by adapting their direction of motion, a process known as \textit{chemotaxis}.
The dynamics of such populations has been analyzed in terms of drift-diffusion models for the agent density coupled to a chemical field, termed \textit{Keller-Segel} (KS) models~\cite{KELLER1971225, Hillen2008, Tindall2008}. 
These studies have identified a plethora of different phenomena -- ranging from aggregation~\cite{Jaeger1992, Herrero1997} to the formation of complex patterns~\cite{Tyson1999, Tello2007, Jin2016, Hillen2008}.
While the role of thermal fluctuations~\cite{Chavanis2008, Grima2004} and fluctuations around a constant background density~\cite{Gelimson2015, Mahdistoliani2021} have been investigated, the role of large-scale demographic noise – which is particularly important close to the extinction threshold ~\cite{Hinrichsen2000, JanssenTauber2005, Tauber2014}  – remains largely unexplored.
In this letter, we investigate how long-ranged chemical signaling affects the collective behaviour of a population consisting of a single type of reproducing agents close to extinction.

% The Model.
We consider a generic model of a population of diffusing cells (agents) and chemicals in terms of two fluctuating density fields $\rho({\bf x},t)$ and $c({\bf x},t)$.
The population dynamics is assumed to follow logistic growth, i.e., cells proliferate at a rate~${\mu}$, die at a rate~${\lambda}$, and resource availability limits population growth to a finite carrying capacity.
In addition, we consider the effect of an \textit{auto-chemotactic interaction},  where each cell is capable of responding to a chemical signal, while simultaneously sourcing it with strength~$\alpha$\@~\cite{Budrene1991,Tweedy2016}.
We are interested in an effective, hydrodynamic description of this system, valid on macroscopic scales and in the presence of demographic noise.
The corresponding Langevin equations are
\begin{align}
    \frac{\mathrm{d}\rho}{\mathrm{d}t}
    &= 
    (D_\rho\boldsymbol{\nabla}^2 +\theta) \, \rho - \gamma \, \rho^2 + \sqrt{2\Lambda \rho} \, \xi  + I[\rho, \boldsymbol\nabla c] \, ,
    \label{eq:LEC}
    \\
    \frac{\mathrm{d}c}{\mathrm{d}t}
    &= (D_{c}\boldsymbol{\nabla}^2-\lambda_c) \, c+\alpha \, \rho  
    \,,
    \label{eq:LEPhi}
\end{align}
where~${\theta = \mu -\lambda}$ is the net growth rate,~$\theta/\gamma$ the carrying capacity,~${\lambda_c}$ the degradation rate of the signaling molecules, and~${D_{\rho,c}}$ are the diffusion constants.
The macroscopically relevant noise is multiplicative with amplitude~${2\Lambda \rho(\boldsymbol{x},t)}$ and Gaussian white noise~${\xi (\boldsymbol{x},t)}$.
Higher order nonlinearities and other noise terms are irrelevant close to the absorbing state~\cite{Note:SM}.
Without the additional interaction $I[\rho, \boldsymbol\nabla c]$, Eq.\@~\eqref{eq:LEC}  corresponds to the noisy \textit{Fisher-Kolmogorov} equation~\cite{FISHER1937, Kolmogorov}, whose universal properties fall into the universality class of \emph{directed percolation} (DP)\@~\cite{Hinrichsen2000, JanssenTauber2005}.

The interaction term $I[\rho, \boldsymbol\nabla c]$ – which we assume to only depend on gradients in $c$ \cite{Note:SM} – accounts for the directed motion of cells along chemical gradients.
Its form not only depends on cellular details but also on the level of coarse graining.
In particular, the absence of global mass-conservation in the population dynamics allows for a nonconservative effective interaction.

At mean-field level, the dynamics exhibits two length scales, a diffusion length of the agents ${\xi_\rho=\sqrt{D_\rho/|\theta|}}$ and of the chemicals ${\xi_c=\sqrt{D_c/\lambda_c}}$.
The latter is linked to the interaction range, since $\lambda_c$ inhibits signal transduction over long distances.
For long-ranged chemotactic interactions (${\xi_c \rightarrow \infty}$, see Appendix A)~\cite{Note:SM}, the only relevant scale is ${\xi_\rho}$.
Below this scale, demographic processes only play a minor role and the chemotactic interaction can be formulated in terms of a conserved current ${I[\rho, \boldsymbol\nabla c] = \boldsymbol\nabla\boldsymbol J}$, where ${\boldsymbol J = \chi[\rho, \boldsymbol\nabla c]\rho\boldsymbol\nabla c}$ and the sensitivity function ${\chi[\rho, \boldsymbol\nabla c]}$ encodes details of the sensing process~\cite{Segel1977, Painter2002, Hillen2008}.

However, close to the extinction threshold the system is dominated by a divergent correlation length ${\xi>\xi_\rho}$ and strongly enhanced fluctuations.
Further, coarse graining to large scales inevitably `mixes' the effects of chemotaxis and birth-death processes.
Whereas the net production by the linear birth-death term is independent of the density distribution, the net degradation due to the growth limiting term is enhanced by density fluctuations.
Thus, the evolution of the total mass is coupled to the chemotactic interaction by the interplay of resource limitation and chemotactic drift, which alters the dynamics of density fluctuations.
Therefore, an explicit coarse graining procedure is needed to determine all the relevant contributions.
This is achieved by a \textit{renormalization group} (RG) analysis~(see \cite{Note:SM}), which reveals that close to the extinction threshold the effective chemotactic interaction, correctly accounting for birth-death processes, is given by
\begin{equation}
    I[\rho, \boldsymbol\nabla c] = \chi_1^{}\boldsymbol\nabla(\rho\boldsymbol\nabla c) + (\chi_2^{}-\chi_1^{})\rho\boldsymbol\nabla^2 c \, .\label{eq:ChemoInteraction}
\end{equation}
It consists of a conservative interaction -- the classical \emph{Keller-Segel} (KS)~\cite{Segel1977} nonlinearity -- and an additional nonconservative term. A dimensional analysis shows that all other contributions are irrelevant at the pertinent length scales~\cite{Note:SM}.
Importantly, Eq.\@~\eqref{eq:ChemoInteraction} does not imply that the chemotactic interaction explicitly breaks particle number conservation.
Rather it accounts for the fact that close to the extinction threshold the interplay between strong density fluctuations, chemotactic drift and population dynamics require an \textit{effective} description of the form \eqref{eq:ChemoInteraction}.
Conversely, if fluctuation corrections are weak, i.e., far away from the extinction threshold, a conserved current yields the proper description.

% Mean-field analysis.
% Figure 1
\begin{figure}
    \includegraphics{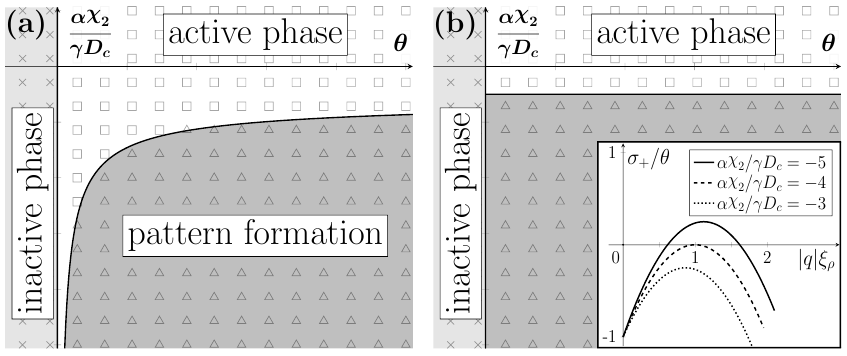}
    \caption{Mean-field phase diagrams of the auto-chemotactic model for~${\lambda_c=0.1}$ (a) and~${\lambda_c=0}$ (b) with an inactive phase (light grey), an active phase (white), and a pattern formation regime (dark grey). 
    The boundary between the active phase and the pattern formation regime is given by Eq.\@~\eqref{eq:InstabilityCon}.
    The states obtained from finite element simulations in one spatial dimension for~${D_\rho=0.1}$,~${\gamma=D_c=1}$,~${\alpha=5}$,~${\theta \in [-0.2, 1]}$ and~${\chi_1=\chi_2 \in [-2, 0.4]}$ are marked by crosses (absorbing state), squares (active state) and triangles (inhomogeneous), respectively.
    The inset shows the largest eigenvalue of the system at~${\lambda_c = \theta}$ and ${D_c=D_\rho}$.}
    \label{fig:PhaseDiagrams}
\end{figure}
To analyze Eqs.~\eqref{eq:LEC} and~\eqref{eq:LEPhi}, we first neglect the noise term and study the resulting mean-field equations.
They yield two homogeneous stationary solutions: the \emph{absorbing state}~${\rho_0 = c_0=0}$ corresponding to the \emph{inactive phase} and a state corresponding to the \textit{active phase} with the agent density equal to the carrying capacity: $\rho_1 = \theta/\gamma$ and~${c_1=\alpha\rho_1 / \lambda_c}$.
From a linear stability analysis of these homogeneous states one infers that there are three distinct phases (Fig.\@~\ref{fig:PhaseDiagrams}).
For~${\theta<0}$, only the absorbing state is stable. 
In contrast, the homogeneous active state is stable for~${\theta>0}$ and 
\begin{align}
    \chi_{2}^{} 
    > 
    -\frac{\gamma D_c}{\alpha } 
    \left( 1+\frac{\lambda_c D_\rho}{\theta D_c}+2\sqrt{\frac{\lambda_c D_\rho}{\theta D_c}}
    \right) 
    \, .
    \label{eq:InstabilityCon}
\end{align}
In the case of~${\theta>0}$ and~${\chi_{2}^{}}$ below this threshold, however, both homogeneous solutions are unstable against spatial perturbations.
This Turing-type~\cite{Turing1952} instability indicates the onset of pattern formation~\cite{Tello2007, Jin2016}, as explicitly confirmed by numerical simulations shown in Fig.\@~\ref{fig:PhaseDiagrams}.

% Critical dynamics.
At~${\theta = 0 }$ one finds a transcritical bifurcation, indicating a continuous, absorbing-state phase transition with~${\theta}$ acting as the control parameter.
Close to the extinction threshold (${\theta \rightarrow 0}$) and for long-ranged interactions (${\xi_c \rightarrow \infty}$) the system becomes intrinsically scale invariant. 
In particular, the correlation length of density fluctuations should diverge as~${\xi \propto \theta^{-\nu}}$, and for a cell cluster emerging from a single seed, its mean-squared radius and survival probability at criticality should scale as~${\langle R^2 \rangle (t) \propto t^{2/z}}$ and~${P(t) \propto t^{-\delta}}$, respectively~\cite{Hinrichsen2000,JanssenTauber2005}.
The mean-field critical exponents are given by~${\nu = 0.5}$,~${z=2}$ and~${\delta = d/4}$.
By dimensional analysis one identifies the following effective parameters:
\begin{align}
	u =  \frac{\gamma\Lambda}{32 \pi^2D_\rho^2} ,
	\ \,
	g_{1,2} = \frac{\alpha \chi_{1,2}^{} \Lambda}{32 \pi^2 D_\rho^2D_c} ,
	\ \,
	w  = \frac{D_c}{D_c+D_\rho}
	\label{eq:effCoupl}
\end{align}
In addition to the DP coupling~$u$ (representing resource limitation) two new chemotactic couplings~$g_1$ and~$g_2$ emerge.
The parameter~${w}$ measures the time delay in the chemotactic interaction due to the finite diffusion speed of the signaling molecules.
Employing field theoretical RG and a systematic perturbation expansion around the upper critical dimension~${d_c=4}$, we derive the flow equations~\cite{Note:SM}
\begin{subequations}
\begin{align}
	\mu \, \frac{\mathrm{d}u}{\mathrm{d}\mu}
	&= -\epsilon u + f_1(u,g_{1,2}, w)
	\, ,
	\label{eq:flowu} 
	\\
	\mu \, \frac{\mathrm{d}g_{1,2}}{\mathrm{d}\mu}
	&= -\epsilon g_{1,2} + f_{2,3}(u,g_{1,2}, w)
	\, , \label{eq:flowg} 
	\\
	\mu \, \frac{\mathrm{d}w}{\mathrm{d}\mu}
	&= -w(1-w)f_4(u,g_{1,2}, w)
	\, . \label{eq:floww}
\end{align}
\end{subequations}
The flow functions~${f_1}$--${f_4}$ contain all information about the dependence of the theory on the arbitrary momentum scale~${\mu}$ 
in~${d=4-\varepsilon}$ dimensions.
Scale invariance is implied by the existence of IR-stable~(${\mu \rightarrow 0}$ stable) fixed points~\cite{Tauber2014}.

In contrast to previous studies~\cite{Gelimson2015},
all calculations are performed by approaching the phase transition from the inactive phase, the full dynamics of the chemical concentration field are taken into account, and the limiting case of DP is correctly recovered.

% Quasi-static limit.
%
Inspecting Eq.\@~\eqref{eq:floww}, one observes that~${w=1}$ is an invariant manifold of the RG flow.
Moreover, systems where~${w\lesssim1}$ only slowly evolve away from this hyperplane.
Therefore, we first focus on this \emph{quasi-static} limit of infinitely fast diffusing chemicals \cite{Note:SM}.

\begin{figure}
  	\includegraphics{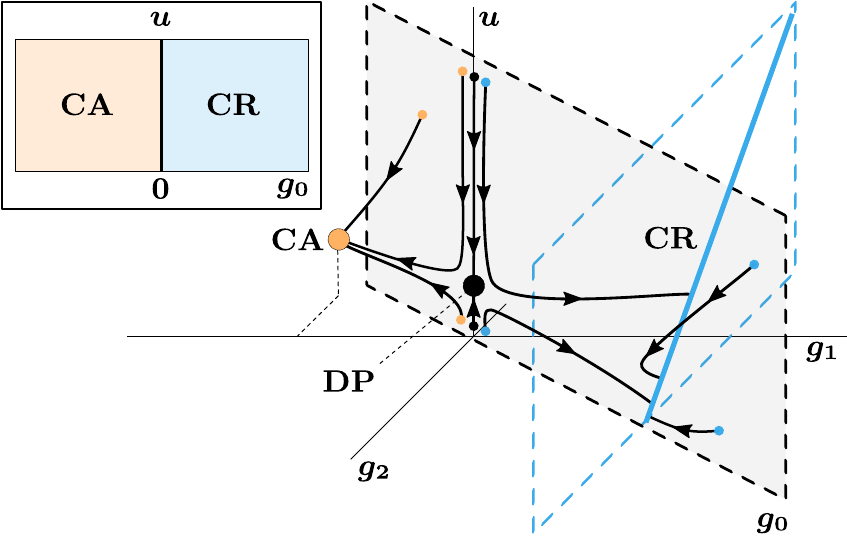}
    \caption{Schematic flow lines at~${w=1}$ for initial conditions sampled from the KS plane~${g_1=g_2=g_0}$ (gray plane). The basins of attraction for the DP (black), CA (orange) and CR (blue) fixed point are shown in the inset. In units of~${\varepsilon}$ the fixed point values~${(u,g_1,g_2)}$ are~${(1/12,0,0)}$,~${(1/9, -1/3,-1/6)}$, and~${(-g_2/2,1/2,g_2)}$, respectively.
    All flow lines starting from ${g_0\neq 0}$ leave the KS plane.}
  	\label{fig:Schematic_Flow_Lines}
\end{figure}

We begin by investigating the case of a classical KS interaction. 
This implies starting the RG coarse graining at a scale where the chemotactic nonlinearities are equal, i.e., ${g_1=g_2=g_0}$ (gray plane in Fig.\@~\ref{fig:Schematic_Flow_Lines}).
In addition to the anticipated Gaussian and DP fixed points, the RG flow exhibits a stable fixed point~(CA) and a stable fixed line~(CR) (Fig.\@~\ref{fig:Schematic_Flow_Lines}).
They represent two different types of scale-invariant dynamics, corresponding to chemo-attractive (CA) and chemo-repellent (CR) systems.
Only if ${g_0=0}$ the flow reaches the DP fixed point, which is unstable under the inclusion of chemotaxis, highlighting the importance of long-ranged interaction for the agents' critical behaviour.
Further, irrespective of the sign of the interaction, the flow leaves the plane of KS interactions and terminates in either the stable subdiffusive CA fixed point~(${z=2+\epsilon/18}$) for chemo-attraction (${g_0<0}$) or the stable superdiffusive CR fixed line~(${z=2-\epsilon/2}$) for chemo-repulsion (${g_0>0}$).

\begin{figure*}
	\begin{minipage}[b]{0.715\textwidth}
	    \label{mini:1}
		\centering
		\includegraphics{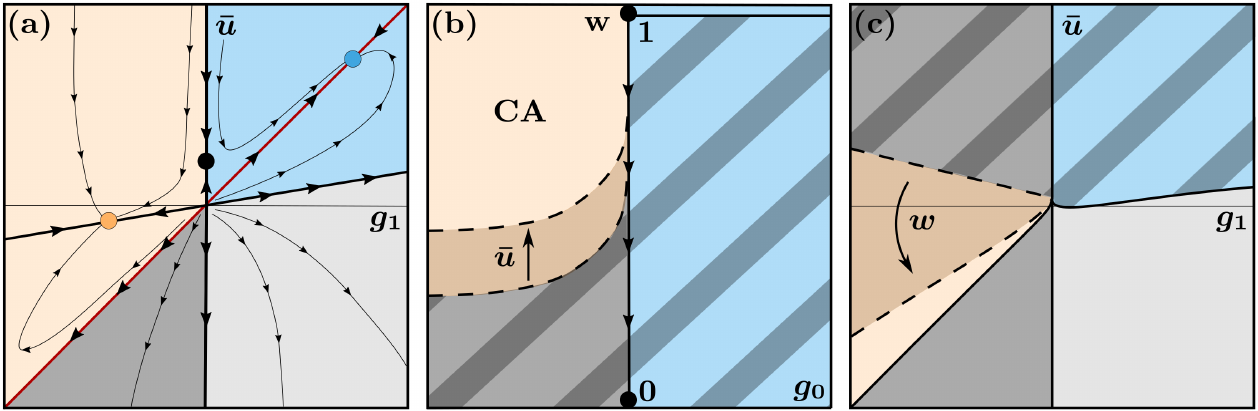}
	\end{minipage}
	\hfill
	\begin{minipage}[b]{0.275\textwidth}
		\centering
		\captionof{table}{Critical Exponents}
		    \begin{ruledtabular}
        		\begin{tabular}{cccc}
                		& $\boldsymbol{\nu}$ & $\boldsymbol{z}$ & $\boldsymbol{\delta}$ \\[1mm]
                		\hline \\[-1mm]
                		\textbf{DP}  & $0.5 + \dfrac{\epsilon}{16}$ & $2-\dfrac{\epsilon}{12}$ & $1-\dfrac{\epsilon}{4}$ \\[3mm]
                		\textbf{CA} & $0.5 + \dfrac{\epsilon}{8}$ & $2 + \dfrac{\epsilon}{18}$ & $1- \dfrac{5\epsilon}{6}$ \\[3mm]
                		\textbf{CR} & $0.5 + \dfrac{\epsilon}{8}$ & $2-\dfrac{\epsilon}{2}$ & $1$ \\[3mm]
                		\textbf{CP} & $0.5 + 0.13\epsilon$ & $2$ & $1-0.93 \epsilon$
                \end{tabular}
            \end{ruledtabular}
		    \label{tab:Exponents}
	\end{minipage}
    \caption{Evolution of initial conditions sampled from different slices of the four dimensional parameter space under the RG flow, which is classified into flow towards the CA fixed point (orange), the CR fixed line (blue), and four possibly different kinds of runaway flow (gray and striped areas).
    The striped areas indicate effects which are only present at~${w<1}$.
    (a) Schematic flow lines in the~${\bar{u}}$-${g_1}$-plane for~${w=1}$ with three invariant manifolds~${g_1=0}$,~${\bar{u} = g_1}$ and~${\bar{u} = g_1/6}$ (bold lines).
    (b) Typical flow behaviors for KS-type models with~${g_1=g_2=g_0}$ at fixed~${u}$ with DP fixed points at~${w=1}$ (unstable) and~${w=0}$ (stable).
    The separatrix (dashed line) introduced by the CP fixed point (not shown) is shifted by increasing~${u}$ (darker orange region).
    The CR fixed line is only stable at~${w=1}$.
    (c) Typical flow behaviors for~${w<1}$ and~${u}$ fixed. The influence of decreasing~${w}$ on the basin of attraction of CA is indicated by dashed lines and the darker orange region.
    The phase boundaries in (b) and (c) were obtained by numerically solving the flow equations~\eqref{eq:flowu}--\eqref{eq:floww}~\cite{Note:SM}.}
	\label{fig:BasinsOfAttraction}
\end{figure*}

We conclude that accounting for long-range chemotactic interactions quantitatively changes the nature of the phase transition compared to DP, leading to two new universality classes of absorbing-state phase transitions.
The values of the associated dynamical exponents~${z}$ (Tab.\@~\ref{tab:Exponents}) match the expectation that chemo-repellent agents accelerate and chemo-attractant agents decelerate colony dispersal compared to DP.

Further, the fact that all flow lines leave the ${g_1=g_2}$ plane confirms that a KS interaction is not sufficient to model the universal dynamics near criticality.
Fluctuation-generated terms are a generic phenomenon close to critical points \cite{Caballero2018,Cavagna2023}. 
Similarly, in our case the nonconservative part of Eq.\@~\eqref{eq:ChemoInteraction} is `generated' even if not included from the beginning and the effective chemotactic interaction can in general not be given in terms of a conserved current. Consequently, close to criticality ${g_1\neq g_2}$ is of great physical interest.
In particular, the question arises how the RG analysis relates to the mean-field analysis, which identified a band of linearly unstable modes for ${g_2<-u}$  (in the long ranged limit).

% Figure 2
%

% Non-Conservative Case
% Figure 3
Indeed, the RG flow equations can be rewritten as a set of only two equations for~${\bar{u} = u + g_2}$ and~${g_1}$:
In the quasi-static limit, the solution of the resulting Poisson equation (see Appendix~B) allows to eliminate the chemical field, leading (among other terms) to an effective growth-limiting term with the shifted coupling constant~${\bar{u} = u + g_2}$~\cite{Note:SM}.

We find that the domain of attraction of the CA and CR fixed points are separated by an invariant manifold at~${g_1=0}$ (Fig.\@~\ref{fig:BasinsOfAttraction}(a)), leading to two different types of dynamical scaling behaviors for~${g_1<0}$ and~${g_1>0}$, respectively.
This further stresses the difference between the two chemotactic couplings:
While the term~${\sim g_2\rho\boldsymbol{\nabla}^2 c}$ can be absorbed into an effective growth-limiting term, only the nonlinearity~${\sim g_1 \boldsymbol{\nabla} \rho \boldsymbol{\nabla} c}$ qualitatively changes the RG flow.
In addition to the separatrix at~${g_1=0}$, the RG flow is organized by the critical manifolds containing the CA and CR fixed points, given (to one-loop order) by the lines~${\bar u = g_1/6}$ and~${\bar u = g_1}$, respectively (Fig.\@~\ref{fig:BasinsOfAttraction}(a)). 
These lines are also the boundaries of the domains of attraction of the CA (orange) and CR (blue) fixed points, separating them from runaway flow.

Given that, in the long-ranged limit, the instability condition~\eqref{eq:InstabilityCon} simplifies to~${\bar{u}<0}$, one might have anticipated runaway flow in this entire region.
Strikingly, the RG analysis predicts scaling for~${g_1<\bar u<0}$, which seems contradictory at first.
However, the linear stability analysis does not allow any conclusions about the steady state of the dynamics.
Crucially,~${g_1}$ does not affect the linear dynamics, but only contributes to nonlinear effects.
In particular, it enters the following exact relation for the time evolution of the average mass (see Appendix C)
\begin{align}
    \frac{\Lambda(\partial_t \langle\bar{\rho}\rangle- \theta\langle\bar{\rho}\rangle)}{32\pi^2 D_\rho^2} = 
    - u\langle  \bar{\rho}^2 \rangle + \frac{g_1-\bar{u}}{|V|} \int_V  \langle \left(\rho-\bar{\rho}\right)^2 \rangle ,
    \label{eq:MassEvolution}
\end{align}
where~${\bar{\rho}}$ indicates a spatial and~${\langle \cdot \rangle}$ an ensemble average with respect to the noise~${\xi}$.
Equation~\eqref{eq:MassEvolution} implies that, depending on the sign of~${g_1 \,{-}\, \bar{u}}$, fluctuations drive the system either toward or away from the absorbing state.
It applies to the dynamics both above and below the absorbing-state phase transition, and especially when approaching the phase transition at~${\theta,\bar\rho\rightarrow 0}$.
This rationalizes why for~${\bar u - g_1 > 0}$ (including all KS models) nonlinear effects combined with demographic noise lead to a continuous absorbing-state phase transition, despite the band of linear unstable modes for~${\bar u < 0}$.
In contrast, for~${g_1>0}$, the system is attracted by the CR fixed point for~${\bar u>g_1/6}$, and exhibits runaway flow when~${\bar u < g_1/6}$ (Fig.\@~\ref{fig:BasinsOfAttraction}(a)).
The region~${0<\bar{u}<g_1}$ is particularly interesting:
Eq.\@~\eqref{eq:MassEvolution} implies that the linear stability of the spatially uniform, active state is counteracted by a nonlinear term~(${\sim g_1 - \bar u}$) disfavoring a homogeneous state.
Our RG analysis indicates that the antagonism between these two effects leads to flow towards the CR fixed point in the regime~${g_1 > \bar u >  g_1/6}$ but to runaway flow for~${\bar u <  g_1/6}$. 
Since the nonlinear instability is dominant in the latter regime, the observed runaway flow is possibly indicative of a fluctuation-driven first-order transition.

The agents' active motion can result in an effective diffusion constant~${D_\rho}$ of similar magnitude as~${D_c}$~\cite{Budrene1991,Lewus2001,Murray2003}.
Therefore, it is crucial to study the case~${w \not \approx 1}$.
In this case, the full flow equations~\eqref{eq:flowu}--\eqref{eq:floww} exhibit an additional fixed point of mixed stability we call critical fixed point~(CP) at~${(u,g_1,g_2, w) \,{=}\, (0.08\varepsilon, -0.45\varepsilon, -0.16\varepsilon, 0.64)}$ and a second DP fixed point at~${w=0}$. 
Our RG analysis shows that the CP fixed point has a dynamic critical exponent~${z=2}$ to all loop orders \cite{Note:SM}, implying purely diffusive dynamics, akin to the critical fixed point characterizing the roughening transition of the Kardar-Parisi-Zhang equation~\cite{Kardar1986, Frey1994, Janssen1999}.
As before, we first consider the case of ${g_1=g_2=g_0}$; the resulting basins of attraction for the various fixed points are depicted in Fig.\@~\ref{fig:BasinsOfAttraction}(b).
All points located on the invariant manifold~${g_0=0}$ flow towards the second DP fixed point at~${w=0}$.
Since the CP fixed point is located at~${w<1}$ and unstable in ${w}$-direction, it separates the parameter space~${g_0<0}$ into two parts.
Points above this separatrix flow to CA, whereas below it the system exhibits a new type of runaway flow (striped dark gray).
In contrast to CA, the basin of attraction of CR does not extend to~${w<1}$.
As pointed out above, a chemo-repellent implies superdiffusive motion~(${z<2}$), equivalent to~${\partial_\mu w < 0}$ near the fixed line (CR).
This renders the fixed line unstable in~${w}$-direction.
However, in the emerging runaway region (striped blue) the projection of the fixed line to~${w<1}$ is still a strong attractor which separates it from other regions of runaway flow (Fig.\@~\ref{fig:BasinsOfAttraction}(b) and~(c)).
The typical shape of the phase diagram for general~${g_1}$ and~${g_2}$, at fixed values of~${u}$ and~${w}$, is shown in Fig.\@~\ref{fig:BasinsOfAttraction}(c).
It features all four, possibly different, kinds of runaway flow and bears a strong resemblance to Fig.\@~\ref{fig:BasinsOfAttraction}(a). 

% Summary and Outlook
Altogether, the analyzed model reveals a correction to the well known Keller-Segel nonlinearity in the presence of large fluctuations and exhibits a rich phase diagram with two new absorbing-state phase transitions and various types of runaway regions.
The emergence of fixed points associated to either a chemo-attractant or -repellent, demonstrates the relevance of auto-chemotactic interactions for the collective behavior of cells at their extinction threshold.
In particular, they highlight the impact of chemotactic signaling for the survival probability and spreading velocity of single colonies (Tab.\@~\ref{tab:Exponents}).
For~${w=1}$ we have presented a possible mechanism by which the runaway flow found in Fig.\@~\ref{fig:BasinsOfAttraction}(a) can be related to a fluctuation-induced first-order transition (cf.\@~\eqref{eq:MassEvolution}).

Further, the emergence of the CP fixed point not only gives rise to an unexpected type of purely diffusive scaling behavior, it also highlights the importance of the time delay introduced by the finite diffusion speed of the signaling substance. 
The reminiscence of the CP fixed point to the critical fixed point describing the roughening transition of the KPZ equation suggests the intriguing scenario of a strong coupling fixed point below the separatrix.

Naturally, the multitude of theoretical predictions presented calls for a numerical study.
Additionally, we hope that our work will stimulate nonperturbative approaches~\cite{Canet2010, Canet2021} that help to unravel the observed anomalous dynamics.
From a broader perspective, our results suggest that by combining known universality classes of nonequilibrium population dynamics~\cite{JanssenTauber2005, Tauber2014} with various types of auto-chemotactic feedbacks, a broad class of novel scale-invariant dynamics could be discovered.

% Acknowledgments
\begin{acknowledgments}
    This work was funded by the Deutsche Forschungsgemeinschaft (DFG, German Research Foundation) through the Collaborative Research Center (SFB) 1032 – Project-ID 201269156 – and the Excellence Cluster ORIGINS under Germany’s Excellence Strategy – EXC-2094 – 390783311. A.B. thanks the SERB, DST (India) for partial financial support through the MATRICS scheme [file no.: MTR/2020/000406].
\end{acknowledgments}

%Appendix
\section{Appendix A: Long-ranged limit}

Since scale invariance can only be observed if no length scale is introduced by the chemotactic interaction, the long-ranged limit~${\lambda_c\rightarrow 0}$ is of particular interest.
However, simply inserting ${\lambda_c=0}$ into Eqs.\@~\eqref{eq:LEC} and~\eqref{eq:LEPhi} leads to a  divergent chemical density and a steady state condition ${\rho_1=-D_c\boldsymbol\nabla^2 c_1(\boldsymbol{x})/\alpha}$.
Thus, there would no longer be a homogeneous steady state for the chemical density, which leads to an unphysical shift to the homogeneous steady state density $\rho_1=\theta/(\gamma+\alpha\chi_2/D_c)$ of the agents.
This deviates from the actual carrying capacity $\theta/\gamma$ and shows that one needs to take into account the `charge-neutral' chemical density
\begin{equation}
    \tilde c(\boldsymbol x, t) = c(\boldsymbol x, t) - \alpha\int_0^t \bar\rho(t^\prime) \mathrm{d}t^\prime \, ,    
\end{equation}
where we subtracted the homogeneous, albeit time dependent average production of the signaling molecule with $\bar\rho(t)$ denoting the spatial average of $\rho$ at time $t$.

Importantly, this homogeneous shift does not alter the dynamics of $\rho$. 
However, the evolution of the charge-neutral chemical density is now given by
\begin{equation}
    \frac{\mathrm{d}\tilde c}{\mathrm{d}t}
    = D_{c}\boldsymbol{\nabla}^2 \tilde c+\alpha \, (\rho-\bar\rho),  \label{eq:time_evolution_charge_neutral_c}
\end{equation}
with no overall net production, i.e.,
\begin{equation}
    \frac{\mathrm{d}}{\mathrm{d}t}\int_V \tilde c = 0 \, .
\end{equation}
More details on this limit are provided in the supplemental material~\cite{Note:SM}.

\section{Appendix B: Quasi-static limit}\label{app:B}

Another important limit is the so-called \textit{quasi-static limit}, where $D_c/D_\rho\rightarrow\infty$ and the chemical field thus instantly adjusts to changes in the density field $\rho$.
Assuming that $\alpha/D_c$ remains finite~\cite{Jaeger1992},  Eq.\@~\eqref{eq:time_evolution_charge_neutral_c} leads to the Poisson equation
\begin{equation}
    \boldsymbol\nabla^2\tilde c(\boldsymbol x,t) = -\frac{\alpha}{D_c}(\rho(\boldsymbol x, t)-\bar\rho(t)).\label{eq:charge_neutral_poisson}
\end{equation}
For more details we refer to the supplemental material~\cite{Note:SM}.

\section{Appendix C: Mass evolution}

One way to analyze the impact of different interactions is to study their effect on the time evolution of the average density $\langle\bar\rho\rangle$, where $\langle\cdot\rangle$ signifies an ensemble average with respect to the noise $\xi$.
To derive this evolution we first note that Eqs.\@~\eqref{eq:LEC} and~\eqref{eq:LEPhi} are It\^o Langevin equations and thus 
\begin{equation}
    \int_V\langle \sqrt{2\Lambda\rho}\, \xi\rangle = \int_V\langle \sqrt{2\Lambda\rho}\rangle \langle\xi\rangle = 0.
\end{equation}
Further we split the agents' density into $\rho(\boldsymbol x,t)=\bar\rho(t) + \hat\rho(\boldsymbol{x}, t)$, integrate Eq.\@~\eqref{eq:LEC} over space and insert Eq.\@~\eqref{eq:charge_neutral_poisson}.
For the deterministic terms, this yelds
\begin{alignat}{8}
    \partial_t\bar\rho &=\theta\bar\rho&&-\frac{1}{|V|}\int_V(\bar\rho+\hat\rho)\bigg(\gamma\,(\bar\rho+\hat\rho)+\frac{\alpha\,(\chi_1+\chi_2)}{D_c}\hat\rho\bigg)\notag\\[3mm]
    &=\theta\bar\rho &&+ \frac{32\pi^2D_\rho^2}{\Lambda}\left(-u\bar\rho^2+\frac{g_1-\bar u}{|V|}\int_V\hat\rho\right) \, , \label{eq:avg_rho_app}
\end{alignat}
where we used that $\int_V\hat\rho=0$ and used the definitions~\eqref{eq:effCoupl} of the effective couplings, as well as ${\bar u=u+g_2}$.
Taking the ensemble average of Eq.\@~\eqref{eq:avg_rho_app} leads to the exact result of Eq.\@~\eqref{eq:MassEvolution}.
This result highlights the difference between the linear growth term $\propto \theta$ and the nonlinearity $\propto\gamma$ modelling resource limitation.
While the former contributes a distribution independent term to Eq.\@~\eqref{eq:avg_rho_app}, the latter leads to a mass evolution which is dependent on the density profile.
Thus, it is the resource limitation which makes the mass evolution susceptible to the influence of chemotaxis.

%%% Bilbliography

\newpage 

%%% Appendix
 
 \onecolumngrid
 	\section{Supplemental Material: \\
 	Anomalous collective dynamics of auto-chemotactic populations}

%%% Correct LE

\section{Derivation of Langevin Equations}

The model analyzed in the main text consists of two parts: Diffusive particles $A$ that obey logistic growth dynamics and a chemical which is secreted by $A$-particles and whose gradients influence the motion of $A$-particles.
In order to derive a set of effective equations, we first treat the dynamics of demographic and chemotactic processes separately.
To this end one may think of the following set of \textit{microscopic} reactions
\begin{align}
    A\overset{\lambda}{\rightarrow}\emptyset, \quad A\overset{\mu}{\rightarrow}A+A, \quad A+A\overset{\gamma}{\rightarrow} A. \label{eq:microReactions}
\end{align}
A \textit{coarse-grained} stochastic description in terms of the continuous density $\rho(t)$ can be derived by a Kramers-Moyal expansion~\cite{gardiner2009} of the corresponding master equation. 
This yields
\begin{align}
    \frac{\mathrm{d}\rho}{\mathrm{d}t} =\,(\theta - \gamma \rho) \rho + \sqrt{\Lambda\rho + \gamma\rho^2} \, \xi \, ,
    \label{eq:LangevinDP}
\end{align}
where~$\theta=\mu-\lambda$ denotes the effective growth rate,~$\Lambda=\mu+\lambda-\gamma$ the noise amplitude and~$\xi$ Gaussian white noise.
Equivalently, one may apply operator based approaches~\cite{Doi1976, Peliti1985} and a subsequent Cole-Hopf transformation of the resulting field theory.
Note that while different approaches strictly speaking correspond to different realizations of the stochastic process, all rely on It\^{o} calculus and the underlying master equation.
Beyond the continuous limit, the only approximation involved in all of these approaches is the truncation at second order in fluctuations which enables the description in terms of a Langevin/Fokker-Planck equation.
However, higher orders can be shown to be irrelevant close to the absorbing state (see section `The Response Functional').
\\
While the above equation implies a well-mixed system, any description of chemotaxis requires a spatially extended description.
Since chemotaxis implies that agents ($A$) adjust their motion to their surroundings, some form of active swimming is required.
Even though the microscopic details of the biological processes leading to chemotaxis may vary significantly in different settings – which are of no particular interest to the present study – one may derive an \textit{effective} descriptions for the chemotactic interaction.
One way such an effective interaction can be formulated is to assume a chemotactic drift whose local velocity is given by the product of the local gradient in the chemical density $c(\boldsymbol x,t)$ and the sensitivity function $\chi[\rho, \boldsymbol\nabla c]$.
This results in a generalized form of the stochastic \textit{Keller-Segel} (KS) model~\cite{Chavanis2008} (which in its classical form assumes a constant sensitivity)
\begin{align}
    \frac{\mathrm{d}\rho}{\mathrm{d}t} =& \,
   D_\rho\boldsymbol\nabla^2\rho  + \boldsymbol\nabla\Big(\chi[\rho,\boldsymbol\nabla c]\rho\boldsymbol\nabla c\Big) + \boldsymbol\nabla\Big(\sqrt{2D_\rho\rho} \, \eta_1\Big) \, ,
\label{eq:DeanRho} 
\end{align}
with an effective diffusion constant $D_\rho$ and Gaussian white noise $\eta_1$. 
A more rigorous derivation of  Eq.\@~\eqref{eq:DeanRho} can be given in terms of the \textit{Dean-Kawasaki} approach~\cite{Dean.1996, Kawasaki.1994} or a lattice gas with modified hopping rates.
Further, the dynamics for the chemical density $c(\boldsymbol x,t)$ can straightforwardly be deduced from the microscopic reactions – secretion by $A$ and decay, with rates $\alpha$ and $\lambda_c$, respectively – and is given by
\begin{align}
    \frac{\mathrm{d}c}{\mathrm{d}t} =& \,
    \left( D_c\boldsymbol\nabla^2 - \lambda_c \right) c + \alpha \rho \, + \boldsymbol\nabla\Big(\sqrt{2D_cc} \, \eta_2\Big) + \sqrt{\lambda_c c + \alpha \rho}\,\eta_3,
\label{eq:LangevinC}
\end{align}
with diffusion constant $D_c$ and $\eta_{2,3}$ again representing Gaussian white noises.
To continue, one has to combine equations~{\eqref{eq:LangevinDP}–\eqref{eq:LangevinC}} and, even though the coarse graining procedures used are different and incorporate distinct effects, the first guess is to simply combine all the appearing terms; this yields
\begin{align}
    \frac{\mathrm{d}\rho}{\mathrm{d}t} =& \,
    \Big(D_\rho\boldsymbol\nabla^2 + \theta - \gamma \rho \Big) \rho + \boldsymbol\nabla\Big(\chi[\rho,\boldsymbol\nabla c]\rho\boldsymbol\nabla c\Big) + \sqrt{\Lambda\rho+\gamma\rho^2} \, \xi + \boldsymbol\nabla\Big(\sqrt{2D_\rho\rho} \, \eta_1\Big)\, 
    \label{eq:LangevinRho}
    \\[2mm]
    \frac{\mathrm{d}c}{\mathrm{d}t} =& \,
    \left( D_c\boldsymbol\nabla^2 - \lambda_c \right) c + \alpha \rho \, + \boldsymbol\nabla\Big(\sqrt{2D_cc} \, \eta_2\Big) + \sqrt{\lambda_c c + \alpha \rho}\,\eta_3\, .
    \label{eq:LangevinC2}
\end{align}
It is important to emphasize that this is a highly coarse-grained description: While some parameters – like the growth rate $\theta$ – have a clear interpretation  in terms of a microscopic model, others – especially  the sensitivity function $\chi[\rho, \boldsymbol\nabla c]$ – have no such interpretation and are purely phenomenological.
Therefore, Eqs.~\@\eqref{eq:LangevinRho} and~\eqref{eq:LangevinC2} can at most be valid at a finite range of scales.
\\
By inspecting Eqs.\@~\eqref{eq:LangevinRho} and~\eqref{eq:LangevinC2} one can already identify two important length scales: the diffusion length of the agents ${\xi_\rho=\sqrt{D_\rho/|\theta|}}$ and the chemicals ${\xi_c=\sqrt{D_c/\lambda_c}}$. 
On length scales below $\xi_\rho$, the population dynamics of particles $A$ only play a minor role and can be neglected.
However, since the goal of this manuscript is to extract the critical behavior of the presented model at the largest length (and time) scales and to evolve the dynamics to these scales, one has to be aware that the effective equations of motion might change as one continuously changes the scale; especially since fluctuations play an important role close to the phase transition where the correlation length diverges.
Possible ways how the equations of motion can change upon coarse graining, as long as no global symmetries are violated, are:
\begin{enumerate}
    \item Parameters tend to zero under coarse graining and are \textit{irrelevant} for the critical dynamics. The dimensional analysis detailed in `The Response Functional' shows that this is the cases for the noise terms $\eta_{1,2,3}$ and that the only relevant contribution of the sensitivity function is a constant sensitivity $\chi_0^{}$.
    \item Microscopic and mesoscopic relations between parameters might change. For example the microscopic relation between the reaction rates and the noise amplitude, i.e.,~${\Lambda = \mu+\lambda-\gamma}$  no longer holds for the effective reaction rates and the effective noise amplitude on larger scales.
    \item New types of effective interactions may arise at larger scales due to the presence of strong fluctuations close to criticality~\cite{Caballero2018,Cavagna2023}. A well-known example for this is seen during real space RG schemes for the two dimensional Ising system where effective next-to-nearest neighbor interactions arise~\cite{kardar_2007}.
\end{enumerate}
One example relevant to our model of points 2 and 3 is the effective chemotactic interaction, which, after reducing the sensitivity $\chi[\rho, \boldsymbol\nabla c]$ to $\chi_0^{}$, is given by $\chi_0^{}\boldsymbol\nabla(\rho\boldsymbol\nabla c)$. 
Even though this looks like a single term, one can, tentatively, separate this interaction into two terms: $\chi_0^{(1)}\boldsymbol\nabla\rho\boldsymbol\nabla c$ and $\chi_0^{(2)}\rho\boldsymbol\nabla^2c$.
It might seem that $\chi_0^{(1)}$ has to equal $\chi_0^{(2)}$ on all scales since both terms combined are supposed to model a particle number conserving chemotactic current.
However, this only holds in absence of any processes that explicitly break particle number conservation.
In the combined model particle numbers are not conserved; hence, there is \textit{a priori} no reason to expect the relation ${\chi_0^{(1)} = \chi_0^{(2)}}$, which might hold at some scales, to also be valid at macroscopic scales.
The RG calculations below, indeed, show that in our model $\chi_0^{(1)}$ is in general not equal to $\chi_0^{(2)}$, giving rise to a contribution ${(\chi_0^{(1)}-\chi_0^{(2)})\rho\boldsymbol\nabla^2 c}$ to the effective chemotactic interaction (also see `Confirmation of Non-Conservative Interaction' and Fig.\@~2 in the main text).
Thus, it is prudent to keep both terms separate with coupling parameters $\chi_1^{}$ and $\chi_2^{}$.
\\
Combining all of the above  finally results in the Langevin equations we rely on for the mean-field and RG analysis
\begin{align}
    \frac{\mathrm{d}\rho}{\mathrm{d}t} =& \,
    \left(D_\rho\boldsymbol\nabla^2 + \theta - \gamma \rho \right) \rho  + \chi_1^{}\boldsymbol\nabla\left(\rho \boldsymbol\nabla c\right) + \left(\chi_2^{}-\chi_1^{}\right)  \rho \boldsymbol\nabla^2 c + \sqrt{\Lambda\rho} \, \xi \, ,
    \label{eq:MeanFieldRho1}
    \\[2mm]
    \frac{\mathrm{d}c}{\mathrm{d}t} =& \,
    \left( D_c\boldsymbol\nabla^2 - \lambda_c \right) c + \alpha \rho \, .
    \label{eq:MeanFieldC1}
\end{align}
Equations~\eqref{eq:MeanFieldRho1} and~\eqref{eq:MeanFieldC1} conclude the derivation of the correct macroscopic Langevin equations close to the continuous phase transition.
It is important to emphasize that this is result is not specific to RG.
RG is a systematic way of analyzing how – close to a continuous phase transition – effective interactions change with larger scales; therefore, the observation that the population dynamics and the chemotactic interaction mix upon coarse graining to give rise to an effective non-particle-number-conserving chemotactic interaction strongly suggests that this effect should be accounted for in any coarse-grained description of our model near the critical point.
However, we make no prediction about the strength of this effect away from criticality or at small scales.
In such cases the chemotactic interaction can most likely be formulated in terms of a conserved current and  other noise terms ($\eta_{1,2,3}^{}$) as well as higher orders of the sensitivity function $\chi[\rho, \boldsymbol\nabla c]$ may be important.

%%% MeanField
\section{Mean-field Analysis}\label{sec:LSAFT}

As in the main text, we here analyze a model of chemotactic cells~(${\rho}$) reacting to gradients in a chemical signalling substance~(${c}$) given by the following Langevin equations:
\begin{align}
    \frac{\mathrm{d}\rho}{\mathrm{d}t} =& \,
    \left(D_\rho\boldsymbol\nabla^2 + \theta - \gamma \rho \Big) \rho  + \Big( \chi_1^{} \boldsymbol\nabla\rho \boldsymbol\nabla c + \chi_2^{}  \rho \boldsymbol\nabla^2 c  \right) + \sqrt{\Lambda\rho} \, \xi \, ,
    \label{eq:MeanFieldRho}
    \\[2mm]
    \frac{\mathrm{d}c}{\mathrm{d}t} =& \,
    \left( D_c\boldsymbol\nabla^2 - \lambda_c \right) c + \alpha \rho \, .
    \label{eq:MeanFieldC}
\end{align}
Here,~${D_\rho}$ represents the diffusion constant of the cells,~${\theta}$ is the effective linear growth rate,~${\gamma}$ models a competition for resources,~${\chi_1^{}}$ and~${\chi_2^{}}$ are the chemotactic response parameters, and~${\Lambda}$ the noise amplitude.
The dynamics of the chemical is characterized by its diffusion constant~${D_c}$, its degradation rate~${\lambda_c}$, and its production by the cells at rate~${\alpha}$.
The noise term~${\xi}$ has zero mean and is delta-correlated in time and space, i.e.\@~${\langle \xi(\boldsymbol{x},t)\rangle = 0}$ and~${\langle \xi(\boldsymbol{x},t) \, \xi(\boldsymbol{y},t^\prime) \rangle = 2 \, \delta(\boldsymbol{x}-\boldsymbol{y}) \, \delta (t-t^\prime)}$.

In this section, we first study the mean-field behavior of the model by ignoring the impact of the demographic noise.
In this case, one finds two qualitatively different homogeneous steady states.
The trivial solution~${\rho_0=c_0=0}$ is coined the \textit{absorbing phase} and the non-trivial, \textit{active phase} solution is given by~${\rho_1=\theta/\gamma, c_1=\alpha\rho_1/\gamma\lambda_c }$.
To determine their stability against small density fluctuations~${\delta\rho(x,t)}$ and~${\delta c(x,t)}$, we expand around the homogeneous states such that
\begin{equation} 
    \begin{pmatrix}
    \rho(x,t)
    \\[2mm]
    c(x,t)
    \end{pmatrix}=
    \begin{pmatrix}
    \rho_{0,1}+\delta \rho(x,t)\\[2mm]
    c_{0,1}+\delta c(x,t)
    \end{pmatrix}.
    \label{eq:MFPerturbation}
\end{equation}
Transforming to momentum space one finds an equation of the form~${\partial_t \phi= \hat{A} \cdot \phi}$, with~${\phi=(\delta\rho,\delta c)}$ and 
\begin{align}
    \hat{A}_{0,1}(k)=\left(\begin{matrix}-D_\rho k^2+\theta-2\gamma\rho_{0,1}&-\chi_2^{}\rho_{0,1}k^2\\[3mm] \alpha&-D_c k^2-\lambda_c\end{matrix}\right)\label{eq:MeanFieldMatrix}.
\end{align}
Note that only~${\chi_2^{}}$ but not~${\chi_1^{}}$ enters the equation at linear order in the perturbations.
For the homogeneous solutions to be stable, we require ~${\Re(\sigma_{\pm}(k))<0\,\,\forall k}$ for the eigenvalues~${\sigma_{\pm}(k)}$ of~${\hat{A}}$.
Inserting~${\rho_0=c_0=0}$, one finds
\begin{align}
    \hat{A}_0(k)=\left(\begin{matrix}-D_\rho k^2+\theta&0\\[2mm] \alpha&-D_c k^2-\lambda_c\end{matrix}\right),
\end{align}
from which we can read off the eigenvalues
${\sigma_+(k)=-D_\rho k^2+\theta}$ and~${\sigma_-(k)=-D_ck^2-\lambda_c}$.
Trivially,~${\sigma_{\pm}<0}$ if and only if~${\theta<0}$ -- i.e.\@ the homogeneous absorbing state is always linearly stable if death dominates birth.
In the case of the active solution, one gets
\begin{align}
    \hat{A}_{1}(k)
    =
    \left(
    \begin{matrix}
    -(D_\rho k^2+\theta) 
    &-\dfrac{\theta \chi_2^{}}{\gamma}k^2 
    \\[3mm] 
    \alpha &-\big(D_c k^2+\lambda_c\big)
    \end{matrix}
    \right).
\end{align}
By calculating the eigenvalues of the zero mode, i.e.\@  of~${\hat{A}_1(k=0)}$, 
${\sigma_+(k=0)=-\theta}$ and~${\sigma_-(k=0)=-\lambda_c}$, 
one finds the active state to be unstable if~${\theta<0}$. 
For general~${k}$ the eigenvalues read
\begin{alignat}{6}
    \sigma_\pm=\dfrac{1}{2}\Big(-b(k)\pm\sqrt{b(k)^2-4c(k)}\Big),
\end{alignat}
where the auxiliary functions~${b}$ and~${c}$ are given by
\begin{alignat}{6}
    b(k) &= \ (D_\rho+D_c)k^2+\theta+\lambda_c \, ,
    \\
    c(k) &= \ (D_c k^2+\lambda_c)(D_\rho k^2+\theta)+\dfrac{\alpha\chi_2^{}\theta}{\gamma}k^2 \, 
    \notag \\
    &= \ Ak^4 + Bk^2 + C \, .
\end{alignat}
For~${\rho_1}$ to be stable in a certain parameter regime, one requires~${\Re(\sigma_{+}(k))<0\,\,\forall k}$, which is equivalent to~${c(k)>0\,\, \forall k}$.
Thus, for~${\theta>0}$ the active state is unstable if and only if~${c(k)}$ has a real root.
This is the case only if~${B < 0}$ and~${B^2 > 4AC}$, which is equivalent to
\begin{equation}
    \frac{\alpha\chi_2^{}}{\gamma D_c}<-\left(1+\frac{\lambda_c D_\rho}{\theta D_c}+2\sqrt{\frac{\lambda_c D_\rho}{\theta D_c}}\,\right). \label{eq:full_stability_condition}
\end{equation}
When this condition is satisfied, there exists a~${k}$ for which~${\sigma_+}$ is positive and where the homogeneously active phase is thus unstable.

As we already argued for in the main part,~${\theta \approx \lambda_c \approx 0}$ is a necessary condition for scale invariant dynamics.
However, it is not intuitively clear how to treat the fraction~${\lambda_c/\theta}$ in the instability condition~\eqref{eq:full_stability_condition} in this limit.
Accounting for fluctuation-induced shifts to the transition values~${\theta^*, \lambda_c^*}$, we write
\begin{align}
	\frac{\lambda_c}{\theta} = \frac{\lambda+\lambda_c^*}{\tau + \theta^*},
	\label{eq:ShiftedTemperature}
\end{align}
where we introduced the relative control parameters~${\tau = \theta - \theta^*}$ and~${\lambda = \lambda_c - \lambda_c^*}$
Right at the transition this Eq.\@~\eqref{eq:ShiftedTemperature} reduces to~${\theta^*/\lambda^*}$.
As we show in the perturbative analysis, there is no diagram contributing to the renormalization of~${\lambda_c}$.
Thus,~${\lambda_c^*=0}$, whereas~${\rho}$ experiences a finite shift in its critical temperature.
In this sense~${\lambda_c/\theta \rightarrow 0}$ holds as one approaches the transition and equation~\eqref{eq:full_stability_condition} reduces to~${\alpha\chi_2^{}<-\gamma D_c}$.

% Limiting Cases
\subsection{The Long-Ranged and Quasi-Static Limit}

For the chemotactic interaction to be intrinsically long-ranged -- and thus for the system's dynamics to exhibit scale invariance -- one needs to be in the limit where the decay rate of the signalling molecules becomes vanishingly small, i.e.~${\lambda_c \rightarrow 0}$. 
However, this limit is more subtle than one might expect.
From equation~\eqref{eq:MeanFieldC}, a formal solution for the time evolution of the chemical density~${c}$ can be given in terms of the spatially Fourier-transformed density
\begin{align}
	c_k(t) 
	= 
	\alpha
	\int\limits_{-\infty}^t \hspace{-1mm}\text{d}s  \, \exp\left[{-}\left(D_ck^2+\lambda_c\right)\big(t-s\big)\right] \rho_k(s).	\label{GeneralSolution}
\end{align}
If we now take~${\lambda_c \rightarrow 0}$, we obtain for the homogeneous mode
\begin{align}
	c_{0}(t) = \alpha\int\limits_{-\infty}^t \hspace{-1mm}\text{d}s \, \rho_{0}(s), \label{eq:c0}
\end{align}
which is in general divergent for~${t \rightarrow \infty}$.
Note that this is the same divergence one encounters for the stationary solution~${c_1}$ when taking~${\lambda_c \rightarrow 0}$.
Fortunately, the dynamics of~${\rho}$ do not depend on the homogeneous mode of the chemical.
This allows us to use the reduced quantity~${\tilde{c}}$ which obeys
\begin{align}
	\frac{\text{d} \tilde{c}}{\text{d} t} (t) 
	= \frac{\text{d} c}{\text{d} t} (t) - \alpha\bar{\rho}(t)
	= D_c\boldsymbol\nabla^2  c + \alpha \left(\rho-\bar{\rho}\right)
\end{align}
which differs from the dynamics of~${c}$ only by a homogeneous, albeit time dependent, term.
Note that this homogeneous shift changes the steady state density to~${\tilde c_1 = 0}$ but does not alter the above linear stability analysis.
By setting~${\lambda_c=0}$ in the instability condition~\eqref{eq:full_stability_condition}, one obtains the simpler condition
\begin{equation}
    \frac{\alpha\chi_2^{}}{\gamma D_c} < -1. \label{eq:InstabilitySimple}
\end{equation}
In addition to the long-ranged limit, we are also interested in the \emph{quasi-static limit}~${D_c/D_\rho \rightarrow \infty}$.
Assuming~${D_c \gg D_\rho}$, the approximate solution of Eq.\@~\eqref{eq:MeanFieldC} reads
\begin{align}
	\tilde{c}_k(t) 
	&= 
	\alpha\int\limits_{-\infty}^t \hspace{-1mm}\text{d}s  \, \exp\left[{-}\left(\frac{D_c}{D_\rho} D_\rho k^2\right)\big(t-s\big)\right] \hat\rho_k(s) 
	\notag \\
	&\approx \frac{\alpha}{D_c} \frac{\hat\rho_k(t)}{k^2},
\end{align}
with~${\hat\rho = \rho-\bar\rho}$ and the average density~${\bar \rho}$.
In the second line we replaced the argument of~${\rho_k}$ with~${t}$ -- since the exponential, for~${D_c/D_\rho \rightarrow \infty}$,  only contributes to the integral at~${s=t}$  -- and then performed the integral.
Given that~${\alpha/D_c}$ is finite,~${\tilde{c}_k}$ is a solution to the Poisson equation
\begin{align}
	\boldsymbol\nabla^2\tilde{c}(x,t) = -\frac{\alpha}{D_c} \hat{\rho}(x,t).
\end{align}
This implies that~${\tilde{c}(x,t)}$ is \textit{quasi-stationary} as it instantly adjusts to~${\hat{\rho}(x,t)}$.
Further, we notice that in order to make sense of this limit, one has to simultaneously assume that~${\alpha \sim D_c}$ (as also pointed out in Ref.~\cite{Jaeger1992}).

To see that it is indeed necessary to handle the divergence in Eq.\@~\eqref{eq:c0} with care, one can impose that the quasi-static limit must not alter the active steady state density.
However, taking the limit without shifting to~${\tilde c}$ would result in~${\boldsymbol\nabla^2 c = -\alpha \rho/D_c}$, thereby shifting~${\rho_1}$ to~${\theta/(\gamma+\alpha\chi_2^{}/D_c)}$.

%%% Action Relevance
\section{The Response Functional} \label{Action}

To investigate the equations of motion (Eqs.\@~\eqref{eq:MeanFieldRho} and~\eqref{eq:MeanFieldC}) beyond their mean field behavior -- i.e.\@ including the demographic noise term -- we employ the dynamical renormalization group in form of the response functional formalism~\cite{BauschWagnerJanssen1976, Janssen1976, deDominics1976, MartinSiggiaRose1973}. 
Following the approach displayed in~\cite{Tauber2014}, all moments of the fields can be written in the form of a path integral
\begin{align}
	\langle A[\rho, c] \rangle 
	= \int \mathcal{D}\left[\bar{\rho}, \rho, \bar{c}, c\right] \  A[\rho, c] \exp \left\{- S\left[\bar{\rho}, \rho, \bar{c}, c\right] \right\}, 
\end{align}
where we introduced the response fields~${\bar{\rho}}$ (not to be confused with the average density) and~${\bar{c}}$.
The statistical weight is given by the action~${S = S_0 + S_{\text{int}}}$ which consists of the Gaussian part
\begin{align}
	S_0\left[\bar{\rho}, \rho, \bar{c}, c\right] 
	= \int_{x,t} &\bar{\rho} \left[ \partial_t - D_\rho  \nabla ^2 - \theta \right] \rho 
	+ \bar{c} \left[ \partial_t - D_c  \nabla ^2 + \lambda_c\right] c
	\label{eq:GaussianAction}
\end{align}
and the nonlinear interaction term 
\begin{align}
	S_{\text{int}}\left[\bar{\rho}, \rho, \bar{c}, c\right] 
	= \int_{x,t} &\bar{\rho} \left[ \gamma \rho - \Lambda \bar{\rho} \right] \rho 
	\, - \, \alpha \bar{c} \rho 
	- \bar{\rho} \left[ \chi_1 \nabla \rho  \nabla  c 
	+ \chi_2 \rho  \nabla ^2 c \right],
	\label{eq:NonlinearActionTest}
\end{align}
where all terms after the integral signs are integrated over.
Note that we included the linear term~${\alpha\bar c\rho}$ into~${\mathcal{S}_{\text{int}}}$ rather than treating it as part of~${S_0}$. 
This is not necessary, but simplifies identifying the effective couplings and calculating the Feynman diagrams.
Our goal is to show that the terms included in Eq.\@~\eqref{eq:NonlinearActionTest}  are the only relevant interactions for the renormalization procedure.

To determine whether an interaction is relevant or irrelevant, we introduce the momentum scale~${\mu}$ and calculate the \textit{naive scale dependence} of the coupling parameters.
From Eq.\@~\eqref{eq:GaussianAction} it follows that~${\left[\rho\bar{\rho}\right] = \left[c\bar{c}\right] = \mu^d}$.
By rescaling~${\rho \rightarrow \mu^\xi \rho }$ and~${\bar{\rho} \rightarrow \mu^{-\xi} \bar{\rho}}$ some freedom in choosing the individual dimensions of the fields is left.
Indeed, this freedom implies that it is still possible to choose the naive scale dependence of the different interaction terms; a seeming contradiction to the fact that the relevance and irrelevance of couplings cannot be arbitrary.
However, this is only at first glance contradictory since it is \textit{a priori} not obvious how the different vertices -- and therefore the different field dimensions -- have to be combined to create valid Feynman diagrams.
For the action~\eqref{eq:NonlinearActionTest} we find that a~${\gamma}$-vertex can only appear together with a~${\Lambda}$-vertex and~${\chi_{1,2}}$-vertices only in combination with an~${\alpha}$- and a~${\Lambda}$-vertex.
Consequently, only specific combinations of vertices have a defined relevance/irrelevance under the RG procedure. 
To correctly analyze which interactions are relevant for the RG analysis, it is therefore prudent to rescale the action in such a way that it contains as few dimensionfull parameters as possible.
We choose
\begin{align}
	\rho \rightarrow s_1 \rho ,\quad
	\bar{\rho} \rightarrow s_1^{-1} \bar{\rho},	 \quad
	c \rightarrow s_1 s_2 c, \quad
	\bar{c} \rightarrow \left(s_1 s_2\right)^{-1} \bar{c},
	\label{eq:Rescaling}
\end{align}
where~${s_1 = \sqrt{\Lambda \gamma^{-1}}}$ and~${s_2 = \alpha D_c^{-1}}$.
Additionally we introduce the effective couplings 
\begin{align}
	u =  \frac{\gamma\Lambda}{32 \pi^2D_\rho^2},
	\ \, 
	g_{1,2} = \frac{\alpha \chi_{1,2}^{} \Lambda}{32 \pi^2 D_\rho^2D_c},
	\ \, 
	w  = \frac{D_c}{D_c+D_\rho},
	\label{eq:APPeffCoupl}
\end{align}
which should, respectively, be interpreted as the standard coupling from \textit{directed percolation}, the strength of the two different chemotactic interactions and a measure of the interaction time delay introduced
by the finite diffusion speed of the chemical.
Then the action reads
\begin{align}
	\label{eq:RescaledAction}
	S\left[\bar{\rho}, \rho, \bar{c}, c\right] 
	&= \int_{x,t}\bar{\rho} \left[ \partial_t -  D_\rho\nabla ^2 - \theta \right] \rho \, 
	+ \, \bar{c} \left[ \partial_tc - D_c \left(\nabla ^2 c - \frac{\lambda_c}{D_c}c + \rho\right)\right]\,
\notag \\[4mm]
	&+ \int_{x,t} \sqrt{32\pi^2u}D_\rho \, \bar{\rho} \left[\rho - \bar{\rho} \right] \rho 
	+\,\sqrt{32\pi^2u^{-1}}D_\rho \,\bar{\rho}\left[ g_1 \nabla \rho  \nabla  c 
	+ g_2 \rho  \nabla ^2 c \right].
\end{align}
Note that by rescaling time as ${t\rightarrow tD^{-1}_\rho}$ and writing ${D_c D_\rho^{-1} = w(1-w)^{-1}}$ the action can be expressed as a function of $u$, $g_{1/2}$ and $w$ only.
We choose not to do so here, since this simplifies introducing all required renormalization factors.
Note, however, that the perturbation series will only depend on $w$ and not on the individual diffusion constants.
Due to the specific form of the rescaling chosen in Eq.\@~\eqref{eq:Rescaling} the field dimensions can now be determined as
\begin{align}
	\left[\rho \right] = \left[\bar{\rho}\right] = \mu^{d/2}, \ 
	\left[c\right] = \mu^{d/2-2},\ 
	\left[\bar{c}\right] = \mu^{d/2+2}.
	\label{eq:FieldDimensions}
\end{align}
Accordingly, the dimensions of the remaining (effective) couplings are
$
	\left[u\right] = \left[g_1\right] = \left[g_2\right] = \mu^{4-d}.
$
Thus, we can identify the upper critical dimension, i.e.\@  the dimension where all couplings are marginal, as~${d_c=4}$.
Note that~${\left[w\right] = \mu^0}$ is not a problem since~${w}$ does not act as a smallness parameter for the perturbation expansion; one should rather interpret it as an interpolation between the two limits~${D_\rho \ll D_c}$ and~${D_c \ll D_\rho}$.
Since all vertex-prefactors are dimensionless at the upper critical dimension, also all combinations that might appear in the Feynman diagrams are dimensionless, showing that all included interactions are equally relevant.
For the same reason, we can now analyze the relevance of other interactions. 
If a certain vertex has a prefactor with negative~${\mu}$-scaling at~${d=d_c}$, no other vertex can counteract this scaling and it has to be irrelevant.
Strictly following the calculation displayed in~\cite{Tauber2014}, the action~${S}$ contains a third term of the form
\begin{align}
	S_{\text{1}} = \int_{x,t} -h_1 \bar{\rho}^2 \rho^2 
	+ h_2 \left( \nabla  \bar{\rho}\right)^2 \rho
	\, , 
\end{align}
whose appearance can be traced back to diffusive noise~(${h_2}$) and a higher order contribution in the demographic noise~(${h_1}$).
Dimensional analysis yields~${\left[h_1\right] = \mu^{2-d}}$ and~${\left[h_2\right] = \mu^{-d/2}}$, which renders both irrelevant in~${d=4-\varepsilon}$ dimensions.
Thus, our initial choice of neglecting these terms in Eq.\@~\eqref{eq:MeanFieldRho} is justified.
The same holds true for any possible noise term in Eq.\@~\eqref{eq:MeanFieldC}.
Taking into account all the reactions associated to~${c}$ and applying a Kramers-Moyal expansion, one can again follow the derivation of noise terms in~\cite{Tauber2014} to obtain the additional contribution
\begin{align}
	S_{2} = \int_{x,t} -h_3 \bar{c}^2 c - h_4 \bar{c}^2 \rho + h_5 \left(  \nabla  \bar{c}\right)^2 c
	\, .
\end{align}
The first two terms result from demographic noise in~${c}$, the last from diffusional noise. 
One finds~${\left[ h_3\right] = \mu^{-d/2}}$ and~${\left[ h_4\right] = \left[h_5\right] = \mu^{-2-d/2}}$, which are thus all irrelevant in~${d=4-\varepsilon}$ dimensions.
This justifies our assumption of~${c}$ being governed by a completely deterministic equation.

It remains to be shown that, as noted in the main part, only the leading order of the sensitivity function~${\chi[\rho, \nabla c]}$ yields a relevant contribution.
Moreover, we show that renormalizability implies~${\chi[\rho,c]=\chi[\rho, \nabla c]}$.
Given that~${\chi[\rho,c]}$ is an analytic function (which is not necessarily the case~\cite{Tindall2008}), the general contribution to the action reads
\begin{align}
	S_{3}=-\int_{x,t} h_{n_1,n_2,n_3} \, \left(  \nabla \bar{\rho} \right) \left( \rho  \nabla c \right)  \nabla ^{2n_1} \rho^{n_2} c^{n_3}.
\end{align}
The dimension of the coupling is then given by~${\left[h_{n_1,n_2,n_3}\right] = \mu^{f}}$, with
\begin{align}
	f(n_1,n_2,n_3) = \frac{1}{2} \left[(4-d)(1+n_3) - n_2 d- 4n_1\right].
\end{align}
Since, at~${d=4}$,~${f(n_1,n_2,n_3)<0}$ whenever~${n_1}$ or~${n_2}$ are greater than zero, all interactions resulting from such choices of~${n_{1,2}}$ are irrelevant for the RG calculations.
For~${n_1=n_2=0}$, on the other hand, all couplings are marginal at~${d=4}$ dimensions as~${f(0,0,n_3)=0}$, independent of~${n_3}$.
This means that in order to fully renormalize the theory to infinite loop order, one, in principle, has to include infinitely many interactions and therefore infinitely many counter terms (see below).
In that sense the action would no longer be renormalizable and one had to employ nonperturbative methods~\cite{Canet2010, Canet2021} to properly treat this set of relevant interactions.
To avoid this subtlety, one has to make the stronger assumption~${\chi[\rho, c] = \chi[\rho,  \nabla c]}$.
In this case, interactions with higher orders in~${c}$ always come with at least one derivative acting on each~${c}$.
It follows that~${n_3 \leq n_1}$ and~${f<0}$ at~${d=4}$ unless~${n_1=n_2=n_3=0}$, implying that for all renormalizable theories no relevant contributions besides~${\chi[\rho,  \nabla c] = \chi_0}$ exist.
Even though this assumption is unlikely to hold in a strict biological sense, chemotaxis is known to be robust over orders of magnitude of chemical concentration~\cite{Mesibov1973} and thus ~${\chi[\rho, c] = \chi[\rho, \boldsymbol{\nabla} c]}$ may be a reasonable approximation.

\begin{figure*}[b]
		\centering
		\includegraphics[width=14cm]{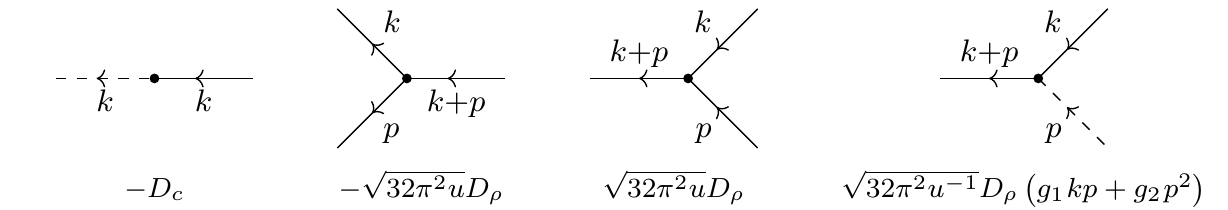}
		\caption{Relevant vertices of the chemotactic model with vertex factors given below. The~${\rho}$ fields are denoted by a solid line, the~${c}$ fields by dashed ones. Arrows indicate causality and the direction of the assigned momenta.}
		\label{fig:TabularVerticesChemotaxis}
\end{figure*}

Note that the above analysis completely relies on the naive scaling dimension of the involved coupling parameters.
For instance, care has to be taken for the predictions in~${d=2}$, since the vertex associated to~${h_1}$ is relevant for~${d \leq2}$.
Altogether, we have that the actions displayed in~\eqref{eq:GaussianAction},~\eqref{eq:NonlinearActionTest} and~\eqref{eq:RescaledAction} indeed contain all relevant vertices (displayed in Fig.\@~\ref{fig:TabularVerticesChemotaxis}) and yield a minimal model for bacterial chemotaxis.

%%% Diagram Calculations
\newpage
\section{Renormalization}

Having identified the relevant minimal action, one can apply a standard graphical perturbation expansion in terms of Feynman diagrams to calculate the flow equations to first non-trivial order.
Details of such a calculation can be found in~\cite{Tauber2014,Zinn-Justin2002} or any textbook on quantum-field theory.
Note that we perform all calculations in the absorbing state~(${\theta<0}$) close to the transition.

The central element of any perturbation expansion are the bare two-point Green's functions or \emph{propagators}
\begin{align}
	\langle \rho(q,\omega) \bar{\rho}(q^\prime,\omega^\prime) \rangle_0
	=\dfrac{\delta^{(d)}(q+q^\prime) \ \delta(\omega+\omega^\prime)}{-i\omega+D_\rho q^2-\theta}, \qquad
	\langle c(q,\omega) \bar{c}(q^\prime,\omega^\prime) \rangle_0  
	=\dfrac{\delta^{(d)}(q+q^\prime) \ \delta(\omega+\omega^\prime)}{-i\omega+D_cq^2+\lambda_c}
\end{align}
associated to the Gaussian part of the action~\eqref{eq:GaussianAction}, as well as the \emph{vertices} displayed in Fig.\@~\ref{fig:TabularVerticesChemotaxis}.
However, Green's functions calculated from this action turn out to be divergent.
To renormalize the above theory, we rely on a multiplicative renormalization scheme, where we introduce the~${Z}$-factors for the quadratic part of the action as
\begin{align}
	\bar{\rho} = \bar{Z}\bar{\rho}_R, \quad 
	\rho = Z \rho_R, \quad
	\tilde{c} = Z^{-1} \tilde{c}_R, \quad
	c = Z c_R, \quad 
	D_{\rho}  =  (Z\tilde{Z})^{-1} Z_D \, D_{\rho R}, \quad 
	\theta = (Z\tilde{Z})^{-1}(\delta+Z_\theta\, \theta_R)
	\, .
\label{eq:ZfactorsLinear}
\end{align}
The nonlinear terms on the other hand, are renormalized via the choice
\begin{align}
	u =  \mu^{\varepsilon} \, Z_u^2 Z_D^{-2} u_R, \quad
	g_1 = {\mu}^{\varepsilon} \, Z_u Z_{g_1} Z^{-1} Z_D^{-2}g_{1R}, \quad 
	g_2 = {\mu}^{\varepsilon} \, Z_u Z_{g_2} Z^{-1} Z_D^{-2}g_{2R} \, . \label{eq:ZfactorsNonLinear}
\end{align}
This leaves us with a total of seven independent renormalization factors.
The subscript~${R}$ indicates dimensionless renormalized parameters (except for the temperatures~${\theta}$ and~${\lambda_c}$, which still have a dimension).
Note that from now on we will drop this subscript~${R}$ as we  exclusively work with the renormalized couplings.
Following the usual steps, introducing the ${Z}$-factors gives rises to counter-terms which take the same form as the vertices shown in Fig.\@~\ref{fig:TabularVerticesChemotaxis}.\\
In principle the ${Z}$-factors can be determined iteratively to arbitrary loop orders by requiring every relevant vertex function to be finite.
Yet, there is some freedom as one can always add finite terms to impose additional renormalization constraints.
We choose not to do so and employ the so-called \emph{minimal subtraction} (MS) scheme.
Determining the ${Z}$-factors correctly to one loop order requires the calculation of all possible one loop Feynman diagrams.
To keep the calculations as concise as possible, we use a list of frequently used standard integrals, identities and abbreviations listed in the section \emph{Standard Integrals and Identities}.
\subsection{Propagator Renormalization}
\begin{figure}
    \begin{tabular}{cc}
             \includegraphics[width=0.22\textwidth]{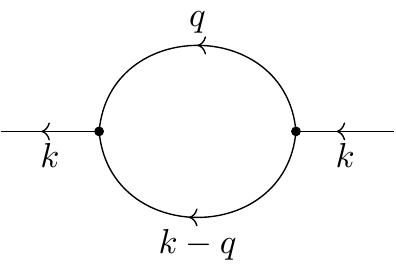} & \includegraphics[width=0.22\textwidth]{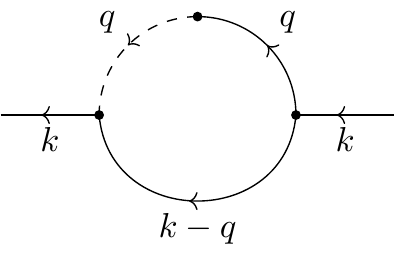} \\
    \end{tabular}
    \caption{The two diagrams contributing to the renormalization of the propagator:~${I^P_1}$ (left) and~${I^P_2}$ (right).}
    \label{fig:PropDiagrams}
\end{figure}

In the following, we give a summary of the calculations of all one loop Feynman diagrams.
More precisely, we determine their contributions to the renormalization of the corresponding Green's functions of the renormalized parameters.
Consequently, we only give the divergent parts of the integrals since all finite contributions are irrelevant for the renormalization.
We start with the renormalization of the propagator, which requires the calculation of the two diagrams shown in Fig.\@~\ref{fig:PropDiagrams}.
To calculate the first diagram~${I_1^P}$ we only have to identify the combinatorial factor associated with the diagram, the involved coupling constants and the propagators. Then we perform the~${\omega}$-integration and use Tab.\@~\ref{tab:StandardResults} to give the final result.
Thus,~${I_1^P}$ can be evaluated to
\begin{alignat}{2}
    I^P_1 &= -2\cdot 32\pi^2D_{\rho}^2 \mu^{\varepsilon}u\int_{q,\omega}\frac{1}{-i\omega+D_\rho q^2-\theta}\,\frac{1}{i(\omega-\Omega)+D_\rho(k-q)^2-\theta}\notag\\[1mm]
    &= -32\pi^2D_{\rho} \mu^{\varepsilon}u\int_q \frac{1}{(q-\frac{k}{2})^2+\Delta}\notag\\[1mm]
    &= -32\pi^2D_{\rho} u\cdot I_{0,1}(\Delta)\notag\\[1mm]
    &=\ \left(D_\rho k^2-2i\Omega-4\theta\right)\frac{u}{\varepsilon} \ ,\label{Result21}
\end{alignat}
where we used the abbreviation
${\Delta=\frac{k^2}{4}-\frac{i\Omega}{2D_\rho}-\frac{\theta}{D_\rho}}$
and employed the result for~${I_{0,1}(\Delta)}$ in the last line.
The calculation of~${I^P_2}$ needs more work and is given here step by step. 
The first few steps, albeit more tedious, are the same as before
\begin{align*}
    I^P_2 &= 2\cdot 32\pi^2 D_c D_{\rho}^2  \mu^{\varepsilon} \int_{q,\omega} \frac{-g_1(k-q)q-g_2q^2}{-i\omega+D_\rho q^2-\theta} \, \frac{1}{-i\omega+D_c q^2+\lambda_c} \, \frac{1}{i(\omega-\Omega)+D_\rho(k-q)^2-\theta}\\[4mm]
    &= 64\pi^2 D_c D_{\rho}^2 \mu^{\varepsilon} \int_q \frac{(g_1-g_2)q^2-g_1 kq}{2D_\rho q^2-2D_\rho kq+D_\rho k^2-i\Omega-2\theta} \, \frac{1}{(D_\rho +D_c)q^2-2D_\rho kq+D_\rho k^2-i\Omega-\theta+\lambda_c}\\[4mm]
    &= 32\pi^2 w D_\rho \mu^{\varepsilon} \int_q \frac{(g_1-g_2)q^2-g_1 kq}{q^2-kq+\frac{k^2}{2}-\frac{i\Omega}{2D_\rho }-\frac{\theta}{D_\rho }} \, \frac{1}{q^2-2(1-w)kq+(1-w)k^2+\frac{1-w}{D_\rho}(-i\Omega-\theta+\lambda_c)}\\[5mm]
    &=  32\pi^2 w D_\rho \mu^{\varepsilon} \int_{q,x} \frac{(g_1-g_2)q^2-g_1 kq}{\left(f(q,k,\Omega,x)\right)^2}.
\end{align*}
Since two propagators are left after the frequency integration one needs to employ the Feynman parameter identity Eq.\@~\ref{eq:FeynmanTrick}. This was done in the last line with~${\overline{x} = 1-x}$ and the use of the abbreviation
\begin{alignat*}{2}
    f &= \overline{x}\left(q^2-kq+\frac{k^2}{2}-\frac{i\Omega}{2D_\rho }-\frac{\theta}{D_\rho }\right)
    + x\left(q^2-2(1-w)kq+(1-w)k^2\right)
    + \frac{x(1-w)}{D_\rho}(-i\Omega-\theta+\lambda_c)\\
    &= q^2 - \Big(\overline{x}+2(1-w)x\Big)kq + \Big(\overline{x}+2(1-w)x\Big)\frac{k^2}{2}
    - \Big(\overline{x}+2(1-w)x \Big) \frac{i\Omega}{2D_\rho} -\Big(\overline{x}+(1-w)x\Big)\frac{\theta}{D_\rho} 
    + \frac{(1-w)x\lambda_c}{D_\rho}
    \\[2mm]
    &= \Big(q-\delta(x)\Big)^2 + \Delta(x).
\end{alignat*}
Here we have defined the auxiliary functions~${\delta(x)=\left(\frac{\overline{x}}{2}+(1-w)x\right)k}$ and
\begin{align*}
    \Delta(x) = -\delta^2 + \left(\frac{\overline{x}}{2}+(1-w)x\right)k^2 + \frac{(1-w)x\lambda_c}{D_\rho} 
	- \Big(\overline{x}+2(1-w)x \Big) \frac{i\Omega}{2D_\rho}-\Big(\overline{x}+(1-w)x\Big)\frac{\theta}{D_\rho}.
\end{align*}
Hence, the whole integral can be written as
\begin{alignat*}{2}
    I^P_2 &= 32\pi^2 wD_\rho \mu^{\varepsilon} \int_{q,x}\frac{(g_1-g_2)(q+\delta)^2-g_1k (q+\delta)}{(q^2+\Delta(x))^2}
    \\[2mm]
    &= 32\pi^2 wD_\rho \mu^{\varepsilon} \int_{q,x}\frac{(g_1-g_2)\left(q^2+\delta^2\right)-g_1\delta k }{(q^2+\Delta(x))^2}
    \\[2mm]
    &=  32\pi^2 wD_\rho \left(I_{0,2}(\Delta)\int_x \left((g_1-g_2)\delta^2-g_1 \delta k\right)
+(g_1-g_2)\int_x I_{2,2}(\Delta(x))\right).
\end{alignat*}
In the second line we got rid of all anti-symmetric parts of the integral, as well as the non divergent contributions. In the last line we made use of the fact that the divergent part of~${I_{0,2}(\Delta)}$ does not depend on~${\Delta}$ and thus can be pulled out of the~${x}$ integral.
To get to the final result the following intermediate integrals need to be calculated:
\begin{alignat*}{5}
    &\int_x \delta(x) = \frac{(3-2w)k}{4}, \qquad\int_x \delta^2(x) =  \frac{k^2}{12}\,\Big(7-10w+4w^2\Big), \\
    &\int_x \Delta(x) = \frac{k^2}{6}\Big(1+2w-2w^2\Big) + \Big(-3+2w\Big)\frac{i\Omega}{4D_\rho} + \Big(-2+w\Big)\frac{\theta}{2D_\rho}+\Big(1-w\Big)\frac{\lambda_c}{2D_\rho}.
\end{alignat*}
Additionally inserting~${I_{0,2}(\Delta)=\frac{1}{8\pi^2}\frac{1}{\varepsilon}+\mathcal{O}(1)}$ and~${I_{2,2}(\Delta)=-\frac{1}{4\pi^2}\frac{1}{\varepsilon}\Delta+\mathcal{O}(1)}$ at~${d=4-\varepsilon}$, gives
\begin{alignat}{4}
    I^P_2 
    =\ & \frac{4w}{\epsilon} &&\bigg\{2\big(g_2-g_1\big)\Bigg(\,\frac{(1+2w-2w^2)D_\rho k^2}{6} 
    +\frac{(-3+2w)i\Omega}{4}+\frac{(-2+w)\theta}{2}+\frac{(1-w)\lambda_c}{2}\Bigg)
    \notag\\[2mm]
    & &&+\frac{(-7+10w-4w^2)g_2D_\rho k^2}{12}+\frac{(-1-2w+2w^2)g_1D_\rho k^2}{6} \bigg\} 
    \notag\\[2mm]
    =\ & \frac{w}{\epsilon} &&\bigg\{\big(-2-4w+4w^2\big)g_1 D_\rho k^2 +\big(-1+6w-4w^2\big)g_2D_\rho k^2 
    \notag\\[2mm]
    & &&+ \big(g_2-g_1\big)\Big((-6+4w)i\Omega + (-8+4w)\theta +  (4-4w)\lambda_c\Big)\bigg\} \ .
\end{alignat}
\subsection{Renormalization of DP couplings}
\begin{figure}[t]
    \begin{tabular}{cccccc}
          \includegraphics[width=0.15\textwidth]{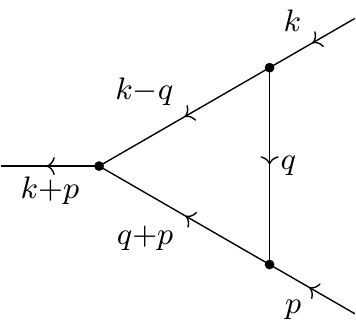} & 
          \includegraphics[width=0.15\textwidth]{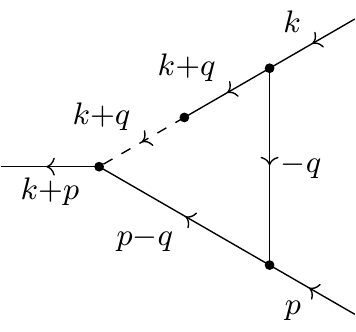} & 
          \includegraphics[width=0.15\textwidth]{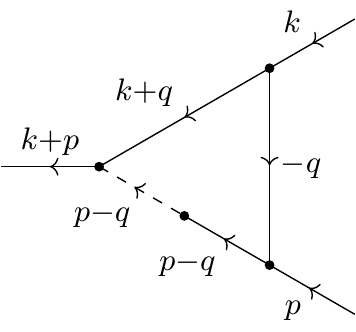} &
          \includegraphics[width=0.15\textwidth]{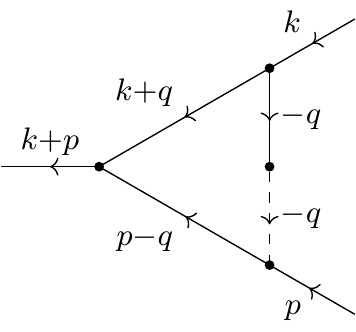} & 
          \includegraphics[width=0.15\textwidth]{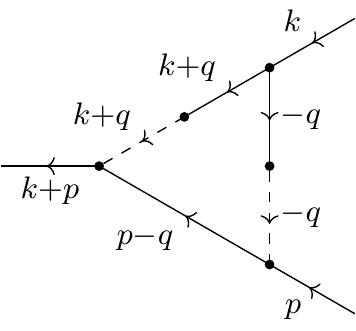} &
          \includegraphics[width=0.15\textwidth]{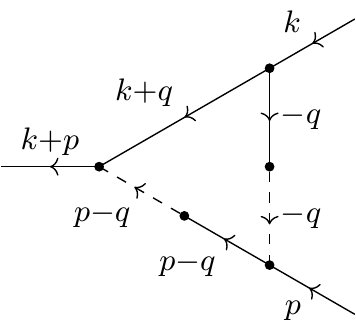}
    \end{tabular}
    \caption{All diagrams contributing to the renormalization of the~${\bar{\rho}\rho\rho}$ vertex. From left to right:~${I^u_1}$,~${I^u_2}$,~${I^u_3}$,~${I^u_4}$,~${I^u_5}$ and~${I^u_6}$.}
    \label{fig:DiagramsGamma}
\end{figure}
\noindent For the renormalization of the~${\bar{\rho}\rho\rho}$-vertex we need to consider six diagrams (cf.\@~Fig.\@~\ref{fig:DiagramsGamma}). 
Their respective contributions are calculated in the following. 
Note that we set all external momenta and frequencies to zero since the DP vertices are momentum and frequency independent.
From a dimensional analysis it can be further deduced that none of the divergences depend on the terms of~${\mathcal{O}(q)}$ and lower in the propagators (which would contribute to the momentum shift~${\delta}$ and the argument of the standard integrals~${\Delta}$ in previous calculations). 
These terms are abbreviated as~${\boldsymbol{\cdot}}$. \\
Also note that none of the Feynman parameter integrals depend on the Feynman parameters and the additional constants introduced by Eq.\@~\eqref{eq:FeynmanTrick} and~\eqref{eq:ParameterInts} always cancel each other. 
Having this in mind, the following calculations are performed in four steps: First all vertex factors, combinatorical prefactors and propagators are introduced; then the frequency integral is performed by identifying the poles of the propagators; subsequently all diffusion constants are extracted and the appropriate standard integral from Tab.\@~\ref{tab:StandardResults} introduced; finally the result for the integral is inserted and the prefactors used to formulate the result in terms of the effective coupling constants.
Following these steps one can give the results for the contributions of the~${\bar{\rho}\rho\rho}$-vertex renormalization as:
\begingroup
\allowdisplaybreaks
\begin{align}
    I^u_1 &= 
    16 \cdot (32 \pi^2 \mu^{\varepsilon}u)^{3/2} D_\rho^3 \int_{q,\omega}\frac{1}{i\omega+D_\rho q^2-\theta} \,\frac{1}{\left(-i\omega+D_\rho q^2-\theta\right)^2} \\
    &= 16 (32 \pi^2 \mu^{\varepsilon}u)^{3/2} D_\rho^3 \int_q\frac{1}{2D_\rho q^2 + \boldsymbol{\cdot}} \ \frac{1}{2D_\rho q^2 + \boldsymbol{\cdot}}
    \notag \\[2mm]
    &= 4 (32 \pi^2 u)^{3/2} D_\rho \mu^{\varepsilon/2}\ I_{0,2} \notag \\[2mm]
    &= \frac{16u \, \sqrt{32\pi^2\mu^{\varepsilon}u} D_\rho}{\varepsilon} 
    \span \span \\[5mm]
    I^u_2 &= 
    8\cdot 32\pi^2 \sqrt{32\pi^2\mu^{\varepsilon}u} \, D_c D_{\rho}^3 \mu^{\varepsilon}\int_{q,\omega} \frac{\left(-g_1 q^2+g_2q^2\right)}{-i\omega+D_\rho q^2-\theta} \, \frac{1}{-i\omega+D_c q^2+\lambda_c} \frac{1}{i\omega+D_\rho q^2-\theta} \, \frac{1}{i\omega+D_\rho q^2-\theta} \notag \\[2mm]
    &= 8\cdot 32\pi^2 \sqrt{32\pi^2\mu^{\varepsilon}u} \,  D_c D_{\rho}^3 \mu^{\varepsilon}\Bigg(\int_q \frac{(g_2-g_1)q^2}{(D_c-D_\rho)q^2 + \boldsymbol{\cdot}}\, \frac{1}{\left(2D_\rho q^2+\boldsymbol{\cdot}\right)^2} + \int_q \frac{(g_2-g_1)q^2}{(D_\rho-D_c)q^2 + \boldsymbol{\cdot}}\, \frac{1}{\left((D_\rho+D_c) q^2+\boldsymbol{\cdot}\right)^2}\Bigg)
    \notag\\[2mm]
    &= \frac{8\cdot 32\pi^2 \sqrt{32\pi^2\mu^{\varepsilon}u} \,  D_c D_{\rho}^3 (g_2-g_1)}{D_\rho-D_c}\left(\frac{1}{(D_\rho+D_c)^2}-\frac{1}{4D_\rho^2}\right)I_{2,3}
    \span \span \notag \\[2mm]
    &= \frac{8(g_1-g_2)\,w\,(-3+2w)\,\sqrt{32\pi^2\mu^{\varepsilon}u}D_\rho}{\varepsilon}
    \span \span \\[5mm]
    I^u_3 &= 
    8\cdot 32\pi^2 \sqrt{32\pi^2\mu^{\varepsilon}u} \, D_c D_{\rho}^3 \mu^{\varepsilon} \int_{q,\omega} 
    \frac{\left(-g_1 q^2+g_2q^2\right)}{-i\omega+D_\rho q^2-\theta}
    \frac{1}{\left(i\omega+D_\rho q^2-\theta\right)^2} \, \frac{1}{i\omega+D_c q^2+\lambda_c}
     \notag\\[2mm]
    &= 8\cdot 32\pi^2 \sqrt{32\pi^2\mu^{\varepsilon}u} \, D_c D_{\rho}^3 \mu^{\varepsilon} \int_q \frac{\left(g_2-g_1\right)q^2}{(D_\rho+D_c)q^2+\boldsymbol{\cdot}} \, \frac{1}{\left(2D_\rho q^2 + \boldsymbol{\cdot}\right)^2}
    \notag \\[2mm]
    &= 2\cdot 32\pi^2 \sqrt{32\pi^2\mu^{\varepsilon}u}D_{\rho} \, w (g_2-g_1)\, I_{2,3}
    \notag  \span \span\\[2mm]
    &= \frac{8(g_2-g_1)w \,\sqrt{32\pi^2\mu^{\varepsilon}u}D_{\rho}}{\varepsilon}
    \span \span\\[5mm]
    I^u_4 &= 
    8\cdot 32\pi^2 \sqrt{32\pi^2\mu^{\varepsilon}u} \, D_c D_{\rho}^3 \mu^{\varepsilon} \int_{q,\omega} \frac{g_2q^2}{-i\omega+D_\rho q^2-\theta} 
    \frac{1}{i\omega+D_c q^2+\lambda_c} \, \frac{1}{\left(i\omega+D_\rho q^2-\theta\right)^2} 
    \notag\\[2mm]
    &= 8\cdot 32\pi^2 \sqrt{32\pi^2\mu^{\varepsilon}u} \, D_c D_{\rho}^3 \mu^{\varepsilon}\int_q \frac{g_2 q^2}{(D_\rho+D_c)q^2 + \boldsymbol{\cdot}} \, \frac{1}{\left(2D_\rho q^2 + \boldsymbol{\cdot}\right)^2}
    \notag\\[2mm]
    &= 2\cdot 32\pi^2 \sqrt{32\pi^2\mu^{\varepsilon}u} D_{\rho} w g_2\, I_{2,3} \notag \span \span \\*[2mm]
    &= \frac{8g_2 w \, \sqrt{32\pi^2\mu^{\varepsilon}u}D_\rho}{\varepsilon}
    \span \span \\[5mm]
    I^u_5 &=
    4 \cdot 32\pi^2 \sqrt{32\pi^2\mu^{\varepsilon}u^{-1}} \, D_c^2 D_{\rho}^3 \mu^{\varepsilon} \int_{q,\omega}\frac{g_2q^2\left(-g_1q^2+g_2q^2\right)}{-i\omega+D_\rho q^2-\theta} \, \frac{1}{-i\omega+D_c q^2+\lambda_c} \frac{1}{\left(i\omega+D_\rho q^2-\theta\right)^2} \, \frac{1}{i\omega+D_c q^2+\lambda_c}
    \notag\\[2mm]
    &= 4 \cdot 32\pi^2 \sqrt{32\pi^2\mu^{\varepsilon}u^{-1}} \, D_c^2 D_{\rho}^3 \mu^{\varepsilon}\Bigg(\int_q \frac{g_2(g_2-g_1)q^4}{(D_\rho+D_c)q^2+\boldsymbol{\cdot}} \, \frac{1}{\left(2D_\rho q^2 + \boldsymbol{\cdot}\right)^2} \frac{1}{(D_c-D_\rho)q^2+\boldsymbol{\cdot}} \notag\\[2mm]
	 &\phantom{=4 \cdot 32\pi^2 \sqrt{32\pi^2\mu^{\varepsilon}u^{-1}} \, D_c^2 D_{\rho}^3 \mu^{\varepsilon}}+ \int_q \frac{g_2(g_2-g_1)q^4}{2D_c q^2+\boldsymbol{\cdot}} \, \frac{1}{\left((D_\rho+D_c) q^2 + \boldsymbol{\cdot}\right)^2} \frac{1}{(D_\rho-D_c)q^2+\boldsymbol{\cdot}}\Bigg)
    \notag\\*[2mm]
    &= \frac{2 \cdot 32\pi^2 \sqrt{32\pi^2\mu^{\varepsilon}u^{-1}} \, D_c^2 D_{\rho}^3 \, g_2(g_2-g_1)}{(D_\rho-D_c)(D_\rho+D_c)} \left(\frac{1}{D_c(D_\rho+D_c)}-\frac{1}{2D_\rho^2}\right)I_{4,4}
    \notag \span \span\\[2mm]
    &= \frac{4g_2(g_2-g_1)w(2-w)}{u}\,\frac{\sqrt{32\pi^2\mu^{\varepsilon}u}D_\rho}{\varepsilon}
    \span \span\\[5mm]
    I^u_6 &=
    4 \cdot 32\pi^2 \sqrt{32\pi^2\mu^{\varepsilon}u^{-1}} \, D_c^2 D_{\rho}^3 \mu^{\varepsilon}\int_{q,\omega} \frac{g_2q^2\left(-g_1q^2 + g_2 q^2\right)}{-i\omega+D_\rho q^2-\theta} \frac{1}{\left(i\omega +D_c q^2+\lambda_c\right)^2} \, \frac{1}{\left(i\omega+D_\rho q^2-\theta\right)^2}
     \notag\\*[2mm]
     &= 4 \cdot 32\pi^2 \sqrt{32\pi^2\mu^{\varepsilon}u^{-1}} \, D_c^2 D_{\rho}^3 \mu^{\varepsilon} \int_q \frac{g_2(g_2-g_1)q^4}{\left(2D_\rho q^2 + \boldsymbol{\cdot}\right)^2} \, \frac{1}{\left((D_\rho+D_c)q^2 + \boldsymbol{\cdot}\right)^2}
     \notag \\*[2mm]
    &= 32\pi^2 \sqrt{32\pi^2\mu^{\varepsilon}u^{-1}} D_{\rho} \, w^2g_2(g_2-g_1) \, I_{4,4}
    \notag \span \span\\*[2mm] 
    &= \frac{4g_2(g_2-g_1)w^2}{u}\frac{\sqrt{32\pi^2\mu^{\varepsilon}u}D_\rho}{\epsilon}
    \span \span
\end{align}
\endgroup
To calculate the results for the~$\bar{\rho}\bar{\rho}\rho$-renormalization, we point out that there are three contributing diagrams which can be obtained by replacing the~$\bar{\rho}\rho\rho$-vertex with a~$\bar{\rho}\bar{\rho}\rho$-vertex in~${I^u_1}$,~${I^u_2}$ and~${I^u_3}$.
Hence, the analytical results for the diagrams can be retrieved by just adding a minus sign in the respective calculations of the~$\bar{\rho}\rho\rho$ renormalization.
\begin{figure}[h]
    \begin{tabular}{ccc}
         \includegraphics[width=0.2\textwidth]{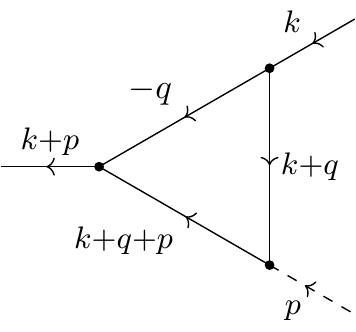} & 
         \includegraphics[width=0.2\textwidth]{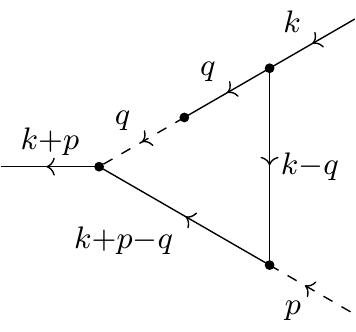} & 
         \includegraphics[width=0.2\textwidth]{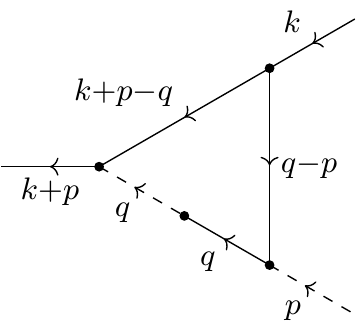}
    \end{tabular}
    \caption{The three diagrams renormalizing the chemotactic couplings. From left to right:~${I^g_1}$,~${I^g_2}$ and~${I^g_3}$.}
    \label{fig:ChemoDiagrams}
\end{figure}\noindent
\subsection{Renormalization of Chemotactic Couplings}

The only remaining diagrams are the ones required for the renormalization of the chemotactic couplings (Fig.\@~\ref{fig:ChemoDiagrams}).
Since these diagrams depend on external momenta, one cannot set them to zero and has to keep track of their contributions to the terms proportional to the loop momentum~${q}$ in the propagators.
The terms of order~${\mathcal{O}(1)}$ in~${q}$ are again denoted by~${\boldsymbol{\cdot}}$ and can be neglected.
To calculate the contributions of~${I_1^g}$ we, write down all the coupling constants and propagators, perform the frequency integration and introduce the Feynman parameters~${x}$ and~${\overline{x} = 1-x}$~\eqref{eq:FeynmanTrick}:
\pagebreak
\begingroup
\allowdisplaybreaks
\begin{align}
     I^g_1&=
     4 \cdot 32\pi^2 \sqrt{32\pi^2\mu^{\varepsilon}u} D_{\rho}^3 \mu^{\varepsilon} \int_{q,\omega}\frac{\left(g_1(k+q)p+g_2 p^2\right)}{-i\omega+D_\rho  (k+p+q)^2-\theta} \frac{1}{-i\omega+D_\rho  (k+q)^2-\theta} \, \frac{1}{i\omega+D_\rho  q^2-\theta} \notag \\*[2mm]
     &= 4 \cdot 32\pi^2 \sqrt{32\pi^2\mu^{\varepsilon}u} D_{\rho}^3 \mu^{\varepsilon} \int_q\frac{g_1(k+q)p + g_2 p^2}{2D_\rho\left( q^2 +kq\right) + \boldsymbol{\cdot}} \, \frac{1}{2D_\rho \left(q^2 + (k+p)q\right) + \boldsymbol{\cdot}}
     \notag \\*[2mm]
     &= 32\pi^2 \sqrt{32\pi^2\mu^{\varepsilon}u} D_{\rho} \mu^{\varepsilon} \int_{q,x} \frac{g_1(k+q)p + g_2 p^2}{\big(q^2 + (xk+\overline{x}(k+p))q + \boldsymbol{\cdot}\big)^2} \notag\\
     &= 32\pi^2 \sqrt{32\pi^2\mu^{\varepsilon}u} D_{\rho} \mu^{\varepsilon} \int_{q,x}\frac{g_1(k+q+\delta)p + g_2 p^2}{(q+\boldsymbol\cdot)^2} 
     \notag \\*[2mm]
     &=  32\pi^2 \sqrt{32\pi^2\mu^{\varepsilon}u} D_{\rho} \left(g_1 kp+g_2 p^2+g_1 p\int_x \delta\right)I_{0,2}
     \notag \\[2mm]
     &= \sqrt{32\pi^2u^{-1}} \mu^{\varepsilon/2}D_\rho\left( \frac{2u}{\varepsilon} g_1 \cdot(kp) + \left(4-\frac{g_1}{g_2}\right)\frac{u}{\varepsilon} g_2 \cdot p^2\right), 
     \span \span \label{eq:Ichi1}
\end{align}
\endgroup
Where, in the fourth line, we defined~${\delta(x) = -\frac{1}{2}(k+\overline{x}p)}$ and shifted~${q\rightarrow q+\delta}$.
We now turn our attention to~${I^g_2}$. 
Simply inserting all the coupling constants and propagators yields
\begin{align*}
   I^g_2 =
   2\cdot 32\pi^2 \sqrt{32\pi^2\mu^{\varepsilon}u^{-1}} D_{\rho}^3 D_c \mu^{\varepsilon} \int_{q,\omega}\frac{A(q,k,p)}{-i\omega+D_\rho q^2-\theta}\frac{1}{-i\omega+D_cq^2+\lambda_c}
   \frac{1}{i\omega+D_\rho(k-q)^2-\theta} \frac{1}{i\omega+D_\rho(k+p-q)^2-\theta}.
\end{align*}
Here we introduced~${A(q,k,p)}$ as the product of the vertex factors for the two~${\Tilde{\rho}\rho c}$-vertices in the diagram:
\begin{equation*}
    A(q,k,p) = \big(g_1 (k-q)p+g_2 p^2\big)\big(g_1(k+p-q)q+g_2 q^2\big)
\end{equation*}
Note that no terms in~${A(q,k,p)}$ are of order~${\mathcal{O}(q^4)}$. Additionally, there are four propagators in the diagram and after performing the~${\omega}$ integral three will be left. 
Thus, all diverging parts are proportional to~${I_{2,3}(\Delta)}$ or~${\Tilde{I}_{2,3}(\Delta)}$ and, therefore, independent of~${\Delta}$. 
Hence, all the parts contributing to~${\Delta}$ are only denoted by~${\boldsymbol{\cdot}}$ and the~${\Delta}$ dependence of~${I_{2,3}}$ and~${\Tilde{I}_{2,3}}$ dropped in the following calculation. 
Then, performing the frequency integral results in
\begin{align*}
     I^g_2 &= 
     64\pi^2 \sqrt{32\pi^2\mu^{\varepsilon}u^{-1}} D_{\rho}^3 D_c \mu^{\varepsilon} \int_q \frac{A(q,k,p)}{(D_{c}-D_\rho )q^2+\boldsymbol{\cdot}}\,\frac{1}{2D_\rho q^2-2D_\rho kq+\boldsymbol{\cdot}}
     \frac{1}{2D_\rho q^2-2D_\rho (k+p)q+\boldsymbol{\cdot}}\,\\[2mm]
     &+64\pi^2 \sqrt{32\pi^2\mu^{\varepsilon}u^{-1}} D_{\rho}^3 D_c \mu^{\varepsilon} \int_q \frac{A(q,k,p)}{(D_\rho-D_{c})q^2+\boldsymbol{\cdot}}\,\frac{1}{(D_\rho +D_{c})q^2-2D_\rho kq+\boldsymbol{\cdot}} \frac{1}{(D_\rho +D_{c})q^2-2D_\rho(k+p)q+\boldsymbol{\cdot}}.
\end{align*}
We continue by pulling out the diffusion constants and using the Feynman parameter trick:
\begin{align*}
     I^g_2 
     &= \frac{128\pi^2 \sqrt{32\pi^2\mu^{\varepsilon}u^{-1}} D_{\rho}^3 D_c \mu^{\varepsilon}}{D_\rho-D_c} \bigg\{-\frac{1}{4D_\rho^2}\int_{q,x,y,z}\frac{A(q,k,p)}{\big(q^2-2\delta_1(x,y,z)q+\boldsymbol{\cdot}\big)^3} +\frac{1}{(D_\rho+D_c)^2}\int_{q,x,y,z}\frac{A(q,k,p)}{\big(q^2-2\delta_2(x,y,z)q+\boldsymbol{\cdot}\big)^3}\bigg\}\\[2mm]
     &= \frac{128\pi^2 \sqrt{32\pi^2\mu^{\varepsilon}u^{-1}} D_{\rho}^3 D_c \mu^{\varepsilon}}{D_\rho-D_c} \bigg\{-\frac{1}{4D_\rho^2}\int_{q,x,y,z}\frac{A(q+\delta_1)}{\big(q^2+\boldsymbol{\cdot}\big)^3} +\frac{1}{(D_\rho+D_c)^2}\int_{q,x,y,z}\frac{A(q+\delta_2)}{\big(q^2+\boldsymbol{\cdot}\big)^3}\bigg\}
\end{align*}
In the last line we shifted the loop momentum in both integrals by~${\delta_1}$ and~${\delta_2}$, respectively, where 
\begin{align*}
    \delta_1(x,y,z) &= \frac{1}{2}\big(xk+y(k+p)\big) \\
    \delta_2(x,y,z) &= (1-w)\big(xk+y(k+p)\big) = 2(1-w)\,\delta_1.
\end{align*}
Now, we need to calculate the shifted numerator~${A(q+\delta)}$:
\begin{align*}
    A(q+\delta) 
    &= (g_1^2-g_1g_2)\big((\delta p)\,q^2+2(\delta q)(pq)\big)
    -g_1^2(pq)(kq) +(g_1g_2-g_1^2)(kp)\,q^2 \\
    &-g_1^2(pq)^2+(g_2^2-g_1g_2)p^2q^2
\end{align*}
Inserting~${\delta_2 = 2(1-w)\,\delta_1}$ into the integral, one can anticipate the appearance of the following expressions:
\begin{align*}
    -\frac{1}{4D_\rho^2}+\frac{1}{(D_\rho+D_c)^2} &= \frac{(D_\rho-D_c)}{4D_\rho^2D_c}\,w(3-2w) \\
    -\frac{1}{4D_\rho^2}+\frac{2(1-w)}{(D_\rho+D_c)^2} &= \frac{(D_\rho-D_c)}{4D_\rho^2D_c}w(7-10w+4w^2).
\end{align*}
With these results and~${\int_{x,y,z} 1 = 1/2}$, one can separate the~${\delta_1}$ dependent terms and get
\begin{align*}
    I^g_2 = 
     32\pi^2 \sqrt{32\pi^2\mu^{\varepsilon}u^{-1}} D_{\rho} \, w&\bigg\{\frac{3-2w}{2} \Big((g_1g_2 - g_1^2) I_{2,3}(kp)-g_1^2\Tilde{I}_{2,3}(k,p) +(g_2^2 - g_1g_2)I_{2,3}\,p^2 - g_1^2\Tilde{I}_{2,3}(p,p)\Big)\\[2mm]
    &+(7-10w+4w^2)(g_1^2-g_1g_2)\cdot\left(\int_{x,y,z}2\Tilde{I}_{2,3}(p,\delta_1) + I_{2,3}(p\delta_1)\right)\bigg\}.
\end{align*}
Inserting the results for~${I_{2,3}}$,~${\Tilde{I}_{2,3}}$ and~${\int_{x,y,z}\delta_1 = \frac{1}{12}(2k+p)}$ yields:
\begin{align*}
    I^g_2 
    = \frac{\sqrt{32\pi^2\mu^{\varepsilon}u^{-1}} D_{\rho} w\mu^{\varepsilon/2}}{2\varepsilon} \, &\bigg\{ (7-10w+4w^2)(g_1^2-g_1g_2)(2k+p)p \\
    &+ (3-2w)\Big((-5g_1^2+4g_1g_2)kp+(-g_1^2-4g_1g_2+4g_2^2)p^2\Big)\bigg\}
\end{align*}
Collecting and grouping all the contributions yields the final result:
\begin{align}
    I^g_2 = 
    &\bigg\{\Big((4-8w+4w^2)\frac{g_1^2}{g_2}+(-19+18w-4w^2)g_1
    +(12-8w)g_2\Big)g_2p^2 \notag \\ 
    &+\Big((-1-10w+8w^2)g_1+(-2+12w-8w^2)g_2\Big)g_1 (kp)\bigg\} \, \frac{w \sqrt{32\pi^2 u^{-1}} D_{\rho}\mu^{\varepsilon/2}}{2\varepsilon}
\end{align}
Finally, we need to calculate
\begin{align*}
     I^g_3 =
     2\cdot 32\pi^2\sqrt{32\pi^2 \mu^{\varepsilon} u^{-1}} D_\rho^3D_c \mu^{\varepsilon}\int_{q,\omega} \frac{A(q,k,p)}{i\omega+D_\rho (k+p-q)^2-\theta}\, \frac{1}{-i\omega+D_\rho q^2-\theta} \frac{1}{-i\omega+D_{c}q^2+\lambda_c} \, \frac{1}{-i\omega+D_\rho (q-p)^2-\theta},
\end{align*}
where~${A(q, k, p)}$ again denotes the product of the chemotactic vertices.
\begin{align*}
    A(q,k,p) = \left(g_1(q-p)p+g_2 p^2\right)\left(g_1(k+p-q)q+g_2q^2\right)
\end{align*}
As before, we first calculate the frequency integral and introduce the Feynman parameters:
\begin{align*}
    I^g_3 &=
    16\pi^2\sqrt{32\pi^2\mu^{\varepsilon}u^{-1}} D_\rho w \mu^{\varepsilon} \int_q \frac{A(q,k,p)}{q^2-2(1-w)(k+p)q+\boldsymbol\cdot}
    \frac{1}{q^2-(k+2p)q+\boldsymbol\cdot} \, \frac{1}{q^2-(k+p)q+\boldsymbol\cdot}
    \\[2mm]
    &= 32\pi^2\sqrt{32\pi^2\mu^{\varepsilon}u^{-1}} D_\rho w \mu^{\varepsilon} \int_{q,x,y,z} \frac{A(q,k,p)}{((q-\delta(x,y,z))^2 + \boldsymbol\cdot)^3}
\end{align*}
With~${\delta}$ given by
\begin{equation*}
    \delta(x,y,z) = \frac{1}{2}\left(k+2p\right)x+\frac{1}{2}(k+p)y+(1-w)(k+p)z.
\end{equation*}
Shifting~${q\rightarrow q+\delta}$ and only keeping terms of order~${\mathcal{O}(q^2)}$ in~${A(q+\delta,k,p)}$ gives:
\begin{align*}
    A(q+\delta,k,p) = \left(g_1g_2-g_1^2\right)\left(2(pq)(\delta q)+(\delta p) q^2\right)
    + g_1^2(pq)(kq) + g_1^2(pq)^2 + (g_1-g_2)^2q^2 p^2
\end{align*}
Now we can collect all the terms, insert the values of~${I_{2,3}}$ and~${\Tilde I_{2,3}}$ and give the final result as
\begin{align}
     I_3^g &=
     16\pi^2\sqrt{32\pi^2\mu^{\varepsilon}u^{-1}} D_\rho w \bigg\{ (g_1-g_2)^2p^2I_{2,3} 
	 +2\left(g_1g_2-g_1^2\right)\left(\int_{x,y,z} (\delta p) \, I_{2,3}+2\Tilde{I}_{2,3}(p,\delta)\right) \notag\\
	 &\phantom{=16\pi^2\sqrt{32\pi^2\mu^{\varepsilon}u^{-1}} D_\rho w}
	 +g_1^2\left(\Tilde{I}_{2,3}(p,p)+\Tilde{I}_{2,3}(k,p)\right)\bigg\} \notag\\
     &= \frac{\sqrt{32\pi^2  u^{-1}} D_\rho \mu^{\varepsilon/2} \, w}{2\varepsilon}\bigg\{\Big((-3+2w)g_1+(4-2w)g_2\Big)\,g_1\, (kp) 
	 + \Big( \frac{2wg_1^2}{g_2}-(3+2w)g_1+4g_2\Big)\,g_2\,p^2\bigg\} \ .
\end{align}
%

%%% FlowEquations

\section{Z-Factors and flow equations}

To determine expressions for the the~${Z}$-factors in Eqs.\@~\eqref{eq:ZfactorsLinear} and~\eqref{eq:ZfactorsNonLinear}, we rely on the \emph{minimal subtraction} (MS) scheme, meaning we give the minimal choice of~${Z}$ without imposing any further renormalization conditions.
This requires adding all the previous (diverging) results for the propagator and the different three-point functions together.
Care has to be taken with respect to the sign of the different interactions in the action and the multiplicity of the counter terms.
Keeping this in mind, one can read off~${\delta=4w(1-w)(g_1-g_2)\lambda_c}$ and
\begingroup
\allowdisplaybreaks
\begin{align}
    Z &= 1+ \Big\{u+wg_1(2w-3) + wg_2(5-2w)+2wg_2u^{-1}(g_2-g_1)\Big\}\,\frac{1}{\varepsilon}
    \\[2mm]
    \tilde{Z} &= 1 + \Big\{u+wg_1(2w-3) + wg_2(1-2w) - 2wg_2u^{-1}(g_2-g_1)\Big\}\,\frac{1}{\varepsilon}
    \\[2mm]
    Z_D &= 1 + \Big\{ u - 2wg_1(-2w^2+2w+1) + wg_2 (-4w^2+6w-1)\Big\}\,\frac{1}{\varepsilon} 
    \\[2mm]
    Z_\theta &= 1+ \Big\{ 4u+ 4w(g_2-g_1) (-w+2)\Big\}\,\frac{1}{\varepsilon}
    \\[2mm]
    Z_u &= 1+ \Big\{7u + wg_1(6w-13) + wg_2(15-6w) + 2w(g_2-g_1)g_2u^{-1}\Big\}\,\frac{1}{\varepsilon}
    \\[2mm]
    Z_{g_1^{}} &= 1+ \Big\{2u+wg_1(4w^2-4w-2)+wg_2(-4w^2+5w+1)\Big\}\,\frac{1}{\varepsilon}
    \\[2mm]
    Z_{g_2^{}} &= 1+ \Big\{4u-\frac{ug_1}{g_2}+wg_1\{-2w^2+8w-11) +wg_2(-4w+8) +\frac{wg_1^2}{g_2}(2w^2-3w+2)\Big\}\,\frac{1}{\varepsilon}
\end{align}
\endgroup

\subsection{Flow Equations}

The above~${Z}$-factors can be used to determine how the system, defined by the various vertex-functions, behaves at different length scales, in particular in the IR-limit.
To this end we relate the bare and renormalized vertex functions as~${\Gamma^{(n,\tilde{n})}=Z^n \tilde Z^{\tilde n}\Gamma_B^{(n,\tilde{n})}}$, where~${n}$ and~${\tilde{n}}$ denote the multiplicity of density and response fields, respectively.
Utilizing that bare quantities are independent of the scale parameter~${\mu}$ one obtains the \emph{Callan-Symanzik}~(CZ) equations
\begin{align}
    \left(\mu\partial_\mu + \beta_\theta\theta\partial_\theta+\beta_DD_\rho\partial_{D_\rho}+\beta_{\lambda_i}\partial_{\lambda_i} - n \sigma -\tilde{n} \tilde{\sigma}\right)\Gamma_R^{(n,\tilde{n})} = 0\label{eq:CZ},
\end{align}
where~${\{\lambda_i\}_i}$ is the collection of effective coupling parameters and the \emph{flow functions}~${\beta_i}$ are given by
\begin{align}
    \beta_{\lambda_i} =\frac{\mathrm{d} \ln(\lambda_i)}{\mathrm{d}\ln\mu}, \quad
    \beta_D =\frac{\mathrm{d} \ln(D_\rho)}{\mathrm{d}\ln\mu}, \quad 
    \beta_\theta =\frac{\mathrm{d}\ln(\theta)}{\mathrm{d}\ln\mu}.
    \label{eq:Betafunctions}
\end{align}
Moreover, we use the abbreviations~${\sigma = \mathrm{d} \ln(Z)/{\mathrm{d} \ln(\mu)}}$ and~${\tilde{\sigma} = \mathrm{d} \ln(\tilde{Z})/{\mathrm{d} \ln(\mu)}.}$
Note that the beta functions~${\beta_{\lambda_i}}$ contain all the scale dependence of our theory.
From the~${Z}$ factors we infer
\begingroup
\allowdisplaybreaks
\begin{align}
    \frac{\mathrm{d} w}{\mathrm{d}\ln\mu} &= w(1-w)\Big(u-(4w^2-8w+4)wg_1 + (4 w^2-10w+7) wg_2\Big),
    \label{eq:AppFloww}\\[2mm]
    \frac{\mathrm{d} u}{\mathrm{d}\ln\mu} &= -\varepsilon u + 12u^2 - (8w^2-20w+22) wug_1 +(8w^2-24w+32)wug_2 +4wg_2(g_2-g_1),
    \\[2mm]
    \frac{\mathrm{d} g_1}{\mathrm{d}\ln\mu} &= -\varepsilon g_1 +6ug_1 -4(w^2-2w+2)wg_1^2 +(4w^2-11w+13)wg_1g_2, 
    \label{eq:AppFlowG1}\\[2mm]
    \frac{\mathrm{d} g_2}{\mathrm{d}\ln\mu} &= -\varepsilon g_2 -ug_1 +8ug_2 + (2w^2-3w+2)wg_1^2 +(-10w^2+20w-17)wg_1g_2 + 4(2w^2-5w+5)wg_2^2. 
    \label{eq:AppFlowG2}
\end{align}
\endgroup
%
%% Fixed Point Table
%%%%%%%%%%%%%%%%%%%%%%%%%%%%%%%%%%%%%
\begin{table*}[tb]
	\begin{ruledtabular}
	\begin{tabular}{c|cccc|cccc|>{$}c<{$}>{$}c<{$}>{$}c<{$}>{$}c<{$}}
		 & $\mathbf{u^*}$ & $\mathbf{g_1^*}$  & $\mathbf{g_2^*}$ & $\mathbf{w^*}$ & $\mathbf{\Sigma_1}$ & $\mathbf{\Sigma_{2}}$ & $\mathbf{\Sigma_{3}}$ & $\mathbf{\Sigma_4}$ & \boldsymbol{\eta} & \boldsymbol{\nu} & \mathbf{z} & \boldsymbol{\delta} \\[1mm]
		 \hline &&&&&&&&&&&& \\[-3mm]
		\textbf{GA} & 0 & 0 & 0 & $w^*$ & $-\varepsilon$ & $-\varepsilon$ & $-\varepsilon$ & $0$ & 0 & \dfrac{1}{2} & 2 & 1-\dfrac{\varepsilon}{4} \\[4mm]
		\textbf{DP0} & $\dfrac{\varepsilon}{12}$ & 0 & 0 & 0 & $\varepsilon$ & $-\dfrac{\varepsilon}{2}$  & $-\dfrac{\varepsilon}{3}$  & $\dfrac{\varepsilon}{12}$ & -\dfrac{\varepsilon}{12} & \dfrac{1}{2} + \dfrac{\varepsilon}{16} & 2-\dfrac{\varepsilon}{12} & 1-\dfrac{\varepsilon}{4} \\[4mm]
		\textbf{DP1} & $\dfrac{\varepsilon}{12}$ & 0 & 0 & 1 & $\varepsilon$ & $-\dfrac{\varepsilon}{2}$  & $-\dfrac{\varepsilon}{3}$  & $-\dfrac{\varepsilon}{12}$ & -\dfrac{\varepsilon}{12} & \dfrac{1}{2} + \dfrac{\varepsilon}{16} & 2-\dfrac{\varepsilon}{12} & 1-\dfrac{\varepsilon}{4}\\[4mm]
		\textbf{CA} & $\dfrac{\varepsilon}{9}$ & $-\dfrac{\varepsilon}3{}$ & $-\dfrac{\varepsilon}{6}$ & 1 & $\dfrac{5\varepsilon}{3}$ & $\varepsilon$ & $\dfrac{5\varepsilon}{9}$ & $\dfrac{\varepsilon}{18}$ & -\dfrac{11\varepsilon}{18} & \dfrac{1}{2} + \dfrac{\varepsilon}{8} & 2 + \dfrac{\varepsilon}{18} & 1- \dfrac{5\varepsilon}{6}\\[4mm]
		\textbf{CR} & $-g_2 + \dfrac{\varepsilon}{2}$ & $\dfrac{\varepsilon}{2}$ & $g_2$ & 1 & $\dfrac{5\varepsilon}{2}$ & $\varepsilon$ & $0$ & $-\dfrac{\varepsilon}{2}$ & \dfrac{\varepsilon}{2}& \dfrac{1}{2}+ \dfrac{\varepsilon}{8} &2-\dfrac{\varepsilon}{2} & 1\\[4mm]
		\textbf{CP} & $0.079\varepsilon$ & $-0.45\varepsilon$ & $-0.16\varepsilon$ & $0.64$ & $2.06\varepsilon$ & $1.0\varepsilon$ & $0.59\varepsilon$ & $-0.11\varepsilon$ & -0.80\varepsilon & \dfrac{1}{2}+0.13\varepsilon & 2 & 1-0.93 \varepsilon 
	\end{tabular}
	\end{ruledtabular}
	\caption{
	A Table containing all fixed points together with their location~(${u^*}$,~${g_1^*}$,~${g_2^*}$ and~${w^*}$), eigenvalues~(${\Sigma_1}$--${\Sigma_4}$) and associated critical exponents.
	Except for the CP fixed point (whose values were obtained numerically) all values are exact to first loop order.}
	\label{tab:EigenvaluesTable}
\end{table*}\noindent
%%%%%%%%%%%%%%%%%%%%%%%%%%%%%%%%%%%%%
%
An \emph{IR-stable} fixed point (stable in the limit~${\mu\rightarrow0}$) of these equations gives rise to the notion of scale invariance.
It implies that the effective parameters of our theory no longer change as one transitions to larger and larger scales.
At each fixed point a set of scaling exponents can be derived by solving the CZ-equations~\eqref{eq:CZ} using the \emph{method of characteristics}.
This necessitates introducing a dimensionless line parameter~${l}$ that relates to the momentum scale~${\mu}$ of the system as~${\mu(l)=\mu l}$.
To illustrate how this can be used to extract scaling exponents, we solve Eq.\@~\eqref{eq:CZ} for~${n=\tilde{n}=1}$, i.e.\@ for the two point vertex.
Employing the method of characteristics, one obtains
\begin{align}
	\Gamma^{(1,1)}(l) 
	= \exp \left\{\int_1^l \frac{\mathrm{d} l}{l} \, \big(\sigma(l)+\tilde{\sigma}(l)\big) \right\} \Gamma^{(1,1)}\big(t,D_\rho,q,\mu,\theta,\{ \lambda_i \}\big).
	\label{wq:Vert}
\end{align}
The beta functions evaluate to
\begin{align}
	\lambda_i(l) = \exp \left\{\int_1^l \frac{\mathrm{d} l}{l} \, \beta_{\lambda_i}(l) \right\}\, \lambda_i, \qquad
	D_\rho(l) = \exp \left\{\int_1^l \frac{\mathrm{d} l}{l} \, \beta_{D}(l) \right\}\, D_\rho, \qquad
	\theta(l) = \exp \left\{\int_1^l \frac{\mathrm{d} l}{l} \, \beta_{\theta}(l) \right\}\, \theta.
\end{align}
Inverting this, in principle, yields an exact solution with the additional line parameter~${l}$.
Employing a dimensional analysis, we infer that at the upper critical dimension
\begin{align}
	\Gamma^{(1,1)}\big(t,D_\rho,q,\mu,\theta,\{ \lambda_i \}\big) 
	=\, q^2\, D_\rho(l) \, \exp \left\{ - \int_1^l \frac{\mathrm{d} l}{l} \, \big(\sigma(l)+\tilde{\sigma}(l)\big) \right\}
	\cdot \Phi\left(t\mu^2(l)D_\rho(l),\frac{q}{\mu(l)}, \frac{\theta(l)}{\mu^2(l) D_\rho(l)},\{ \lambda_i(l) \}\right)
	\label{eq:TwoPointFinalScaling}	
\end{align}
has to hold.
In this expression, one can safely take the IR-limit~${q \rightarrow 0}$ by simultaneously taking~${l \rightarrow 0}$ such that their ratio remains fixed at 
\begin{align}
	\lim\limits_{q \rightarrow0} \lim\limits_{l\rightarrow 0	} \frac{q}{\mu l} =1.
	\label{eq:LimitDef}
\end{align}
This requires the existence of an IR-stable fixed point~${\{\lambda_i^*\}}$.
Otherwise, the effective coupling constants~${\{ \lambda_i(l) \}}$ never stop running in the limit~${l \rightarrow 0}$  and the scaling function in~\eqref{eq:TwoPointFinalScaling} contains diverging elements.
Given the existence of such a fixed point, we can expand the beta functions around it to obtain
\begin{align}
	\Gamma^{(1,1)}\big(t,D_\rho,q,\mu,\theta,\{ \lambda_i \}\big) 
	= q^{2-\eta} \, \hat{g}\left( \frac{D_\rho t}{\xi^z} ,q\xi,\{ \lambda_i^* \} \right)\, ,
	\label{eq:ScalingRelation2pt}
\end{align}
where we defined the scaling exponents
\begin{align}
	\eta = \sigma^* + \tilde{\sigma}^*-\beta_D^*
	\, , \quad
	z = 2+ \beta_D^*
	\, , \quad 
	\nu^{-1} = 2+\beta_D^* - \beta_\theta^*
	\label{eq:DefintionExponents}
\end{align}
and the renormalized correlation length
\begin{align}
	\xi =  \mu^{-1}\left( \frac{\theta}{D_\rho\mu^2} \right)^{-\nu}.
	\label{eq:CorrelationLength}
\end{align}
The same procedure can be applied to all Green- and vertex functions to derive expressions for other scaling exponents.
For the survival probability~${P(t) \sim G^{(0, 1)}}$~\cite{Janssen2005} we find
\begin{align}
    G^{(0,1)}\big(t,D_\rho,q,\mu,\theta,\{ \lambda_i \}\big)
    &= (D_\rho t)^{-\delta} \, \Phi^{(2)}\left(q\xi, \frac{D_\rho t}{\xi^z} \right), \qquad \delta = \frac{\sigma^*+d}{2z}.
	\label{eq:alphascaling} 
\end{align}

\subsection{Flow Equations II}

\begin{figure*}
    \centering
    \includegraphics[width=\textwidth]{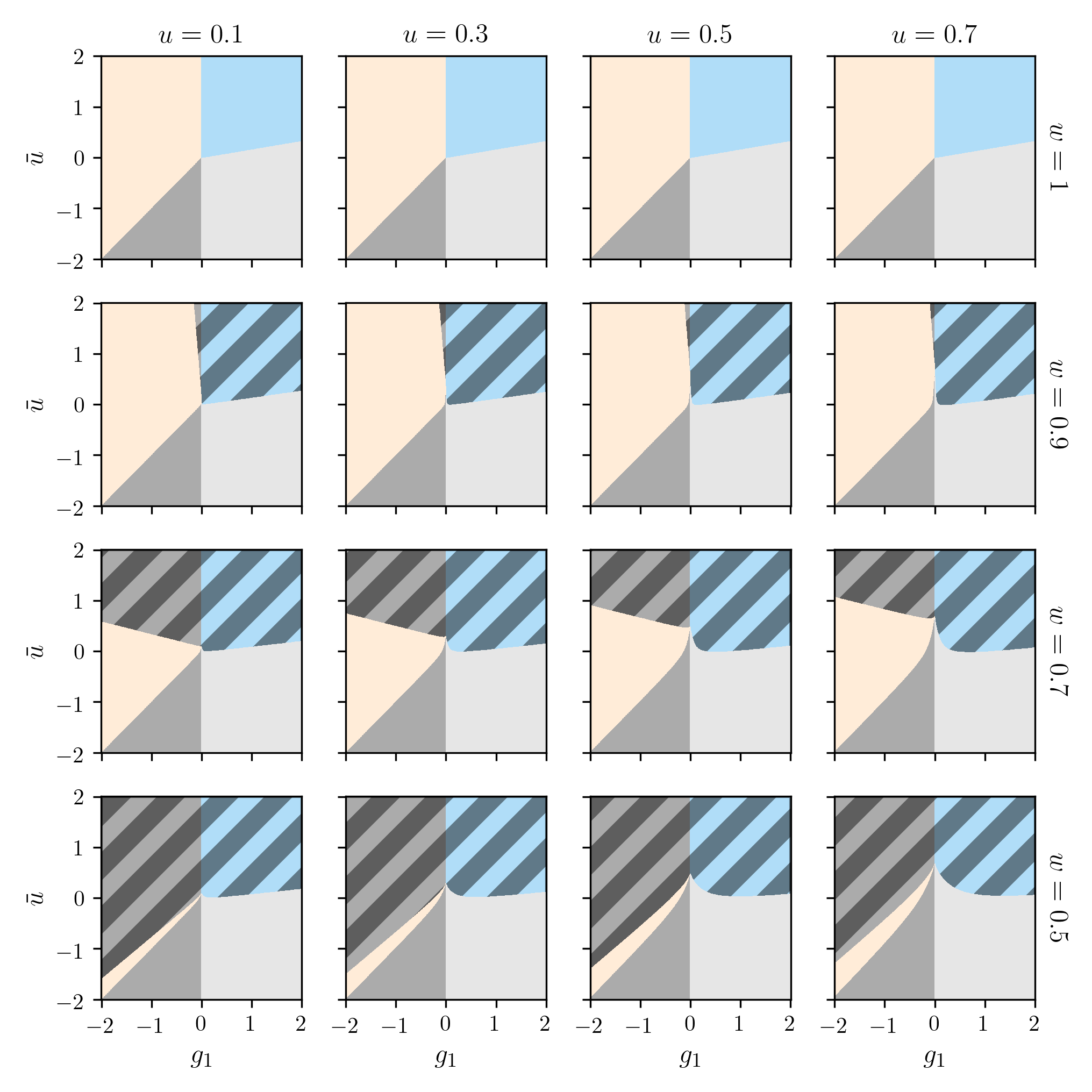}
    \caption{
        Basins of attraction in the~${\bar u}$-${g_1}$ plane for varying~${u}$ and~${w}$.
        Different flow behaviors are color-coded: Orange and blue points flow to the CA and CR fixed point, respectively.
        Gray and striped regions indicate runaway flow.
        All results were obtained by a numerical analysis.}
    \label{fig:Full4dFlow}
\end{figure*}

\begin{figure*}
    \centering
    \includegraphics[width=0.5\textwidth]{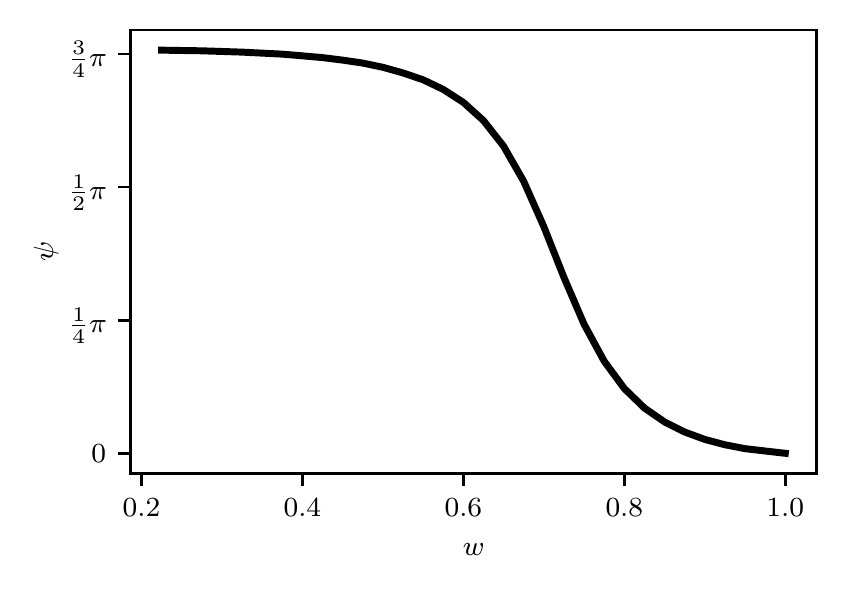}
    \caption{
        The behavior of the angle~${\psi}$ between the line defined by~${g_1=0}$ and the boundary between the runaway flow given by the gray striped region and the CA basin of attraction in Fig.\@~\ref{fig:Full4dFlow} as~${w}$ is varied. We note that the~${\psi}$ interpolates between~${\psi=3/4\pi}$, implying no orange region, at small~${w}$ and~${\psi=0}$, implying no runaway, at large~${w}$. In must be remarked that~${\psi}$ is defined in the case of~${u=1}$ where the boundary can be approximated by a straight line and the angle can thus be taken as a good representation of the size of the striped runaway region. However, as in Fig.\@~\ref{fig:Full4dFlow} it is clear that the growth of the region is very similar for all~${u}$, we believe the behavior to hold qualitatively in general. }
    \label{fig:BoundaryAngle}
\end{figure*}

All fixed points, together with the associated eigenvalues and scaling exponents are shown in Tab.\@~\ref{tab:EigenvaluesTable}.
Notably, the exponents do not vary along the fixed line (CR).
However, this was to be expected, since the fixed line collapses to a fixed point upon a change of variables, as shown in the main text.
As the flow equations describe a four dimensional space, it is in general not possible to give an exhaustive visualization of the flow of the system.
Only in specific cases does the flow remain in a lower dimensional hyperplane (such as the~${g_1}$-${\bar{u}}$-plane at~${w=1}$ as described in the main text).
In the rest of the cases we are obliged to ignore the exact flow behavior and focus on where a flow line starting at a point with initial coordinates~${(u,g_1,g_2,w)_i}$ ends up.
The collection of points that flow towards a certain fixed point makes up its \textit{basin of attraction}, and we can visualize two-dimensional slices of this space to ascertain which regions in parameter space are controlled by which fixed point. 
In the main text we extensively treat several cases. 
In Fig.\@\ref{fig:Full4dFlow} we plot the basins of attraction and the topology of the four dimension flow in different~${g_1}$-${\bar{u}}$-planes at various fixed values of~${u}$ and~${w}$. 
Notably, the topology of the flow diagram is independent of~${u}$ for~${w=1}$.
This relates to the fact that in this limit the proper effective variable is given by~${\bar u = g_2+u}$ as argued for in the main part of this letter.
One observes that once the~(${w=1}$)-plane is left, a new region of runaway flow appears at~${g_1<0}$.
Interestingly, this region grows in a winding fashion, increasing in size as~${w}$ decreases, at the same time causing the basin of attraction of the CA fixed point to shrink. 
The dependence of the size of this `wedge' of runaway flow on the parameter~${w}$ can be studied by defining an angle~${\psi}$ between the line defined by~${g_1=0}$ and the boundary between the runaway flow and the basin of attraction of the CA fixed point. 
In Fig.\@~\ref{fig:BoundaryAngle} we observe that for a relatively large range of~${w}$ this angle is very small indicating a negligible region of runaway flow and a phase diagram that is not very different from that at~${w=1}$. 
The CR fixed line becomes unstable for~${w<1}$, but as this instability is relatively weak, one can expect the large scale behavior to be similar to that of~${w=1}$ on \emph{both} sides of the~${g_1=0}$ invariant manifold.
For smaller~${w}$ the influence of~${u}$ starts to become more pronounced, distorting the boundaries between the different runaway regions.

Note that apart from the basins of attraction of the CA and CR fixed points we define four different types of runaway flow.
The dark gray region is defined as the runaway that lies below the basin of attraction of CA for~${g_1<0}$. The light gray region and the striped blue region lie in the~${g_1>0}$ plane, and are divided on the basis of flow behavior. Flow in the striped blue region is affected by the attractive nature of the projection of the CR fixed point below~${w<1}$, whereas flow originating in the light gray region is not and is, therefore, associated to the runaway flow already present at~${w=1}$.
The gray striped region corresponds to runaway linked to the CP fixed point, and is given by the runaway that lies above the basin of attraction of the CA fixed point for~${g_1<0}$.

From Fig.\@~\ref{fig:Full4dFlow} it is obvious that the CR fixed line becomes unstable for~${w<1}$.
By noting that we can relate the flow equation of~${w}$ to the one of~${D_\rho}$ via
\begin{align}
    \frac{\mathrm{d} w}{\mathrm{d} \ln(\mu)} = - w(1-w) \, \frac{\mathrm{d} \ln(D_\rho)}{\mathrm{d} \ln(\mu)} = -w(1-w)\beta_D,
    \label{eq:DflowDemo}
\end{align}
we can easily explain why this was to be expected.
In the vicinity of a super-diffusive fixed point~(${\beta_D^*<0}$), Eq.\@~\eqref{eq:DflowDemo} implies that~${w}$ decreases as~${\mu \rightarrow 0}$.
By the same argument, the opposite is to be expected at a sub-diffusive fixed point.
Therefore, the super-diffusive fixed line CR has to be unstable in~${w}$-direction.
On the other hand, the sub-diffusive fixed point CA is expected to be stable in~${w}$-direction, explaining why its basin of attraction extends in the~${w}$-direction.
Moreover, it is apparent that~${z=2}$, i.e.\@~${\beta_D=0}$ has to hold for any fixed point at~${w\neq [0,1]}$.
Thus it is clear that the CP fixed points obeys~${z=2}$ to all loop orders.

Apart from the fixed points, another important feature of the flow equations~\eqref{eq:AppFloww}--\eqref{eq:AppFlowG2} are its invariant manifolds.
Inspecting Eq.\@~\eqref{eq:AppFlowG1} it is clear that~${\partial g_1/\partial\text{ln}\mu|_{g_1=0}=0}$; thus,~${g_1=0}$ is such an invariant manifold.
Moreover, one can show that this result is true to any loop order which can be understood by inspecting~\eqref{eq:Ichi1}: The part of the result contributing to the~${g_1}$ renormalization is proportional to~${g_1}$, whereas the contribution to the~${g_2}$ renormalization is not proportional to~${g_2}$.
Consequently, dividing by~${g_1}$ and multiplying with~${g_1}$ (which essentially leads to the contribution of this diagram to Eq.\@~\eqref{eq:AppFlowG1}) results in something proportional to~${g_1}$.
Repeating this procedure for~${g_2}$, one realizes that this results in a contribution to the~${g_2}$-flow that is not proportional to~${g_2}$, thus allowing the flow to cross the~${g_2=0}$ hyperplane. 
Therefore it is sufficient to show that to all loop orders all divergences proportional to~${kp}$ are also proportional to~${g_1}$.
This is the case, because one had to take the~${g_2}$ term of the vertex factor of every chemotactic vertex for the contrary to be possible.
In particular this includes the~${g_2 p^2}$ part from the vertex where the incoming~${c}$-field connects with the rest of the diagram.
However, since this is already proportional to~${p^2}$, it can no longer renormalize~${g_1}$, proving that at least one factor of~${g_1}$ is included in every~${kp}$-divergence.
Hence, the~${g_1=0}$ hyperplane can never be crossed.
%

%%% howg1notg2happens
% \begingroup \color{blue}
\section{Confirmation of Nonconservative Interaction}

In section `Derivation of Langevin Equations' the different impacts of the RG flow on the effective equations of motion are explained and a non-particle-number-conserving effective chemotactic interaction is proposed.
In this section we demonstrate how the generalized chemotactic interaction ${\chi_1^{}\boldsymbol\nabla(\rho\boldsymbol\nabla c) + (\chi_2^{}-\chi_1^{})\rho\boldsymbol\nabla^2 c}$ arises already at one-loop level.
To this end, we analyze how a conserved chemotactic interaction is modified during the RG step.
Starting from the classical Keller-Segel nonlinearity $\chi_0^{}\boldsymbol\nabla(\rho\boldsymbol\nabla c)$, one can derive the RG flow functions by inserting ${g_1=g_2=g_0}$ into Eqs.\@~\eqref{eq:AppFloww}--\eqref{eq:AppFlowG2}.
If the resulting flow equations for $g_1$ and $g_2$ are identical, i.e., the ${g_1=g_2}$ hyperplane constitutes an invariant manifold, it is possible to renormalize the theory with a single effective coupling constant.
There are three Feynman diagrams that contribute to these flow functions and their respective values for ${g_1=g_2=g_0}$ are
\begin{align}
    I_1^g\Big|_{g_0} &= \frac{\sqrt{32\pi^2u^{-1}}D_\rho\mu^{\varepsilon/2}ug_0}{\varepsilon}\Big(2(kp) + 3p^2\Big)
    \\[2mm]
    I_2^g\Big|_{g_0} &= \frac{\sqrt{32\pi^2u^{-1}}D_\rho w\mu^{\varepsilon/2}g_0^2}{2\varepsilon}(-3+2w)\Big(kp + p^2\Big)
    \\[2mm]
    I_3^g\Big|_{g_0} &= \frac{\sqrt{32\pi^2u^{-1}}D_\rho w\mu^{\varepsilon/2}g_0^2}{2\varepsilon}\Big(kp + p^2\Big)\, .
\end{align}
Here, all the terms proportional to $kp$  and $p^2$ contribute to the renormalization of $g_1$ and $g_2$, respectively.
Importantly, one recognizes that while $I_2^g$  and $I_3^g$ contribute equally to both flow equations – being consistent with the \textit{Keller-Segel} nonlinearity -- only the diagram $I_1^g$ which couples the chemotactic vertex with the resource limiting nonlinearity breaks this relation.
This is crucial, since it shows that performing a single RG step in the presence of resource limitation \textit{generates} a nonconservative contribution to the chemotactic interaction also if it is not included from the beginning.
Thus, a consistent coarse graining of the theory with a conservative effective chemotactic interaction close to criticality is not possible and an additional term needs to be included.
% \endgroup

%%% StandardIntegrals
\section{Standard Integrals and Identities}\noindent
Here we give a short overview of some integrals that frequently appear during the calculation of Feynman diagrams.
One important group of integrals is of the form 
\begin{equation*}
		I_{n,a}(\Delta)=\mu^\varepsilon\int_{-\infty}^{+\infty}\frac{\mathrm{d}^{d}q}{(2\pi)^d} 
		\frac{q^n}{(q^2+\Delta)^a} , 
\end{equation*}
This integral can be solved as follows:
\begin{align}
	I_{n,a}(\Delta)
	&= \mu^\varepsilon \int_{-\infty}^{+\infty}\frac{\mathrm{d}^{d}q}{(2\pi)^d} 
		\frac{q^n}{(q^2+\Delta)^a}
		\notag \\[2mm]
	&= \mu^\varepsilon \int_{0}^{+\infty}\mathrm{d}q\frac{q^{d+n-1}}{(q^2+\Delta)^a}\int\frac{d
		\Omega_d}{(2\pi)^d}
		\notag\\[2mm]
	&= \frac{2\mu^\varepsilon}{(4\pi)^\frac{d}{2}\Gamma(\frac{d}{2})}\int_0^\infty
		\mathrm{d} q\frac{q^{d+n-1}}{(q^2+\Delta)^a}
		\notag\\[2mm]
	&= \frac{\mu^\varepsilon}{(4\pi)^\frac{d}{2}\Gamma(\frac{d}{2})}\int_0^\infty
		\mathrm{d} l\frac{l^{\frac{d+n}{2}-1}}{(l+\Delta)^a}
		\notag \\[2mm]
	&= \frac{\mu^\varepsilon}{(4\pi)^\frac{d}{2}\Gamma(\frac{d}{2})}\int_0^1
		\mathrm{d} x \, \bar{x}^{\frac{d+n}{2}-1}x^{a-\frac{d+n}{2}-1}\Delta^{\frac{d+n}{2}-a}
		\notag\\[2mm]
	&= \frac{\mu^\varepsilon}{(4\pi)^\frac{d}{2}\Gamma(\frac{d}{2})}
		\frac{\Gamma(\frac{d+n}{2})\Gamma(a-\frac{d+n}{2})}{\Gamma(a)}\Delta^{\frac{d}{2}-2}
		\Delta^{\frac{n}{2}+2-a} 
\end{align}
Here we first changed the integration variable to~${l=q^2}$ and then to~${x=\frac{\Delta}{l+\Delta}}$ and used the Euler beta function
\begin{equation*}
	B(\alpha, \beta) = \int_0^1 \, \mathrm{d} x \, \bar{x}^{\alpha-1} x^{\beta-1} = 
	\frac{\Gamma(\alpha)\Gamma(\beta)}{\Gamma(\alpha+\beta)}\ .
\end{equation*}
The results for specific values of~${n}$ and~${a}$, after inserting~${d=4-\epsilon}$ and performing a Taylor expansion around~${\epsilon=0}$ are given in Tab.\@~\ref{tab:StandardResults}.
%
%% Integral Table
%%%%%%%%%%%%%%%%%%%%%%%%%%%%%%%%%%
\begin{table}[tb]
    \centering
    \begin{ruledtabular}
    \begin{tabular}{cccccc}
        n &  a & $I_{n,a}(\Delta)$ & n &  a & $I_{n,a}(\Delta)$\\[2mm]\hline 
        & & & & \\[-2mm]
        0 & 1 & $\displaystyle-\frac{\Delta}{(4\pi)^2} \left( \frac{2}{\varepsilon}+1-\gamma-\ln\left(\frac{\Delta}{4\pi\mu^2}\right)\right)$ &
        2 & 3 & $\displaystyle\frac{1}{(4\pi)^2} \left( \frac{2}{\varepsilon}-\frac{1}{2}-\gamma-\ln\left(\frac{\Delta}{4\pi\mu^2}\right)\right)$\\[2mm]
        0 & 2 & $\displaystyle\frac{1}{(4\pi)^2} \left( \frac{2}{\varepsilon}-\gamma-\ln\left(\frac{\Delta}{4\pi\mu^2}\right)\right)$ &
        4 & 4 & $\displaystyle\frac{1}{(4\pi)^2} \left( \frac{2}{\varepsilon}-\frac{5}{6}-\gamma-\ln\left(\frac{\Delta}{4\pi\mu^2}\right)\right)$\\[2mm]
        2 & 2 & $\displaystyle-\frac{\Delta}{(4\pi)^2}  \left( \frac{4}{\varepsilon}+1-2\gamma-2\ln\left(\frac{\Delta}{4\pi\mu^2}\right)\right)$ &
        & &\\[2mm]
    \end{tabular}
    \end{ruledtabular}
    \caption{
    Analytical expression for~${I_{n,a}(\Delta}$) in~${d=4-\varepsilon}$ dimensions at specific choices of~${n}$ and~${a}$.
    The parameter~${\gamma}$ denotes the Euler–Mascheroni constant.}
    \label{tab:StandardResults}
\end{table}
%%%%%%%%%%%%%%%%%%%%%%%%%%%%%%%%%%
%
The second important type of integral is of the form
\begin{align*}
	\tilde{I}_{2,a}(\vec{k},\vec{p},\Delta)=\mu^\varepsilon\int_{-\infty}^{+\infty}\frac{\mathrm{d}^{d}q}{(2\pi)^d}\frac{(\vec{q}\cdot\vec{k})
	(\vec{q}\cdot\vec{p})}{(q^2+\Delta)^a}.
\end{align*}
Since all terms containing~${q_iq_j}$ with~${i\neq j}$ give zero due to their antisymmetry in~${q_i}$ and~${q_j}$, we note that the integral can be rewritten as
\begin{equation*}
	\tilde{I}_{2,a}(\vec{k},\vec{p},\Delta) = \sum_{i=1}^d \mu^\varepsilon \int\frac{\mathrm{d}^{d-1}q}{(2\pi)^{d-1}}
	\int_{-\infty}^{+\infty}
	\frac{\mathrm{d} q_i}{2\pi}\frac{q_i^2k_ip_i}{\Big( q_i^2+{\Delta_i}\Big)^a} \, ,
\end{equation*}
with~${\Delta_i = \Delta + \sum_{n\neq i}^d q_n^2}$.
Now we can solve the integral for each~${i\in \{0,...,d\}}$ separately and add the results:
\begin{align*}
	\tilde{I}_{2,a}(\vec{k},\vec{p},\Delta)
	&= \sum_{i=1}^d\frac{\mu^\varepsilon}{2\pi}\, \int\frac{\mathrm{d}^{d-1}q}{(2\pi)^{d-1}}\Delta_i^{\frac{3}{2}-a} \int_0^1\mathrm{d} x \, \, \bar{x}^\frac{1}{2}x^{a-\frac{5}{2}} \, k_ip_i \notag \\
	& = \sum_{i=1}^d \frac{\mu^\varepsilon}{2\pi}\frac{\Gamma(\frac{3}{2})
	\Gamma(a-\frac{3}{2})}{\Gamma(a)} \int\frac{\mathrm{d}^{d-1}q}{(2\pi)^{d-1}} \frac{k_ip_i}{\Big(\sum_{n\neq i}^d q_n^2+\Delta\Big)^{a-\frac{3}{2}}}	
\end{align*}
Now we use the previous result for~${d-1}$ dimensions and~${n=0}$ to get
\begin{equation*}
	\tilde{I}_{2,a}(\vec{k},\vec{p},\Delta)
	=\frac{\vec{k}\cdot\vec{p}}{32\pi^2}\,\frac{\Gamma(a-\frac{d}{2}-1)}{\Gamma(a)}
	\left(\frac{\Delta}{4\pi\mu^2}\right)^{\frac{d}{2}-2}\Delta^{3-a}.
\end{equation*}
Inserting the case~${a=3}$, which is relevant for our calculations, gives:
\begin{align}
    \tilde{I}_{2,3}(\vec{k},\vec{p},\Delta) 
    = \frac{k\cdot p}{64\pi^2}\left( \frac{2}{\varepsilon}-\gamma-\ln\left(\frac{\Delta}{4\pi\mu^2}\right)\right)
\end{align}
Other important integral identities revolve around the Feynman parameter trick
\begin{align}
    \frac{1}{P_1^{a_1}P_2^{a_2}...P_n^{a_n}}&=\frac{\Gamma(a_1+a_2+...+a_n)}{\Gamma(a_1)\Gamma(a_2)...\Gamma(a_n)} \int_0^1\mathrm{d} x_1..\mathrm{d} x_n \frac{x_1^{a_1-1}...x_n^{a_n-1}\,\delta(1-x_1-...x_n)}{(x_1P_1+...x_nP_n)^{a_1+...+a_n}}\label{eq:FeynmanTrick}.
\end{align}
Through this introduction of the so called Feynman parameters~${x_i}$, also the following integrals appear frequently.
\begin{equation}
    \int_{x,y,z} 1 = \frac{1}{2}, \qquad  \qquad \int_{x,y,z} x =\int_{w,x,y,z} 1 = \frac{1}{6} \label{eq:ParameterInts}
\end{equation}
With the shorthands for Feynman parameter integrals given by
\begin{alignat*}{2}
    &\int_x &&= \ \int_0^1 \ \mathrm{d} x, \\
    &\int_{x,y,z} &&= \ \int_0^1 \delta(1-x-y-z) \ \mathrm{d} x \, \mathrm{d} y \, \mathrm{d} z,\\
    &\int_{x,y,z,w} &&= \ \int_0^1 \delta(1-w-x-y-z) \ \mathrm{d} w \, \mathrm{d} x \, \mathrm{d} y \, \mathrm{d} z.
\end{alignat*}
Additionally, we define the following shorthands for the momentum and frequency integrals:
\begin{align*}
    \int_q =\mu^D\ \int_{-\infty}^{+\infty}\frac{\mathrm{d}^{d}q}{(2\pi)^d}, \qquad
    \int_\omega = \int_{-\infty}^{+\infty}  \mathrm{d}w
\end{align*}
%

%%% NumericalMethods

\section{Numerical Methods}\noindent
Throughout the paper and supplementary information, figures that display basins of attraction where obtained by creating a fine grid of initial conditions and then evolving these according to the flow equations using a fourth order Runge-Kutta method implemented in C++.
After a fixed number of iterations it is checked if the flow is located within a ball of radius~${\delta}$ from any of the fixed points.
If so, the initial condition lies in the basin of attraction of the respective fixed point.
If not, the flow is determined to run away.
In all cases~${\epsilon=1}$.
The finite element simulations performed to obtain the data displayed in Fig.\@~1 of the main part were done using DOLFIN (FENICS project)~\cite{LangtangenLogg2017, Logg2010}, where we implemented a backwards Euler scheme with periodic boundary conditions,~${dt=0.01}$ and at least~${200}$ nodes per unit length.


\begin{thebibliography}{63}%
\makeatletter
\providecommand \@ifxundefined [1]{%
 \@ifx{#1\undefined}
}%
\providecommand \@ifnum [1]{%
 \ifnum #1\expandafter \@firstoftwo
 \else \expandafter \@secondoftwo
 \fi
}%
\providecommand \@ifx [1]{%
 \ifx #1\expandafter \@firstoftwo
 \else \expandafter \@secondoftwo
 \fi
}%
\providecommand \natexlab [1]{#1}%
\providecommand \enquote  [1]{``#1''}%
\providecommand \bibnamefont  [1]{#1}%
\providecommand \bibfnamefont [1]{#1}%
\providecommand \citenamefont [1]{#1}%
\providecommand \href@noop [0]{\@secondoftwo}%
\providecommand \href [0]{\begingroup \@sanitize@url \@href}%
\providecommand \@href[1]{\@@startlink{#1}\@@href}%
\providecommand \@@href[1]{\endgroup#1\@@endlink}%
\providecommand \@sanitize@url [0]{\catcode `\\12\catcode `\$12\catcode
  `\&12\catcode `\#12\catcode `\^12\catcode `\_12\catcode `\%12\relax}%
\providecommand \@@startlink[1]{}%
\providecommand \@@endlink[0]{}%
\providecommand \url  [0]{\begingroup\@sanitize@url \@url }%
\providecommand \@url [1]{\endgroup\@href {#1}{\urlprefix }}%
\providecommand \urlprefix  [0]{URL }%
\providecommand \Eprint [0]{\href }%
\providecommand \doibase [0]{http://dx.doi.org/}%
\providecommand \selectlanguage [0]{\@gobble}%
\providecommand \bibinfo  [0]{\@secondoftwo}%
\providecommand \bibfield  [0]{\@secondoftwo}%
\providecommand \translation [1]{[#1]}%
\providecommand \BibitemOpen [0]{}%
\providecommand \bibitemStop [0]{}%
\providecommand \bibitemNoStop [0]{.\EOS\space}%
\providecommand \EOS [0]{\spacefactor3000\relax}%
\providecommand \BibitemShut  [1]{\csname bibitem#1\endcsname}%
\let\auto@bib@innerbib\@empty
%</preamble>
\bibitem [{\citenamefont {Hinrichsen}(2000)}]{Hinrichsen2000}%
  \BibitemOpen
  \bibfield  {author} {\bibinfo {author} {\bibfnamefont {H.}~\bibnamefont
  {Hinrichsen}},\ }\bibfield  {title} {\enquote {\bibinfo {title}
  {Non-equilibrium critical phenomena and phase transitions into absorbing
  states},}\ }\href {\doibase 10.1080/00018730050198152} {\bibfield  {journal}
  {\bibinfo  {journal} {Advances in Physics}\ }\textbf {\bibinfo {volume}
  {49}},\ \bibinfo {pages} {815--958} (\bibinfo {year} {2000})}\BibitemShut
  {NoStop}%
\bibitem [{\citenamefont {Janssen}\ and\ \citenamefont
  {Täuber}(2005)}]{JanssenTauber2005}%
  \BibitemOpen
  \bibfield  {author} {\bibinfo {author} {\bibfnamefont {H.~K.}\ \bibnamefont
  {Janssen}}\ and\ \bibinfo {author} {\bibfnamefont {U.~C.}\ \bibnamefont
  {Täuber}},\ }\bibfield  {title} {\enquote {\bibinfo {title} {The field
  theory approach to percolation processes},}\ }\href {\doibase
  https://doi.org/10.1016/j.aop.2004.09.011} {\bibfield  {journal} {\bibinfo
  {journal} {Annals of Physics}\ }\textbf {\bibinfo {volume} {315}},\ \bibinfo
  {pages} {147 -- 192} (\bibinfo {year} {2005})},\ \bibinfo {note} {special
  Issue}\BibitemShut {NoStop}%
\bibitem [{\citenamefont {Halpin-Healy}\ and\ \citenamefont
  {Zhang}(1995)}]{Halpin1995}%
  \BibitemOpen
  \bibfield  {author} {\bibinfo {author} {\bibfnamefont {T.}~\bibnamefont
  {Halpin-Healy}}\ and\ \bibinfo {author} {\bibfnamefont {Y.~C.}\ \bibnamefont
  {Zhang}},\ }\bibfield  {title} {\enquote {\bibinfo {title} {Kinetic
  roughening phenomena, stochastic growth, directed polymers and all that.
  {A}spects of multidisciplinary statistical mechanics},}\ }\href {\doibase
  https://doi.org/10.1016/0370-1573(94)00087-J} {\bibfield  {journal} {\bibinfo
   {journal} {Physics Reports}\ }\textbf {\bibinfo {volume} {254}},\ \bibinfo
  {pages} {215--414} (\bibinfo {year} {1995})}\BibitemShut {NoStop}%
\bibitem [{\citenamefont {Kardar}\ \emph {et~al.}(1986)\citenamefont {Kardar},
  \citenamefont {Parisi},\ and\ \citenamefont {Zhang}}]{Kardar1986}%
  \BibitemOpen
  \bibfield  {author} {\bibinfo {author} {\bibfnamefont {M.}~\bibnamefont
  {Kardar}}, \bibinfo {author} {\bibfnamefont {G.}~\bibnamefont {Parisi}}, \
  and\ \bibinfo {author} {\bibfnamefont {Y.}~\bibnamefont {Zhang}},\ }\bibfield
   {title} {\enquote {\bibinfo {title} {Dynamic scaling of growing
  interfaces},}\ }\href {\doibase 10.1103/physrevlett.56.889} {\bibfield
  {journal} {\bibinfo  {journal} {Physical Review Letters}\ }\textbf {\bibinfo
  {volume} {56}},\ \bibinfo {pages} {889--892} (\bibinfo {year}
  {1986})}\BibitemShut {NoStop}%
\bibitem [{\citenamefont {Ramaswamy}(2010)}]{Ramaswamy2010}%
  \BibitemOpen
  \bibfield  {author} {\bibinfo {author} {\bibfnamefont {S.}~\bibnamefont
  {Ramaswamy}},\ }\bibfield  {title} {\enquote {\bibinfo {title} {The mechanics
  and statistics of active matter},}\ }\href {\doibase
  10.1146/annurev-conmatphys-070909-104101} {\bibfield  {journal} {\bibinfo
  {journal} {Annual Review of Condensed Matter Physics}\ }\textbf {\bibinfo
  {volume} {1}},\ \bibinfo {pages} {323--345} (\bibinfo {year}
  {2010})}\BibitemShut {NoStop}%
\bibitem [{\citenamefont {Marchetti}\ \emph {et~al.}(2013)\citenamefont
  {Marchetti}, \citenamefont {Joanny}, \citenamefont {Ramaswamy}, \citenamefont
  {Liverpool}, \citenamefont {Prost}, \citenamefont {Rao},\ and\ \citenamefont
  {Simha}}]{Marchetti2013}%
  \BibitemOpen
  \bibfield  {author} {\bibinfo {author} {\bibfnamefont {M.~C.}\ \bibnamefont
  {Marchetti}}, \bibinfo {author} {\bibfnamefont {J.~F.}\ \bibnamefont
  {Joanny}}, \bibinfo {author} {\bibfnamefont {S.}~\bibnamefont {Ramaswamy}},
  \bibinfo {author} {\bibfnamefont {T.~B.}\ \bibnamefont {Liverpool}}, \bibinfo
  {author} {\bibfnamefont {J.}~\bibnamefont {Prost}}, \bibinfo {author}
  {\bibfnamefont {M.}~\bibnamefont {Rao}}, \ and\ \bibinfo {author}
  {\bibfnamefont {R.~A.}\ \bibnamefont {Simha}},\ }\bibfield  {title} {\enquote
  {\bibinfo {title} {Hydrodynamics of soft active matter},}\ }\href {\doibase
  10.1103/RevModPhys.85.1143} {\bibfield  {journal} {\bibinfo  {journal} {Rev.
  Mod. Phys.}\ }\textbf {\bibinfo {volume} {85}},\ \bibinfo {pages}
  {1143--1189} (\bibinfo {year} {2013})}\BibitemShut {NoStop}%
\bibitem [{\citenamefont {Ziepke}\ \emph {et~al.}(2022)\citenamefont {Ziepke},
  \citenamefont {Maryshev}, \citenamefont {Aranson},\ and\ \citenamefont
  {Frey}}]{Ziepke2022}%
  \BibitemOpen
  \bibfield  {author} {\bibinfo {author} {\bibfnamefont {A.}~\bibnamefont
  {Ziepke}}, \bibinfo {author} {\bibfnamefont {I.}~\bibnamefont {Maryshev}},
  \bibinfo {author} {\bibfnamefont {I.S.}\ \bibnamefont {Aranson}}, \ and\
  \bibinfo {author} {\bibfnamefont {Erwin}\ \bibnamefont {Frey}},\ }\bibfield
  {title} {\enquote {\bibinfo {title} {Multi-scale organization in
  communicating active matter},}\ }\href {\doibase 10.1038/s41467-022-34484-2}
  {\bibfield  {journal} {\bibinfo  {journal} {Nature Communications}\ }\textbf
  {\bibinfo {volume} {13}} (\bibinfo {year} {2022}),\
  10.1038/s41467-022-34484-2}\BibitemShut {NoStop}%
\bibitem [{\citenamefont {Parent}\ and\ \citenamefont
  {Devreotes}(1999)}]{Parent1999}%
  \BibitemOpen
  \bibfield  {author} {\bibinfo {author} {\bibfnamefont {C.~A.}\ \bibnamefont
  {Parent}}\ and\ \bibinfo {author} {\bibfnamefont {P.~N.}\ \bibnamefont
  {Devreotes}},\ }\bibfield  {title} {\enquote {\bibinfo {title} {A cell's
  sense of direction},}\ }\href {\doibase 10.1126/science.284.5415.765}
  {\bibfield  {journal} {\bibinfo  {journal} {Science}\ }\textbf {\bibinfo
  {volume} {284}},\ \bibinfo {pages} {765--770} (\bibinfo {year}
  {1999})}\BibitemShut {NoStop}%
\bibitem [{\citenamefont {Bauer}\ \emph {et~al.}(2017)\citenamefont {Bauer},
  \citenamefont {Knebel}, \citenamefont {Lechner}, \citenamefont {Pickl},\ and\
  \citenamefont {Frey}}]{Bauer2017}%
  \BibitemOpen
  \bibfield  {author} {\bibinfo {author} {\bibfnamefont {M.}~\bibnamefont
  {Bauer}}, \bibinfo {author} {\bibfnamefont {J.}~\bibnamefont {Knebel}},
  \bibinfo {author} {\bibfnamefont {M.}~\bibnamefont {Lechner}}, \bibinfo
  {author} {\bibfnamefont {P.}~\bibnamefont {Pickl}}, \ and\ \bibinfo {author}
  {\bibfnamefont {E.}~\bibnamefont {Frey}},\ }\bibfield  {title} {\enquote
  {\bibinfo {title} {Ecological feedback in quorum-sensing microbial
  populations can induce heterogeneous production of autoinducers},}\ }\href
  {https://elifesciences.org/articles/25773} {\bibfield  {journal} {\bibinfo
  {journal} {{eLife}}\ }\textbf {\bibinfo {volume} {6}} (\bibinfo {year}
  {2017})}\BibitemShut {NoStop}%
\bibitem [{\citenamefont {Katzschmann}\ \emph {et~al.}(2018)\citenamefont
  {Katzschmann}, \citenamefont {DelPreto}, \citenamefont {MacCurdy},\ and\
  \citenamefont {Rus}}]{Katzschmann2018}%
  \BibitemOpen
  \bibfield  {author} {\bibinfo {author} {\bibfnamefont {R.~K.}\ \bibnamefont
  {Katzschmann}}, \bibinfo {author} {\bibfnamefont {J.}~\bibnamefont
  {DelPreto}}, \bibinfo {author} {\bibfnamefont {R.}~\bibnamefont {MacCurdy}},
  \ and\ \bibinfo {author} {\bibfnamefont {D.}~\bibnamefont {Rus}},\ }\bibfield
   {title} {\enquote {\bibinfo {title} {Exploration of underwater life with an
  acoustically controlled soft robotic fish},}\ }\href
  {https://www.science.org/doi/10.1126/scirobotics.aar3449} {\bibfield
  {journal} {\bibinfo  {journal} {Science Robotics}\ }\textbf {\bibinfo
  {volume} {3}} (\bibinfo {year} {2018})}\BibitemShut {NoStop}%
\bibitem [{\citenamefont {Fisher}\ \emph {et~al.}(1972)\citenamefont {Fisher},
  \citenamefont {Ma},\ and\ \citenamefont {Nickel}}]{Fisher1972}%
  \BibitemOpen
  \bibfield  {author} {\bibinfo {author} {\bibfnamefont {M.~E.}\ \bibnamefont
  {Fisher}}, \bibinfo {author} {\bibfnamefont {S.~K.}\ \bibnamefont {Ma}}, \
  and\ \bibinfo {author} {\bibfnamefont {B.~G.}\ \bibnamefont {Nickel}},\
  }\bibfield  {title} {\enquote {\bibinfo {title} {Critical exponents for
  long-range interactions},}\ }\href {\doibase 10.1103/PhysRevLett.29.917}
  {\bibfield  {journal} {\bibinfo  {journal} {Phys. Rev. Lett.}\ }\textbf
  {\bibinfo {volume} {29}},\ \bibinfo {pages} {917--920} (\bibinfo {year}
  {1972})}\BibitemShut {NoStop}%
\bibitem [{\citenamefont {Frey}\ and\ \citenamefont
  {Schwabl}(1994)}]{Frey1994b}%
  \BibitemOpen
  \bibfield  {author} {\bibinfo {author} {\bibfnamefont {E.}~\bibnamefont
  {Frey}}\ and\ \bibinfo {author} {\bibfnamefont {F.}~\bibnamefont {Schwabl}},\
  }\bibfield  {title} {\enquote {\bibinfo {title} {Critical dynamics of
  magnets},}\ }\href {https://doi.org/10.1080/00018739400101535} {\bibfield
  {journal} {\bibinfo  {journal} {Advances in Physics}\ }\textbf {\bibinfo
  {volume} {43}},\ \bibinfo {pages} {577--683} (\bibinfo {year}
  {1994})}\BibitemShut {NoStop}%
\bibitem [{\citenamefont {Bayong}\ \emph {et~al.}(1999)\citenamefont {Bayong},
  \citenamefont {Diep},\ and\ \citenamefont {Truong}}]{Truong1999}%
  \BibitemOpen
  \bibfield  {author} {\bibinfo {author} {\bibfnamefont {E.}~\bibnamefont
  {Bayong}}, \bibinfo {author} {\bibfnamefont {H.~T.}\ \bibnamefont {Diep}}, \
  and\ \bibinfo {author} {\bibfnamefont {T.~T.}\ \bibnamefont {Truong}},\
  }\bibfield  {title} {\enquote {\bibinfo {title} {Phase transition in a
  general continuous {Ising} model with long-range interactions},}\ }\href
  {\doibase 10.1063/1.370270} {\bibfield  {journal} {\bibinfo  {journal}
  {Journal of Applied Physics}\ }\textbf {\bibinfo {volume} {85}},\ \bibinfo
  {pages} {6088--6090} (\bibinfo {year} {1999})},\ \Eprint
  {http://arxiv.org/abs/https://doi.org/10.1063/1.370270}
  {https://doi.org/10.1063/1.370270} \BibitemShut {NoStop}%
\bibitem [{\citenamefont {Janssen}\ \emph
  {et~al.}(1999{\natexlab{a}})\citenamefont {Janssen}, \citenamefont {Oerding},
  \citenamefont {van Wijland.},\ and\ \citenamefont {Hilhorst}}]{Janssen1999b}%
  \BibitemOpen
  \bibfield  {author} {\bibinfo {author} {\bibfnamefont {H.~K.}\ \bibnamefont
  {Janssen}}, \bibinfo {author} {\bibfnamefont {K.}~\bibnamefont {Oerding}},
  \bibinfo {author} {\bibfnamefont {F.}~\bibnamefont {van Wijland.}}, \ and\
  \bibinfo {author} {\bibfnamefont {H.~J.}\ \bibnamefont {Hilhorst}},\
  }\bibfield  {title} {\enquote {\bibinfo {title} {{Lévy}-flight spreading of
  epidemic processes leading to percolating clusters},}\ }\href {\doibase
  10.1007/s100510050596} {\bibfield  {journal} {\bibinfo  {journal} {The
  European Physical Journal B - Condensed Matter and Complex Systems}\ }\textbf
  {\bibinfo {volume} {7}},\ \bibinfo {pages} {137--145} (\bibinfo {year}
  {1999}{\natexlab{a}})}\BibitemShut {NoStop}%
\bibitem [{\citenamefont {Hinrichsen}(2007)}]{Hinrichsen2007}%
  \BibitemOpen
  \bibfield  {author} {\bibinfo {author} {\bibfnamefont {H.}~\bibnamefont
  {Hinrichsen}},\ }\bibfield  {title} {\enquote {\bibinfo {title}
  {Non-equilibrium phase transitions with long-range interactions},}\ }\href
  {\doibase 10.1088/1742-5468/2007/07/p07006} {\bibfield  {journal} {\bibinfo
  {journal} {Journal of Statistical Mechanics: Theory and Experiment}\ }\textbf
  {\bibinfo {volume} {2007}},\ \bibinfo {pages} {P07006--P07006} (\bibinfo
  {year} {2007})}\BibitemShut {NoStop}%
\bibitem [{\citenamefont {Argolo}\ \emph {et~al.}(2013)\citenamefont {Argolo},
  \citenamefont {Quintino}, \citenamefont {Barros},\ and\ \citenamefont
  {Lyra}}]{Lyra2013}%
  \BibitemOpen
  \bibfield  {author} {\bibinfo {author} {\bibfnamefont {C.}~\bibnamefont
  {Argolo}}, \bibinfo {author} {\bibfnamefont {Y.}~\bibnamefont {Quintino}},
  \bibinfo {author} {\bibfnamefont {P.~H.}\ \bibnamefont {Barros}}, \ and\
  \bibinfo {author} {\bibfnamefont {M.~L.}\ \bibnamefont {Lyra}},\ }\bibfield
  {title} {\enquote {\bibinfo {title} {Vanishing order-parameter critical
  fluctuations of an absorbing-state transition driven by long-range
  interactions},}\ }\href {\doibase 10.1103/PhysRevE.87.032141} {\bibfield
  {journal} {\bibinfo  {journal} {Phys. Rev. E}\ }\textbf {\bibinfo {volume}
  {87}},\ \bibinfo {pages} {032141} (\bibinfo {year} {2013})}\BibitemShut
  {NoStop}%
\bibitem [{\citenamefont {Reia}\ and\ \citenamefont
  {Fontanari}(2016)}]{Fontanari2016}%
  \BibitemOpen
  \bibfield  {author} {\bibinfo {author} {\bibfnamefont {S.~M.}\ \bibnamefont
  {Reia}}\ and\ \bibinfo {author} {\bibfnamefont {J.~F.}\ \bibnamefont
  {Fontanari}},\ }\bibfield  {title} {\enquote {\bibinfo {title} {Effect of
  long-range interactions on the phase transition of {Axelrod's} model},}\
  }\href {\doibase 10.1103/PhysRevE.94.052149} {\bibfield  {journal} {\bibinfo
  {journal} {Phys. Rev. E}\ }\textbf {\bibinfo {volume} {94}},\ \bibinfo
  {pages} {052149} (\bibinfo {year} {2016})}\BibitemShut {NoStop}%
\bibitem [{\citenamefont {Keller}\ and\ \citenamefont
  {Segel}(1971)}]{KELLER1971225}%
  \BibitemOpen
  \bibfield  {author} {\bibinfo {author} {\bibfnamefont {E.~F.}\ \bibnamefont
  {Keller}}\ and\ \bibinfo {author} {\bibfnamefont {L.~A.}\ \bibnamefont
  {Segel}},\ }\bibfield  {title} {\enquote {\bibinfo {title} {Model for
  chemotaxis},}\ }\href {\doibase https://doi.org/10.1016/0022-5193(71)90050-6}
  {\bibfield  {journal} {\bibinfo  {journal} {Journal of Theoretical Biology}\
  }\textbf {\bibinfo {volume} {30}},\ \bibinfo {pages} {225 -- 234} (\bibinfo
  {year} {1971})}\BibitemShut {NoStop}%
\bibitem [{\citenamefont {Hillen}\ and\ \citenamefont
  {Painter}(2008)}]{Hillen2008}%
  \BibitemOpen
  \bibfield  {author} {\bibinfo {author} {\bibfnamefont {T.}~\bibnamefont
  {Hillen}}\ and\ \bibinfo {author} {\bibfnamefont {K.~J.}\ \bibnamefont
  {Painter}},\ }\bibfield  {title} {\enquote {\bibinfo {title} {A user's guide
  to {PDE} models for chemotaxis},}\ }\href {\doibase
  10.1007/s00285-008-0201-3} {\bibfield  {journal} {\bibinfo  {journal}
  {Journal of Mathematical Biology}\ }\textbf {\bibinfo {volume} {58}},\
  \bibinfo {pages} {183--217} (\bibinfo {year} {2008})}\BibitemShut {NoStop}%
\bibitem [{\citenamefont {Tindall}\ \emph {et~al.}(2008)\citenamefont
  {Tindall}, \citenamefont {Maini}, \citenamefont {Porter},\ and\ \citenamefont
  {Armitage}}]{Tindall2008}%
  \BibitemOpen
  \bibfield  {author} {\bibinfo {author} {\bibfnamefont {M.~J.}\ \bibnamefont
  {Tindall}}, \bibinfo {author} {\bibfnamefont {P.~K.}\ \bibnamefont {Maini}},
  \bibinfo {author} {\bibfnamefont {S.~L.}\ \bibnamefont {Porter}}, \ and\
  \bibinfo {author} {\bibfnamefont {J.~P.}\ \bibnamefont {Armitage}},\
  }\bibfield  {title} {\enquote {\bibinfo {title} {Overview of mathematical
  approaches used to model bacterial chemotaxis {II}: Bacterial populations},}\
  }\href {\doibase 10.1007/s11538-008-9322-5} {\bibfield  {journal} {\bibinfo
  {journal} {Bulletin of Mathematical Biology}\ }\textbf {\bibinfo {volume}
  {70}},\ \bibinfo {pages} {1570--1607} (\bibinfo {year} {2008})}\BibitemShut
  {NoStop}%
\bibitem [{\citenamefont {Jäger}\ and\ \citenamefont
  {Luckhaus}(1992)}]{Jaeger1992}%
  \BibitemOpen
  \bibfield  {author} {\bibinfo {author} {\bibfnamefont {W.}~\bibnamefont
  {Jäger}}\ and\ \bibinfo {author} {\bibfnamefont {S.}~\bibnamefont
  {Luckhaus}},\ }\bibfield  {title} {\enquote {\bibinfo {title} {On explosions
  of solutions to a system of partial differential equations modelling
  chemotaxis},}\ }\href {\doibase 10.1090/s0002-9947-1992-1046835-6} {\bibfield
   {journal} {\bibinfo  {journal} {Transactions of the American Mathematical
  Society}\ }\textbf {\bibinfo {volume} {329}},\ \bibinfo {pages} {819--824}
  (\bibinfo {year} {1992})}\BibitemShut {NoStop}%
\bibitem [{\citenamefont {Herrero}\ and\ \citenamefont
  {Vel{\'a}zquez}(1997)}]{Herrero1997}%
  \BibitemOpen
  \bibfield  {author} {\bibinfo {author} {\bibfnamefont {M.~A.}\ \bibnamefont
  {Herrero}}\ and\ \bibinfo {author} {\bibfnamefont {J.~J.~L.}\ \bibnamefont
  {Vel{\'a}zquez}},\ }\bibfield  {title} {\enquote {\bibinfo {title} {A blow-up
  mechanism for a chemotaxis model},}\ }\href
  {http://www.numdam.org/item/ASNSP_1997_4_24_4_633_0/} {\bibfield  {journal}
  {\bibinfo  {journal} {Annali Della Scuola Normale Superiore Di Pisa-classe Di
  Scienze}\ }\textbf {\bibinfo {volume} {24}},\ \bibinfo {pages} {633--683}
  (\bibinfo {year} {1997})}\BibitemShut {NoStop}%
\bibitem [{\citenamefont {Tyson}\ \emph {et~al.}(1999)\citenamefont {Tyson},
  \citenamefont {Lubkin},\ and\ \citenamefont {Murray}}]{Tyson1999}%
  \BibitemOpen
  \bibfield  {author} {\bibinfo {author} {\bibfnamefont {R.}~\bibnamefont
  {Tyson}}, \bibinfo {author} {\bibfnamefont {S.~R.}\ \bibnamefont {Lubkin}}, \
  and\ \bibinfo {author} {\bibfnamefont {J.~D.}\ \bibnamefont {Murray}},\
  }\bibfield  {title} {\enquote {\bibinfo {title} {Model and analysis of
  chemotactic bacterial patterns in a liquid medium},}\ }\href {\doibase
  10.1007/s002850050153} {\bibfield  {journal} {\bibinfo  {journal} {Journal of
  Mathematical Biology}\ }\textbf {\bibinfo {volume} {38}},\ \bibinfo {pages}
  {359--375} (\bibinfo {year} {1999})}\BibitemShut {NoStop}%
\bibitem [{\citenamefont {Tello}\ and\ \citenamefont
  {Winkler}(2007)}]{Tello2007}%
  \BibitemOpen
  \bibfield  {author} {\bibinfo {author} {\bibfnamefont {J.~I.}\ \bibnamefont
  {Tello}}\ and\ \bibinfo {author} {\bibfnamefont {M.}~\bibnamefont
  {Winkler}},\ }\bibfield  {title} {\enquote {\bibinfo {title} {A chemotaxis
  system with logistic source},}\ }\href {\doibase 10.1080/03605300701319003}
  {\bibfield  {journal} {\bibinfo  {journal} {Communications in Partial
  Differential Equations}\ }\textbf {\bibinfo {volume} {32}},\ \bibinfo {pages}
  {849--877} (\bibinfo {year} {2007})}\BibitemShut {NoStop}%
\bibitem [{\citenamefont {Jin}\ \emph {et~al.}(2016)\citenamefont {Jin},
  \citenamefont {Wang},\ and\ \citenamefont {Zhang}}]{Jin2016}%
  \BibitemOpen
  \bibfield  {author} {\bibinfo {author} {\bibfnamefont {L.}~\bibnamefont
  {Jin}}, \bibinfo {author} {\bibfnamefont {Q.}~\bibnamefont {Wang}}, \ and\
  \bibinfo {author} {\bibfnamefont {Z.}~\bibnamefont {Zhang}},\ }\bibfield
  {title} {\enquote {\bibinfo {title} {Pattern formation in
  {K}eller{\textendash}{S}egel chemotaxis models with logistic growth},}\
  }\href {https://www.worldscientific.com/doi/abs/10.1142/S0218127416500334}
  {\bibfield  {journal} {\bibinfo  {journal} {International Journal of
  Bifurcation and Chaos}\ }\textbf {\bibinfo {volume} {26}},\ \bibinfo {pages}
  {1650033} (\bibinfo {year} {2016})}\BibitemShut {NoStop}%
\bibitem [{\citenamefont {Chavanis}(2008)}]{Chavanis2008}%
  \BibitemOpen
  \bibfield  {author} {\bibinfo {author} {\bibfnamefont {P.~H.}\ \bibnamefont
  {Chavanis}},\ }\bibfield  {title} {\enquote {\bibinfo {title} {A stochastic
  {Keller}-{Segel} model of chemotaxis},}\ }\href
  {https://doi.org/10.1016/j.cnsns.2008.09.002} {\bibfield  {journal} {\bibinfo
   {journal} {Communications in Nonlinear Science and Numerical Simulation}\
  }\textbf {\bibinfo {volume} {15}},\ \bibinfo {pages} {60--70} (\bibinfo
  {year} {2008})}\BibitemShut {NoStop}%
\bibitem [{\citenamefont {Newman}\ and\ \citenamefont
  {Grima}(2004)}]{Grima2004}%
  \BibitemOpen
  \bibfield  {author} {\bibinfo {author} {\bibfnamefont {T.~J.}\ \bibnamefont
  {Newman}}\ and\ \bibinfo {author} {\bibfnamefont {R.}~\bibnamefont {Grima}},\
  }\bibfield  {title} {\enquote {\bibinfo {title} {Many-body theory of
  chemotactic cell-cell interactions},}\ }\href {\doibase
  10.1103/PhysRevE.70.051916} {\bibfield  {journal} {\bibinfo  {journal} {Phys.
  Rev. E}\ }\textbf {\bibinfo {volume} {70}},\ \bibinfo {pages} {051916}
  (\bibinfo {year} {2004})}\BibitemShut {NoStop}%
\bibitem [{\citenamefont {Gelimson}\ and\ \citenamefont
  {R.Golestanian}(2015)}]{Gelimson2015}%
  \BibitemOpen
  \bibfield  {author} {\bibinfo {author} {\bibfnamefont {A.}~\bibnamefont
  {Gelimson}}\ and\ \bibinfo {author} {\bibnamefont {R.Golestanian}},\
  }\bibfield  {title} {\enquote {\bibinfo {title} {Collective dynamics of
  dividing chemotactic cells},}\ }\href {\doibase
  10.1103/PhysRevLett.114.028101} {\bibfield  {journal} {\bibinfo  {journal}
  {Phys. Rev. Lett.}\ }\textbf {\bibinfo {volume} {114}},\ \bibinfo {pages}
  {028101} (\bibinfo {year} {2015})}\BibitemShut {NoStop}%
\bibitem [{\citenamefont {Mahdisoltani}\ \emph {et~al.}(2021)\citenamefont
  {Mahdisoltani}, \citenamefont {{Ben Al\`{\i} Zinati}}, \citenamefont
  {Duclut}, \citenamefont {Gambassi},\ and\ \citenamefont
  {Golestanian}}]{Mahdistoliani2021}%
  \BibitemOpen
  \bibfield  {author} {\bibinfo {author} {\bibfnamefont {S.}~\bibnamefont
  {Mahdisoltani}}, \bibinfo {author} {\bibfnamefont {R.}~\bibnamefont {{Ben
  Al\`{\i} Zinati}}}, \bibinfo {author} {\bibfnamefont {C.}~\bibnamefont
  {Duclut}}, \bibinfo {author} {\bibfnamefont {A.}~\bibnamefont {Gambassi}}, \
  and\ \bibinfo {author} {\bibfnamefont {R.}~\bibnamefont {Golestanian}},\
  }\bibfield  {title} {\enquote {\bibinfo {title} {Nonequilibrium
  polarity-induced chemotaxis: Emergent {G}alilean symmetry and exact scaling
  exponents},}\ }\href {\doibase 10.1103/PhysRevResearch.3.013100} {\bibfield
  {journal} {\bibinfo  {journal} {Phys. Rev. Research}\ }\textbf {\bibinfo
  {volume} {3}},\ \bibinfo {pages} {013100} (\bibinfo {year}
  {2021})}\BibitemShut {NoStop}%
\bibitem [{\citenamefont {Täuber}(2014)}]{Tauber2014}%
  \BibitemOpen
  \bibfield  {author} {\bibinfo {author} {\bibfnamefont {U.~C.}\ \bibnamefont
  {Täuber}},\ }\href {\doibase 10.1017/CBO9781139046213} {\emph {\bibinfo
  {title} {Critical dynamics: A field theory approach to equilibrium and
  non-equilibrium scaling behavior}}}\ (\bibinfo  {publisher} {Cambridge
  University Press},\ \bibinfo {year} {2014})\BibitemShut {NoStop}%
\bibitem [{\citenamefont {Budrene}\ and\ \citenamefont
  {Berg}(1991)}]{Budrene1991}%
  \BibitemOpen
  \bibfield  {author} {\bibinfo {author} {\bibfnamefont {E.~O.}\ \bibnamefont
  {Budrene}}\ and\ \bibinfo {author} {\bibfnamefont {H.~C.}\ \bibnamefont
  {Berg}},\ }\bibfield  {title} {\enquote {\bibinfo {title} {Complex patterns
  formed by motile cells of {Escherichia} coli},}\ }\href {\doibase
  10.1038/349630a0} {\bibfield  {journal} {\bibinfo  {journal} {Nature}\
  }\textbf {\bibinfo {volume} {349}},\ \bibinfo {pages} {630--633} (\bibinfo
  {year} {1991})}\BibitemShut {NoStop}%
\bibitem [{\citenamefont {Tweedy}\ \emph {et~al.}(2016)\citenamefont {Tweedy},
  \citenamefont {Knecht}, \citenamefont {Mackay},\ and\ \citenamefont
  {Insall}}]{Tweedy2016}%
  \BibitemOpen
  \bibfield  {author} {\bibinfo {author} {\bibfnamefont {L.}~\bibnamefont
  {Tweedy}}, \bibinfo {author} {\bibfnamefont {D.A.}\ \bibnamefont {Knecht}},
  \bibinfo {author} {\bibfnamefont {G.M.}\ \bibnamefont {Mackay}}, \ and\
  \bibinfo {author} {\bibfnamefont {R.H.}\ \bibnamefont {Insall}},\ }\bibfield
  {title} {\enquote {\bibinfo {title} {Self-generated chemoattractant
  gradients: attractant depletion extends the range and robustness of
  chemotaxis},}\ }\href {\doibase 10.1371/journal.pbio.1002404} {\bibfield
  {journal} {\bibinfo  {journal} {{PLOS} Biology}\ }\textbf {\bibinfo {volume}
  {14}},\ \bibinfo {pages} {e1002404} (\bibinfo {year} {2016})}\BibitemShut
  {NoStop}%
\bibitem [{Note:SM()}]{Note:SM}%
  \BibitemOpen
  \bibinfo {note} {See Supplemental Material for detailed calculations and technical background information, which includes
  Refs.~\cite{gardiner2009,Doi1976,Peliti1985,Chavanis2008,Dean.1996,Kawasaki.1994,Caballero2018,Cavagna2023,kardar_2007,Jaeger1992,BauschWagnerJanssen1976,
  Janssen1976, deDominics1976, MartinSiggiaRose1973, Tauber2014, Tindall2008,
  Canet2010, Canet2021, Mesibov1973, Zinn-Justin2002, Janssen2005,
  LangtangenLogg2017, Logg2010}.}\BibitemShut {Stop}%
\bibitem [{\citenamefont {Fisher}(1937)}]{FISHER1937}%
  \BibitemOpen
  \bibfield  {author} {\bibinfo {author} {\bibfnamefont {R.~A.}\ \bibnamefont
  {Fisher}},\ }\bibfield  {title} {\enquote {\bibinfo {title} {The wave of
  advance of advantageous genes},}\ }\href {\doibase
  10.1111/j.1469-1809.1937.tb02153.x} {\bibfield  {journal} {\bibinfo
  {journal} {Annals of Eugenics}\ }\textbf {\bibinfo {volume} {7}},\ \bibinfo
  {pages} {355--369} (\bibinfo {year} {1937})}\BibitemShut {NoStop}%
\bibitem [{\citenamefont {A.Kolmogorov}\ \emph {et~al.}(1937)\citenamefont
  {A.Kolmogorov}, \citenamefont {I.Petrovsky},\ and\ \citenamefont
  {Piskunov}}]{Kolmogorov}%
  \BibitemOpen
  \bibfield  {author} {\bibinfo {author} {\bibnamefont {A.Kolmogorov}},
  \bibinfo {author} {\bibnamefont {I.Petrovsky}}, \ and\ \bibinfo {author}
  {\bibfnamefont {M.}~\bibnamefont {Piskunov}},\ }\bibfield  {title} {\enquote
  {\bibinfo {title} {A study of the diffusion equation with increase in the
  amount of substance, and its application to a biological problem.}}\
  }\href@noop {} {\bibfield  {journal} {\bibinfo  {journal} {Moscow Univ. Bull.
  Math}\ ,\ \bibinfo {pages} {1--26}} (\bibinfo {year} {1937})}\BibitemShut
  {NoStop}%
\bibitem [{\citenamefont {Segel}(1977)}]{Segel1977}%
  \BibitemOpen
  \bibfield  {author} {\bibinfo {author} {\bibfnamefont {L.~A.}\ \bibnamefont
  {Segel}},\ }\bibfield  {title} {\enquote {\bibinfo {title} {A theoretical
  study of receptor mechanisms in bacterial chemotaxis},}\ }\href {\doibase
  10.1137/0132054} {\bibfield  {journal} {\bibinfo  {journal} {{SIAM} Journal
  on Applied Mathematics}\ }\textbf {\bibinfo {volume} {32}},\ \bibinfo {pages}
  {653--665} (\bibinfo {year} {1977})}\BibitemShut {NoStop}%
\bibitem [{\citenamefont {Painter}\ and\ \citenamefont
  {Hillen}(2002)}]{Painter2002}%
  \BibitemOpen
  \bibfield  {author} {\bibinfo {author} {\bibfnamefont {K.~J.}\ \bibnamefont
  {Painter}}\ and\ \bibinfo {author} {\bibfnamefont {T.}~\bibnamefont
  {Hillen}},\ }\bibfield  {title} {\enquote {\bibinfo {title} {Volume-filling
  and quorum-sensing in models for chemosensitive movement},}\ }\href@noop {}
  {\bibfield  {journal} {\bibinfo  {journal} {Canadian Applied Mathematics
  Quarterly}\ }\textbf {\bibinfo {volume} {10}},\ \bibinfo {pages} {501--544}
  (\bibinfo {year} {2002})}\BibitemShut {NoStop}%
\bibitem [{\citenamefont {Turing}(1952)}]{Turing1952}%
  \BibitemOpen
  \bibfield  {author} {\bibinfo {author} {\bibfnamefont {A.}~\bibnamefont
  {Turing}},\ }\bibfield  {title} {\enquote {\bibinfo {title} {The chemical
  basis of morphogenesis},}\ }\href {\doibase 10.1098/rstb.1952.0012}
  {\bibfield  {journal} {\bibinfo  {journal} {Philosophical Transactions of the
  Royal Society of London. Series B, Biological Sciences}\ }\textbf {\bibinfo
  {volume} {237}},\ \bibinfo {pages} {37--72} (\bibinfo {year}
  {1952})}\BibitemShut {NoStop}%
\bibitem [{\citenamefont {Cavagna}\ \emph {et~al.}(2023)\citenamefont
  {Cavagna}, \citenamefont {Carlo}, \citenamefont {Giardina}, \citenamefont
  {Grigera}, \citenamefont {Melillo}, \citenamefont {Parisi}, \citenamefont
  {Pisegna},\ and\ \citenamefont {Scandolo}}]{Cavagna2023}%
  \BibitemOpen
  \bibfield  {author} {\bibinfo {author} {\bibfnamefont {A.}~\bibnamefont
  {Cavagna}}, \bibinfo {author} {\bibfnamefont {L.~Di}\ \bibnamefont {Carlo}},
  \bibinfo {author} {\bibfnamefont {I.}~\bibnamefont {Giardina}}, \bibinfo
  {author} {\bibfnamefont {T.S.}\ \bibnamefont {Grigera}}, \bibinfo {author}
  {\bibfnamefont {S.}~\bibnamefont {Melillo}}, \bibinfo {author} {\bibfnamefont
  {L.}~\bibnamefont {Parisi}}, \bibinfo {author} {\bibfnamefont
  {G.}~\bibnamefont {Pisegna}}, \ and\ \bibinfo {author} {\bibfnamefont
  {M.}~\bibnamefont {Scandolo}},\ }\bibfield  {title} {\enquote {\bibinfo
  {title} {Natural swarms in 3.99 dimensions},}\ }\href {\doibase
  10.1038/s41567-023-02028-0} {\bibfield  {journal} {\bibinfo  {journal}
  {Nature Physics}\ } (\bibinfo {year} {2023}),\
  10.1038/s41567-023-02028-0}\BibitemShut {NoStop}%
\bibitem [{\citenamefont {Lewus}\ and\ \citenamefont {Ford}(2001)}]{Lewus2001}%
  \BibitemOpen
  \bibfield  {author} {\bibinfo {author} {\bibfnamefont {P.}~\bibnamefont
  {Lewus}}\ and\ \bibinfo {author} {\bibfnamefont {R.~M.}\ \bibnamefont
  {Ford}},\ }\bibfield  {title} {\enquote {\bibinfo {title} {Quantification of
  random motility and chemotaxis bacterial transport coefficients using
  individual-cell and population-scale assays},}\ }\href {\doibase
  10.1002/bit.10021} {\bibfield  {journal} {\bibinfo  {journal} {Biotechnology
  and Bioengineering}\ }\textbf {\bibinfo {volume} {75}},\ \bibinfo {pages}
  {292--304} (\bibinfo {year} {2001})}\BibitemShut {NoStop}%
\bibitem [{\citenamefont {Murray}(2003)}]{Murray2003}%
  \BibitemOpen
  \bibfield  {author} {\bibinfo {author} {\bibfnamefont {J.~D.}\ \bibnamefont
  {Murray}},\ }\href {\doibase 10.1007/b98869} {\emph {\bibinfo {title}
  {Mathematical biology}}}\ (\bibinfo  {publisher} {Springer New York},\
  \bibinfo {year} {2003})\BibitemShut {NoStop}%
\bibitem [{\citenamefont {Frey}\ and\ \citenamefont
  {T\"auber}(1994)}]{Frey1994}%
  \BibitemOpen
  \bibfield  {author} {\bibinfo {author} {\bibfnamefont {E.}~\bibnamefont
  {Frey}}\ and\ \bibinfo {author} {\bibfnamefont {U.~C.}\ \bibnamefont
  {T\"auber}},\ }\bibfield  {title} {\enquote {\bibinfo {title} {Two-loop
  renormalization-group analysis of the {Burgers}-{Kardar}-{Parisi}-{Zhang}
  equation},}\ }\href {\doibase 10.1103/PhysRevE.50.1024} {\bibfield  {journal}
  {\bibinfo  {journal} {Phys. Rev. E}\ }\textbf {\bibinfo {volume} {50}},\
  \bibinfo {pages} {1024--1045} (\bibinfo {year} {1994})}\BibitemShut {NoStop}%
\bibitem [{\citenamefont {Janssen}\ \emph
  {et~al.}(1999{\natexlab{b}})\citenamefont {Janssen}, \citenamefont
  {T{\"a}uber},\ and\ \citenamefont {Frey}}]{Janssen1999}%
  \BibitemOpen
  \bibfield  {author} {\bibinfo {author} {\bibfnamefont {H.~K.}\ \bibnamefont
  {Janssen}}, \bibinfo {author} {\bibfnamefont {U.~C.}\ \bibnamefont
  {T{\"a}uber}}, \ and\ \bibinfo {author} {\bibfnamefont {E.}~\bibnamefont
  {Frey}},\ }\bibfield  {title} {\enquote {\bibinfo {title} {Exact results for
  the {Kardar}-{Parisi}-{Zhang} equation with spatially correlated noise},}\
  }\href {\doibase 10.1007/s100510050790} {\bibfield  {journal} {\bibinfo
  {journal} {The European Physical Journal B}\ }\textbf {\bibinfo {volume}
  {9}},\ \bibinfo {pages} {491--511} (\bibinfo {year}
  {1999}{\natexlab{b}})}\BibitemShut {NoStop}%
\bibitem [{\citenamefont {Canet}\ \emph {et~al.}(2010)\citenamefont {Canet},
  \citenamefont {Chat\'e}, \citenamefont {Delamotte},\ and\ \citenamefont
  {Wschebor}}]{Canet2010}%
  \BibitemOpen
  \bibfield  {author} {\bibinfo {author} {\bibfnamefont {L.}~\bibnamefont
  {Canet}}, \bibinfo {author} {\bibfnamefont {H.}~\bibnamefont {Chat\'e}},
  \bibinfo {author} {\bibfnamefont {B.}~\bibnamefont {Delamotte}}, \ and\
  \bibinfo {author} {\bibfnamefont {N.}~\bibnamefont {Wschebor}},\ }\bibfield
  {title} {\enquote {\bibinfo {title} {Nonperturbative renormalization group
  for the {K}ardar-{P}arisi-{Z}hang equation},}\ }\href {\doibase
  10.1103/PhysRevLett.104.150601} {\bibfield  {journal} {\bibinfo  {journal}
  {Phys. Rev. Lett.}\ }\textbf {\bibinfo {volume} {104}},\ \bibinfo {pages}
  {150601} (\bibinfo {year} {2010})}\BibitemShut {NoStop}%
\bibitem [{\citenamefont {Dupuis}\ \emph {et~al.}(2021)\citenamefont {Dupuis},
  \citenamefont {Canet}, \citenamefont {Eichhorn}, \citenamefont {Metzner},
  \citenamefont {Pawlowski}, \citenamefont {Tissier},\ and\ \citenamefont
  {Wschebor}}]{Canet2021}%
  \BibitemOpen
  \bibfield  {author} {\bibinfo {author} {\bibfnamefont {N.}~\bibnamefont
  {Dupuis}}, \bibinfo {author} {\bibfnamefont {L.}~\bibnamefont {Canet}},
  \bibinfo {author} {\bibfnamefont {A.}~\bibnamefont {Eichhorn}}, \bibinfo
  {author} {\bibfnamefont {W.}~\bibnamefont {Metzner}}, \bibinfo {author}
  {\bibfnamefont {J.~M.}\ \bibnamefont {Pawlowski}}, \bibinfo {author}
  {\bibfnamefont {M.}~\bibnamefont {Tissier}}, \ and\ \bibinfo {author}
  {\bibfnamefont {N.}~\bibnamefont {Wschebor}},\ }\bibfield  {title} {\enquote
  {\bibinfo {title} {The nonperturbative functional renormalization group and
  its applications},}\ }\href {\doibase
  https://doi.org/10.1016/j.physrep.2021.01.001} {\bibfield  {journal}
  {\bibinfo  {journal} {Physics Reports}\ }\textbf {\bibinfo {volume} {910}},\
  \bibinfo {pages} {1--114} (\bibinfo {year} {2021})}\BibitemShut {NoStop}%
\bibitem [{\citenamefont {Gardiner}(2009)}]{gardiner2009}%
  \BibitemOpen
  \bibfield  {author} {\bibinfo {author} {\bibfnamefont {C.}~\bibnamefont
  {Gardiner}},\ }\href {https://books.google.de/books?id=otg3PQAACAAJ} {\emph
  {\bibinfo {title} {Stochastic Methods: A Handbook for the Natural and Social
  Sciences}}},\ Springer Series in Synergetics\ (\bibinfo  {publisher}
  {Springer Berlin Heidelberg},\ \bibinfo {year} {2009})\BibitemShut {NoStop}%
\bibitem [{\citenamefont {Doi}(1976)}]{Doi1976}%
  \BibitemOpen
  \bibfield  {author} {\bibinfo {author} {\bibfnamefont {M}~\bibnamefont
  {Doi}},\ }\bibfield  {title} {\enquote {\bibinfo {title} {Second quantization
  representation for classical many-particle system},}\ }\href {\doibase
  10.1088/0305-4470/9/9/008} {\bibfield  {journal} {\bibinfo  {journal}
  {Journal of Physics A: Mathematical and General}\ }\textbf {\bibinfo {volume}
  {9}},\ \bibinfo {pages} {1465} (\bibinfo {year} {1976})}\BibitemShut
  {NoStop}%
\bibitem [{\citenamefont {{Peliti, L.}}(1985)}]{Peliti1985}%
  \BibitemOpen
  \bibfield  {author} {\bibinfo {author} {\bibnamefont {{Peliti, L.}}},\
  }\bibfield  {title} {\enquote {\bibinfo {title} {Path integral approach to
  birth-death processes on a lattice},}\ }\href {\doibase
  10.1051/jphys:019850046090146900} {\bibfield  {journal} {\bibinfo  {journal}
  {J. Phys. France}\ }\textbf {\bibinfo {volume} {46}},\ \bibinfo {pages}
  {1469--1483} (\bibinfo {year} {1985})}\BibitemShut {NoStop}%
\bibitem [{\citenamefont {Dean}(1996)}]{Dean.1996}%
  \BibitemOpen
  \bibfield  {author} {\bibinfo {author} {\bibfnamefont {D.~S.}\ \bibnamefont
  {Dean}},\ }\bibfield  {title} {\enquote {\bibinfo {title} {Langevin equation
  for the density of a system of interacting langevin processes},}\ }\href
  {\doibase 10.1088/0305-4470/29/24/001} {\bibfield  {journal} {\bibinfo
  {journal} {Journal of Physics A: Mathematical and General}\ }\textbf
  {\bibinfo {volume} {29}},\ \bibinfo {pages} {L613--L617} (\bibinfo {year}
  {1996})}\BibitemShut {NoStop}%
\bibitem [{\citenamefont {Kawasaki}(1994)}]{Kawasaki.1994}%
  \BibitemOpen
  \bibfield  {author} {\bibinfo {author} {\bibfnamefont {K.}~\bibnamefont
  {Kawasaki}},\ }\bibfield  {title} {\enquote {\bibinfo {title} {Stochastic
  model of slow dynamics in supercooled liquids and dense colloidal
  suspensions},}\ }\href {\doibase 10.1016/0378-4371(94)90533-9} {\bibfield
  {journal} {\bibinfo  {journal} {Physica A: Statistical Mechanics and its
  Applications}\ }\textbf {\bibinfo {volume} {208}},\ \bibinfo {pages} {35--64}
  (\bibinfo {year} {1994})}\BibitemShut {NoStop}%
\bibitem [{\citenamefont {Caballero}\ \emph
  {et~al.}(2018{\natexlab{b}})\citenamefont {Caballero}, \citenamefont
  {Nardini},\ and\ \citenamefont {Cates}}]{Caballero2018}%
  \BibitemOpen
  \bibfield  {author} {\bibinfo {author} {\bibfnamefont {F.}~\bibnamefont
  {Caballero}}, \bibinfo {author} {\bibfnamefont {C.}~\bibnamefont {Nardini}},
  \ and\ \bibinfo {author} {\bibfnamefont {M.~E.}\ \bibnamefont {Cates}},\
  }\bibfield  {title} {\enquote {\bibinfo {title} {From bulk to microphase
  separation in scalar active matter: a perturbative renormalization group
  analysis},}\ }\href {\doibase 10.1088/1742-5468/aaf321} {\bibfield  {journal}
  {\bibinfo  {journal} {Journal of Statistical Mechanics: Theory and
  Experiment}\ }\textbf {\bibinfo {volume} {2018}},\ \bibinfo {pages} {123208}
  (\bibinfo {year} {2018}{\natexlab{b}})}\BibitemShut {NoStop}%
\bibitem [{\citenamefont {Kardar}(2007)}]{kardar_2007}%
  \BibitemOpen
  \bibfield  {author} {\bibinfo {author} {\bibfnamefont {M.}~\bibnamefont
  {Kardar}},\ }\href {\doibase 10.1017/CBO9780511815881} {\emph {\bibinfo
  {title} {Statistical Physics of Fields}}}\ (\bibinfo  {publisher} {Cambridge
  University Press},\ \bibinfo {year} {2007})\BibitemShut {NoStop}%
\bibitem [{\citenamefont {Bausch}\ \emph {et~al.}(1976)\citenamefont {Bausch},
  \citenamefont {Janssen},\ and\ \citenamefont
  {Wagner}}]{BauschWagnerJanssen1976}%
  \BibitemOpen
  \bibfield  {author} {\bibinfo {author} {\bibfnamefont {R.}~\bibnamefont
  {Bausch}}, \bibinfo {author} {\bibfnamefont {H.~K.}\ \bibnamefont {Janssen}},
  \ and\ \bibinfo {author} {\bibfnamefont {H.}~\bibnamefont {Wagner}},\
  }\bibfield  {title} {\enquote {\bibinfo {title} {Renormalized field theory of
  critical dynamics},}\ }\href {\doibase 10.1007/BF01312880} {\bibfield
  {journal} {\bibinfo  {journal} {Zeitschrift für Physik B Condensed Matter}\
  }\textbf {\bibinfo {volume} {24}},\ \bibinfo {pages} {113--127} (\bibinfo
  {year} {1976})}\BibitemShut {NoStop}%
\bibitem [{\citenamefont {Janssen}(1976)}]{Janssen1976}%
  \BibitemOpen
  \bibfield  {author} {\bibinfo {author} {\bibfnamefont {H.~K.}\ \bibnamefont
  {Janssen}},\ }\bibfield  {title} {\enquote {\bibinfo {title} {On a
  {Lagrangean} for classical field dynamics and renormalization group
  calculations of dynamical critical properties},}\ }\href {\doibase
  10.1007/BF01316547} {\bibfield  {journal} {\bibinfo  {journal} {Zeitschrift
  für Physik B Condensed Matter}\ }\textbf {\bibinfo {volume} {23}},\ \bibinfo
  {pages} {377--380} (\bibinfo {year} {1976})}\BibitemShut {NoStop}%
\bibitem [{\citenamefont {de~Dominicis}(1976)}]{deDominics1976}%
  \BibitemOpen
  \bibfield  {author} {\bibinfo {author} {\bibfnamefont {C.}~\bibnamefont
  {de~Dominicis}},\ }\bibfield  {title} {\enquote {\bibinfo {title} {Techniques
  de renormalisation de la theorie des champs et dynamique des phenomenes
  critques},}\ }\href {https://doi.org/10.1051/jphyscol:1976138} {\bibfield
  {journal} {\bibinfo  {journal} {Journal de Physique Colloques}\ }\textbf
  {\bibinfo {volume} {37}},\ \bibinfo {pages} {247--253} (\bibinfo {year}
  {1976})}\BibitemShut {NoStop}%
\bibitem [{\citenamefont {Martin}\ \emph {et~al.}(1973)\citenamefont {Martin},
  \citenamefont {Siggia},\ and\ \citenamefont {Rose}}]{MartinSiggiaRose1973}%
  \BibitemOpen
  \bibfield  {author} {\bibinfo {author} {\bibfnamefont {P.~C.}\ \bibnamefont
  {Martin}}, \bibinfo {author} {\bibfnamefont {E.~D.}\ \bibnamefont {Siggia}},
  \ and\ \bibinfo {author} {\bibfnamefont {H.~A.}\ \bibnamefont {Rose}},\
  }\bibfield  {title} {\enquote {\bibinfo {title} {Statistical dynamics of
  classical systems},}\ }\href {\doibase 10.1103/PhysRevA.8.423} {\bibfield
  {journal} {\bibinfo  {journal} {Phys. Rev. A}\ }\textbf {\bibinfo {volume}
  {8}},\ \bibinfo {pages} {423--437} (\bibinfo {year} {1973})}\BibitemShut
  {NoStop}%
\bibitem [{\citenamefont {Mesibov}\ \emph {et~al.}(1973)\citenamefont
  {Mesibov}, \citenamefont {Ordal},\ and\ \citenamefont {Adler}}]{Mesibov1973}%
  \BibitemOpen
  \bibfield  {author} {\bibinfo {author} {\bibfnamefont {R.}~\bibnamefont
  {Mesibov}}, \bibinfo {author} {\bibfnamefont {G.~W.}\ \bibnamefont {Ordal}},
  \ and\ \bibinfo {author} {\bibfnamefont {J.}~\bibnamefont {Adler}},\
  }\bibfield  {title} {\enquote {\bibinfo {title} {The range of attractant
  concentrations for bacterial chemotaxis and the threshold and size of
  response over this range},}\ }\href {\doibase 10.1085/jgp.62.2.203}
  {\bibfield  {journal} {\bibinfo  {journal} {The Journal of General
  Physiology}\ }\textbf {\bibinfo {volume} {62}},\ \bibinfo {pages} {203--223}
  (\bibinfo {year} {1973})}\BibitemShut {NoStop}%
\bibitem [{\citenamefont {Zinn-Justin}(2002)}]{Zinn-Justin2002}%
  \BibitemOpen
  \bibfield  {author} {\bibinfo {author} {\bibfnamefont {J.}~\bibnamefont
  {Zinn-Justin}},\ }\href {\doibase 10.1093/acprof:oso/9780198509233.001.0001}
  {\emph {\bibinfo {title} {Quantum field theory and critical phenomena}}}\
  (\bibinfo  {publisher} {Oxford University Press},\ \bibinfo {year}
  {2002})\BibitemShut {NoStop}%
\bibitem [{\citenamefont {Janssen}(2005)}]{Janssen2005}%
  \BibitemOpen
  \bibfield  {author} {\bibinfo {author} {\bibfnamefont {H.~K.}\ \bibnamefont
  {Janssen}},\ }\bibfield  {title} {\enquote {\bibinfo {title} {Survival and
  percolation probabilities in the field theory of growth models},}\ }\href
  {\doibase 10.1088/0953-8984/17/20/021} {\bibfield  {journal} {\bibinfo
  {journal} {Journal of Physics: Condensed Matter}\ }\textbf {\bibinfo {volume}
  {17}},\ \bibinfo {pages} {S1973--S1993} (\bibinfo {year} {2005})}\BibitemShut
  {NoStop}%
\bibitem [{\citenamefont {Langtangen}\ and\ \citenamefont
  {Logg}(2017)}]{LangtangenLogg2017}%
  \BibitemOpen
  \bibfield  {author} {\bibinfo {author} {\bibfnamefont {H.~P.}\ \bibnamefont
  {Langtangen}}\ and\ \bibinfo {author} {\bibfnamefont {A.}~\bibnamefont
  {Logg}},\ }\href {\doibase 10.1007/978-3-319-52462-7} {\emph {\bibinfo
  {title} {Solving {PDE}s in {Python}}}}\ (\bibinfo  {publisher} {Springer},\
  \bibinfo {year} {2017})\BibitemShut {NoStop}%
\bibitem [{\citenamefont {Logg}\ and\ \citenamefont {Wells}(2010)}]{Logg2010}%
  \BibitemOpen
  \bibfield  {author} {\bibinfo {author} {\bibfnamefont {A.}~\bibnamefont
  {Logg}}\ and\ \bibinfo {author} {\bibfnamefont {G.~N.}\ \bibnamefont
  {Wells}},\ }\bibfield  {title} {\enquote {\bibinfo {title} {{DOLFIN}:
  Automated finite element computing},}\ }\href
  {https://doi.org/10.1145/1731022.1731030} {\bibfield  {journal} {\bibinfo
  {journal} {ACM Trans. Math. Softw.}\ }\textbf {\bibinfo {volume} {37}}
  (\bibinfo {year} {2010})}\BibitemShut {NoStop}%
\end{thebibliography}
\end{document}